\documentclass[]{pasj01}

\usepackage{url}
\newenvironment{contribution}{\section*{Author contributions}\fontsize{8}{11}\selectfont}{\par}

\usepackage{natbib}
\usepackage{rotating} 
\usepackage[flushleft]{threeparttable} 
\usepackage{lscape}

\begin{document}
\Received{2017/07/29}
\Accepted{2017/12/13}

\title{Atomic data and spectral modeling constraints from high-resolution
X-ray observations of the Perseus cluster with Hitomi
\email{sawada@phys.aoyama.ac.jp, L.Gu@sron.nl}
\thanks{Corresponding authors are
Makoto~Sawada,
Liyi~Gu,
Jelle~Kaastra,
Randall~K.~Smith,
Adam~R.~Foster,
Greg~V.~Brown,
Hirokazu~Odaka,
Hiroki~Akamatsu,
and Takayuki~Hayashi.
}
}

\author{Hitomi Collaboration,
Felix \textsc{Aharonian}\altaffilmark{1,2,3},
Hiroki \textsc{Akamatsu}\altaffilmark{4},
Fumie \textsc{Akimoto}\altaffilmark{5},
Steven W. \textsc{Allen}\altaffilmark{6,7,8},
Lorella \textsc{Angelini}\altaffilmark{9},
Marc \textsc{Audard}\altaffilmark{10},
Hisamitsu \textsc{Awaki}\altaffilmark{11},
Magnus \textsc{Axelsson}\altaffilmark{12},
Aya \textsc{Bamba}\altaffilmark{13,14},
Marshall W. \textsc{Bautz}\altaffilmark{15}
% ...
}

\altaffiltext{1}{Dublin Institute for Advanced Studies, 31 Fitzwilliam Place, Dublin 2, Ireland}
\altaffiltext{2}{Max-Planck-Institut f{\"u}r Kernphysik, P.O. Box 103980, 69029 Heidelberg, Germany}
\altaffiltext{3}{Gran Sasso Science Institute, viale Francesco Crispi, 7 67100 L'Aquila (AQ), Italy}
\altaffiltext{4}{SRON Netherlands Institute for Space Research, Sorbonnelaan 2, 3584 CA Utrecht, The Netherlands}
\altaffiltext{5}{Institute for Space-Earth Environmental Research, Nagoya University, Furo-cho, Chikusa-ku, Nagoya, Aichi 464-8601}
\altaffiltext{6}{Kavli Institute for Particle Astrophysics and Cosmology, Stanford University, 452 Lomita Mall, Stanford, CA 94305, USA}
\altaffiltext{7}{Department of Physics, Stanford University, 382 Via Pueblo Mall, Stanford, CA 94305, USA}
\altaffiltext{8}{SLAC National Accelerator Laboratory, 2575 Sand Hill Road, Menlo Park, CA 94025, USA}
\altaffiltext{9}{NASA, Goddard Space Flight Center, 8800 Greenbelt Road, Greenbelt, MD 20771, USA}
\altaffiltext{10}{Department of Astronomy, University of Geneva, ch. d'\'Ecogia 16, CH-1290 Versoix, Switzerland}
\altaffiltext{11}{Department of Physics, Ehime University, Bunkyo-cho, Matsuyama, Ehime 790-8577}
\altaffiltext{12}{Department of Physics and Oskar Klein Center, Stockholm University, 106 91 Stockholm, Sweden}
\altaffiltext{13}{Department of Physics, The University of Tokyo, 7-3-1 Hongo, Bunkyo-ku, Tokyo 113-0033}
\altaffiltext{14}{Research Center for the Early Universe, School of Science, The University of Tokyo, 7-3-1 Hongo, Bunkyo-ku, Tokyo 113-0033}
\altaffiltext{15}{Kavli Institute for Astrophysics and Space Research, Massachusetts Institute of Technology, 77 Massachusetts Avenue, Cambridge, MA 02139, USA}
% ...


\maketitle

\begin{abstract}
The Hitomi SXS spectrum of the Perseus cluster, with $\sim$5~eV resolution in the 2--9~keV band,
offers an unprecedented benchmark of the atomic modeling and database for hot collisional plasmas.
It reveals both successes and challenges of the current atomic codes. The latest versions of
AtomDB/APEC (3.0.8), SPEX (3.03.00), and CHIANTI (8.0) all provide reasonable fits to
the broad-band spectrum, and are in close agreement on best-fit temperature, emission measure,
and abundances of a few elements such as Ni.
For the Fe abundance, the APEC and SPEX measurements differ by 16\%, which is 17 times higher
than the statistical uncertainty. This is mostly attributed to the differences in adopted collisional
excitation and dielectronic recombination rates of the strongest emission lines. We further investigate
and compare the sensitivity of the derived physical parameters to the astrophysical source modeling and
instrumental effects. The Hitomi results show that an accurate atomic code is as important as the
astrophysical modeling and instrumental calibration aspects. Substantial updates of atomic databases and targeted
laboratory measurements are needed to get the current codes ready for the data from the next Hitomi-level mission.
\end{abstract}

\section{Introduction}

Many major achievements in X-ray studies of clusters of galaxies were made possible by
the advent of new X-ray spectroscopic instruments. The proportional counters on the Ariel~V
mission (spectral resolving power $R\equiv E/\Delta E\sim$ 6) revealed the highly ionized
Fe line emission near 7~keV in the Perseus cluster \citep{mitchell1976},
establishing the thermal origin of cluster X-rays.
The CCDs ($R=$ 10--60) onboard the ASCA satellite further identified line emission from
O, Ne, Mg, Si, S, Ar, Ca, and Ni in the hot intracluster medium \citep[ICM:][]{fukazawa1994,mushotzky1996}.
The Reflection Grating Spectrometer (RGS: $R=$ 50--100 for spatially extended sources) onboard XMM-Newton
\citep{peterson2001,tamura2001,kaastra2001} discovered the lack of strong cooling flows in
cool-core clusters. Most recently, the Soft X-ray Spectrometer \citep[SXS:][]{kelley2016}
onboard the Hitomi satellite \citep{takahashi2016} disclosed the low
energy density of turbulent motions in the central region of the Perseus cluster
with the resolving power of $R\sim$ 1250 \citep{hitomi2016}.
Each iteration of higher resolution spectroscopy enhances our understanding of clusters and
other cosmic objects.

As more high-resolution X-ray spectra become available, the X-ray community ---
including observers, theoreticians, and laboratory scientists --- urgently needs
accurate and complete atomic data and plasma models. As a first step in achieving this,
we will compare the current data and models (collectively called ``codes'' hereafter).
The most used plasma codes in X-ray
astronomy are AtomDB/APEC \citep{smith2001,foster2012}, SPEX \citep{kaastra1996}, and CHIANTI \citep{dere1997,delzanna2015}.
The AtomDB code descends from the original work of \citet{raymond1977}, SPEX
started with \citet{mewe1972}, and CHIANTI started with \citet{landini1970}. All
these codes have evolved significantly since their initial beginnings, often
stimulated by the challenges imposed by new generations of instruments.
It is clear that the code comparison is strongly needed to verify the scientific output and
to understand systematic uncertainties in the results originating from the codes and atomic databases.
However, few code comparisons have been done \citep[e.g.,][]{audard2003},
and in particular, so far there is no comparison based on high-resolution X-ray spectra of galaxy clusters.

The Hitomi X-ray observatory was launched on February 17, 2016. Among the main
scientific instruments, the SXS has an unprecedented resolving power of $R\sim$1250 at 6~keV
over a 6$\times$6 pixel array (\timeform{3'}$\times$\timeform{3'}).
It has a near-Gaussian energy response with FWHM$=$4--6~eV over the 0.3--12~keV band \citep{leutenegger2017}.
The X-ray mirror has an angular resolution with a half-power
diameter of \timeform{1.2'} \citep{maeda2017}. A gate valve was in place for early observations to
minimize the risk of contamination from out-gassing of the spacecraft \citep{tsujimoto2016}, which
includes a Be window that absorbs most X-rays below $\sim$2~keV. As the SXS is
a non-dispersive instrument (unlike gratings) it can be used to observe extended
objects without a loss of spectral resolution. This makes the SXS the
best instrument for high-resolution spectroscopic studies of galaxy clusters.
The Perseus cluster was observed as the first-light target of the SXS, and the
first paper showing its spectroscopic capabilities focused on the turbulence
in the Perseus cluster \citep{hitomi2016}.

With these data, we can also measure abundances (\citealt{hitomi-z}: hereafter Z~paper),
temperature structure (\citealt{hitomi-t}: T~paper), and resonance scattering (\citealt{hitomi-res}: RS~paper).
These quantities are essential to understand the origin and evolution of galaxy clusters 
(see review by \citealt{bohringer2010}).
Metal abundances trace products of billions of supernovae explosions integrated over cosmic time and
the measurements are crucial for understanding chemical evolution of ICM as well as the
evolutions and explosions of progenitor stars \citep{werner2008}.
Temperature structure or anisothermality gives an insight into thermodynamics in ICM
and thus important for understanding of the heating mechanism against effective radiative cooling in a dense core region 
\citep{peterson2006}.
Resonance scattering is another, indirect tool to assess turbulence, one of the candidate mechanisms of the ICM heating.
Required precisions to these quantities depend on astrophysical objectives --- for the cosmic star-formation history
the Ni-to-Fe abundance ratio needs to be measured to $\approx$10\% and for detection of
resonance scattering with the Fe He$\alpha$ complex the forbidden-to-resonance (z-to-w) line ratio to a few percent,
for instance (see individual topical papers for details).

In this paper we focus on the atomic physics and modeling aspects of the Perseus spectrum with the Hitomi SXS.
We show that this high-resolution spectrum offers a sensitive
probe of several important aspects of cluster physics including turbulence,
elemental abundance measurements, and structures in temperature and velocity (section~\ref{sect:baseline}).
We investigate the sensitivity of the related derived physical parameters to various aspects
of the spectroscopic codes (section~\ref{sect:plasmacode})
and their underlying atomic data (section~\ref{sect:atomicdata}),
spectral (section~\ref{sect:plasmamodel}) and astrophysical (sections~\ref{sect:astromodel} and \ref{sect:components}) modelings,
as well as fitting techniques (section~\ref{sect:fittingtech}).
By consolidating these systematic factors and by comparing them to statistical uncertainties as well as
the systematic factors due to instrumental calibration effects (appendix~\ref{app:instrumental}),
we can evaluate with what precisions the important quantities can be determined.
This allows us to be optimally prepared for future high-resolution X-ray missions.
We highlight the relative changes to each parameter by using different atomic modelings and so on,
rather than the changes in fitting statistics, since the former is more fundamental for understanding
the systematic uncertainties in the scientific results.
The astrophysical interpretation of our derived parameters is not discussed in
this paper, but will be in a series of separate papers focusing in greater detail on
the relevant astrophysics,
e.g., abundances (Z~paper), temperature structure (T~paper),
resonance scattering (RS~paper), velocity structure (\citealt{hitomi-v}: V~paper),
and the central active galactic nucleus (AGN) of the Perseus cluster (\citealt{hitomi-agn}: AGN~paper).
Also, we do not examine combined effects of different types of
systematic factors (e.g., plasma-code dependence in the detailed
astrophysical modeling like multi-temperature models), which will be
separately discussed in the individual topical papers.

\section{Data reduction\label{sect:reduction}}

In this paper, the cleaned event data in the pipeline products version 03.01.005.005
are analyzed with the Hitomi software version 005a and the calibration database (CALDB) version 005 \citep{angelini2016}\footnote{For the SXS pipeline products and CALDB, these are identical to the latest ones (03.01.006.007 and 007, respectively).}.
There are four Hitomi observations of the Perseus cluster
(name: sequence number $=$ ``Obs\,1'': 100040010, ``Obs\,2'': 100040020,
``Obs\,3'': 100040030--100040050, and ``Obs\,4'': 100040060).
The instrument had nearly reached thermal equilibrium by Obs\,4 \citep{fujimoto2016},
and the calibrations of Obs\,2 and Obs\,3 can be checked against Obs\,4
because of their overlapping fields of view (FOVs),
but the FOV of Obs\,1 does not overlap the others and the instrument was the most out of
equilibrium during that pointing. Hence only the Obs\,2, 3, and 4 are used in this work.

Events registered during low-Earth elevation angles below two degrees and passages of the South Atlantic Anomaly
were already excluded by the pipeline processing which created the cleaned events file.
Events coincident with the particle veto had also already been rejected.
Data were further screened by criteria described as ``recommended screening'' in the Hitomi data reduction
guide\footnote{See $\langle$https://heasarc.gsfc.nasa.gov/docs/hitomi/analysis/$\rangle$.}
to remove those with distorted pulse shapes or coincident events in any two pixels,
which further reduces the background, though the difference is
negligible given the surface brightness of the Perseus cluster.
For all the three observations (Obs\,2--4), only high-resolution primary events
(an event with no pulse in the interval 69.2 ms before or after it) were extracted and used.
This choice is fine because relative ratios are the same between different event types \citep{seta2012, ishisaki2016}.

The line broadening due to the spatial velocity gradient in the ICM is removed,
since it is not relevant to the atomic study. To do this, we apply
an additional energy scale correction \citep[also used in][]{hitomi2016, hitomi-3.5}, forcing the
strong Fe-K lines to appear at the same energy in each pixel, aligned to the same redshift as the central AGN
\citep[$z=$0.01756 or $cz=$5264~km\,s$^{-1}$: ][]{ferruit1997}.
This also removes residual gain errors in the Fe-K band.
The effect of the spatial velocity correction on the baseline-model fitting (section~\ref{sect:baseline}) is discussed in appendix~\ref{app:instrumental}.
A recent measurement of NGC\,1275 indicates an alternative redshift of 0.017284$\pm$0.00005 (V~paper). In this paper, we
do not refer to the new value, since its impact on the other fitting parameters would be washed out by
the non-linear energy-scale correction applied later (appendix~\ref{app:datacor}) or by the redshift component included in the baseline model (section~\ref{sect:baseline}).

The large-size redistribution matrix files (RMFs) for high-primary events created by \texttt{sxsmkrmf} are used
to take into account the main Gaussian component, the low-energy exponential tail, and escape peaks of the line spread function
\citep{leutenegger2017}.
We have also tested two different types of RMFs; one is the small-size RMFs which includes only the Gaussian core,
and the other is the extra-large--size RMFs with all components in the large-size RMFs plus electron-loss continuum.
The effect of changing the RMF type is discussed in appendix~\ref{app:instrumental}.
The ancillary response files (ARFs) are generated
separately for the diffuse emission and the point-source component. To enhance the precision of the diffuse
ARFs, a background-subtracted Chandra image of the Perseus cluster in the 1.8--9.0~keV band
whose AGN core is replaced with the average value of the surrounding regions is used to provide
the spatial distribution of seed photons. Since the effective area is estimated based on the input image
with a radius of \timeform{12'}, which is larger than the detector FOV (\timeform{3'}$\times$\timeform{3'}),
the measured spectral normalization reported in this paper is larger than the actual value.
We do not correct this effect since this paper is focused on the relative uncertainties instead of the absolute values.
We have further tested to use the point-source ARF for the both components, and show the effects in appendix~\ref{app:instrumental}.

The non X-ray background (NXB) of the SXS is much lower than those of the X-ray CCDs thanks to
the anti-coincidence screening, which reduces the NXB rate by a factor of $\approx$10 \citep{kilbourne2017}.
We extract the NXB spectrum from Earth occultation data with \texttt{sxsnxbgen},
and screened with the standard NXB criteria and the same additional screening as the source events.
The NXB spectrum is taken into account as a SPEX \textsl{file} model in the baseline analysis (section~\ref{sect:baseline}).
Other background components, which include the cosmic X-ray background and Galactic foreground emission,
are negligible for the Perseus data.
The relative changes of the baseline parameters for a fitting in absence of the NXB is shown in appendix~\ref{app:instrumental}.

The main remaining issue in the data analysis
is that the planned calibration procedures were not fully available for these early
observations. In particular, the contemporaneous calibration of the energy scale
(or gain) for the detector array was not yet carried out.
The previous Hitomi papers \citep{hitomi2016, hitomi-3.5} focused on a relatively narrow energy range;
in this work we study a wide energy band of 1.9--9.5~keV. This forces us to apply two additional
corrections to the energy scale and effective area as described in appendix~\ref{app:datacor}.

\section{Baseline model\label{sect:baseline}}

The result of a spectral model fit is a list of parameters representing the
source. These parameters depend on several factors, like the statistical quality
of the data, the instrument calibration, background subtraction method,
fitting techniques, spectral model components, physical processes included in the
spectral model, and atomic parameters. All of these factors contribute to the final set of
source parameters that is derived. Apart from the statistical uncertainties, all
other factors act like a kind of systematic uncertainty, and by carefully
analyzing each individual contribution we can assess its contribution to the
final uncertainty.

We proceed as follows. Below we define our baseline best-fit model and explain why
we incorporate each component in the model. We then list the best-fit
parameters with their statistical uncertainties. The effects of the different
systematic factors are in general not excessively large, and therefore  we list
their impact by showing by how much the best-fit parameters are increased or
decreased due to  these factors. Usually the statistical uncertainties on the
best-fit parameters are very similar for all investigated cases, so we only list
the statistical uncertainties of the baseline model.

We use the SPEX package \citep{kaastra1996} to define the baseline model because
it allows us to test the system in a straightforward way.
The version of SPEX that is being used here is 3.03.00.
It calculates all relevant rates, ion concentrations, level populations,
and line emissivities on the fly (see section~\ref{sect:spexv3.03} for more details).

We use optimally binned spectra (using the SPEX \textsl{obin}
command structure; see appendix~\ref{app:binning}) with C-statistics \citep{cash1979}.
This choice will be elaborated later (section~\ref{sect:fittingtech}).

All abundances are relative to the \citet{lodders2009} proto-solar abundances
with free values relative to those abundances for the relevant elements.

The dominant spectral component is a collisionally ionized plasma, with a
temperature of about 4~keV \citep{hitomi2016}, modeled with the SPEX
\textsl{cie} model. For the ionization balance we choose the
\citet{urdampilleta2017} ionization balance (for more detail see
section~\ref{sect:ibal}). The electron temperature,
abundances of Si, S, Ar, Ca, Cr, Mn, Fe, and Ni are free parameters;
the abundances of all other metals (usually with no or very weak lines in the bandpass of the Hitomi SXS)
are tied to the Fe abundance. In addition, we leave the turbulent velocity free;
the value of this turbulent velocity has been discussed in detail in \citet{hitomi2016}.
Although in SPEX the magnitude of turbulence is parameterized by a two-dimensional
root-mean-square velocity $v_{\rm mic}$ assuming isotropic velocity distribution,
we convert it into one-dimensional line-of-sight (LOS) velocity dispersion $\sigma_v$
($=v_{\rm mic}/\sqrt{2}$) and use it throughout this paper to enable direct comparisons
to the previous studies \citep{hitomi2016, hitomi-3.5}.

The Hitomi SXS spectrum of the Perseus cluster shows clear signatures of resonance
scattering (RS~paper); in addition, we may expect absorption of He-like line
emission by Li-like ions \citep{mehdipour2015}. To account for both effects, we
include the absorption from a CIE plasma as modeled by the SPEX \textsl{hot}
model to our model. The \textsl{hot} model calculates the continuum and line
absorption from a plasma with the temperature, chemical composition, turbulent
velocity and outflow velocity as free parameters. This absorption is applied to
all emission components from the cluster. Because the FOV of the Hitomi SXS is
relatively small compared to the size of the Perseus cluster, the effects of
resonance scattering to lowest order imply the removal of photons from the line
of sight towards the cluster core; we do not observe the re-emitted photons
further away from the nucleus. A more sophisticated resonance scattering model
is discussed by RS~paper. In order not to
over-constrain the model, we leave only the column density of the hot absorbing
gas $N_{\rm H,hot}$ free, and tie the other parameters (electron temperature, abundances,
turbulent and outflow velocities) to the values of the main 4-keV emission
component (but see section~\ref{sect:astromodel}).

Our spectrum also contains a contribution from the central AGN of NGC\,1275. This is
modeled by a powerlaw (SPEX component \textsl{pow}) plus two Gaussians (\textsl{gaus})
for the neutral Fe K$\alpha$ lines. We use the powerlaw model which has a 2--10-keV luminosity
of 2.4$\times$10$^{36}$~W or a flux of 3.5$\times$10$^{-14}$~W~m$^{-2}$,
almost one fifth of the total 2--10-keV luminosity of the observed field,
and a photon index of 1.91.
The Gaussian lines have rest-frame energies of 6.391~keV and 6.404~keV,
an intrinsic FWHM of 25~eV and a total luminosity of 5.6$\times$10$^{33}$~W
or a total flux of 8.0$\times$10$^{-17}$~W~m$^{-2}$.
We have kept the parameters of the central AGN frozen in our fits to the above values.
The above model and parameter values are from the initial evaluation for
AGN~paper, which have been updated later.
Updating the AGN spectrum modeling results in slightly different best-fit values
of the baseline model (section~\ref{sect:noagn}),
but the changes are so small that relative differences in the ICM parameters
due to other systematic factors are unchanged. Thus we use the original AGN model
and parameters throughout this paper except in section~\ref{sect:noagn}.

We apply further the cosmological redshift (SPEX \textsl{reds} component) to the
model,  but leave it as a free parameter for the baseline model to account for
any residual systematic energy scale corrections (either of instrumental or
astrophysical origin; this is not important for the present study).

The last spectral component applied to all spectra is another \textsl{hot}
component to account for the interstellar absorption from our Galaxy; we have
frozen the temperature to 0.5~eV (essentially a neutral plasma), with a column
density of 1.38$\times$10$^{21}$~cm$^{-2}$, following the argumentation in
\cite{hitomi-3.5}. The abundances are frozen to the proto-solar abundances
\citep{lodders2009}.

The model contains further a component of pure neutral Be and a correction factor for
the effective area (see appendix~\ref{app:fudge}); these serve purely as
instrumental effective area corrections and are kept frozen for our modeling.

To summarize, the baseline model starts with a thermal ICM and AGN components,
self-absorbed, redshifted, absorbed again by the foreground, and corrected for instrumental effects.
The free parameters of our model are then the emission measure $Y$ and temperature $kT$
of the hot gas, the turbulent velocity $\sigma_v$ of the hot gas,
the abundances of Si, S, Ar, Ca, Cr, Mn, Fe, and Ni,
the effective absorption column of the hot cluster gas $N_{\rm H,hot}$,
and the overall redshift of the system $z$.
This baseline model achieves a C-statistic value ($C_{\rm stat}$) of 4926 for an expected value of 4876$\pm$99.

\setlength{\tabcolsep}{2pt}
\begin{table*}
\caption{Parameters of the reference model and sensitivity to model assumptions.
The first two lines give the best-fit values with their 1$\sigma$ statistical uncertainty. The next lines show the parameter
differences of the tested models relative to the baseline model. Differences larger than 3$\sigma$ statistical uncertainty are emphasized in boldfaces.}
\label{tab:parameters}
\footnotesize
\begin{tabular}{lcccccccccccccc} 
\hline
Model & $C_{\rm stat}$ &$Y$\footnotemark[$*$]&$kT$\footnotemark[$*$]&$\sigma_v$\footnotemark[$*$]& \multicolumn{8}{c}{Abundance (solar)\footnotemark[$\dagger$]} &  $N_{\rm H,hot}$\footnotemark[$*$] & $cz$\footnotemark[$*$] \\ \cline{6-13}
      &                & (10$^{73}$~m$^{-3}$)& (keV)                &  (km~s$^{-1}$)                &Si & S    & Ar   & Ca   & Cr   & Mn    & Fe   & Ni             &  ($10^{24}$~m$^{-2}$)              & (km~s$^{-1}$)          \\
\hline
Baseline               & 4926.03\footnotemark[$\ddagger$] &  3.73 & 3.969 & 156 & 0.91 & 0.94 & 0.83 & 0.88 & 0.70 & 0.74 & 0.827 & 0.76 & 18.8 & 5264 \\
Stat. error                  &      -- &  0.01 & 0.017 &   3 & 0.05 & 0.03 & 0.04 & 0.04 & 0.10 & 0.15 & 0.008 & 0.05 &  1.3 &  2   \\
\hline\noalign{\smallskip}
\multicolumn{15}{l}{\textsl{Plasma codes (section~\ref{sect:plasmacode}):}} \\
\multicolumn{15}{l}{Old versions of SPEX} \\
% Model                    &   C-stat &              Y &                kT & sigv &      Si &       S &      Ar &      Ca &      Cr &      Mn &       Fe &      Ni & N_H & cz \\
$\ldots$ v2                &  1125.06 &           0.03 &      0.031 & \textbf{14} &$-$0.13 & \textbf{$-$0.14} &$-$0.05 &$-$0.08 & --   & --   & \textbf{$-$0.026} & 0.11 & $-$0.8 & $-$6 \\
$\ldots$ v3.00             &  2372.33 &\textbf{$-$0.08}& \textbf{0.263} & \textbf{12} & 0.03 & 0.09 & 0.10 & 0.06 & $-$0.11 & $-$0.12 & \textbf{$-$0.243} & \textbf{$-$0.28} & \textbf{$-$18.8} & $-$2 \\
\multicolumn{15}{l}{APEC/AtomDB} \\
$\ldots$ v3.0.2          & 670.06  & \textbf{0.07}  & $-$0.039 & \textbf{$-$13} & \textbf{$-$0.24} & \textbf{$-$0.21} & \textbf{$-$0.15} & \textbf{$-$0.13} & $-$0.24 & $-$0.39 & \textbf{$-$0.047} & \textbf{$-$0.17} & $-$2.7 &  1 \\
$\ldots$ v3.0.8          & 22.27    & 0.03 & \textbf{0.071} & \textbf{$-$16} & $-$0.10 & $-$0.07 & $-$0.05 & $-$0.07 & 0.01 & $-$0.05 & \textbf{$-$0.134} & $-$0.05 & \textbf{$-$7.6} & $-$6 \\
CHIANTI v8.0       & 327.44   & 0.01  & 0.002 & 4 & \textbf{$-$0.17} & \textbf{$-$0.12} & \textbf{0.14} & $-$0.08 & --   & --   &  0.011 & $-$0.04 & $-$1.8 & \textbf{8}  \\
Cloudy v13.04         & 21416.07 & \textbf{0.74} & \textbf{$-$0.370} & $-$7 & \textbf{$-$0.54} & \textbf{$-$0.52} & \textbf{$-$0.53} & \textbf{$-$0.46} & \textbf{$-$0.43} & $-$0.15 & \textbf{$-$0.399} & 0.14 & \textbf{$-$18.8} & \textbf{$-$8} \\
\hline\noalign{\smallskip}
\multicolumn{15}{l}{\textsl{Atomic data (section~\ref{sect:atomicdata}):}} \\
Fe\emissiontype{XXV} triplet & $-$10.68 &         0.00 &             0.003 &    1 &    0.00 &    0.00 &    0.00 &    0.00 &    0.00 &    0.00 & $-$0.007 &    0.00 & $-$0.4 & 0 \\
\multicolumn{15}{l}{ionization balance} \\
$\ldots$ AR85              &   104.80 & \textbf{0.13}  &             0.017 & $-$3 & $-$0.02 & $-$0.02 & $-$0.03 & $-$0.02 &    --   &    --   &    0.017 & $-$0.02 & 2.4 & 1 \\
$\ldots$ AR92              &    94.65 & \textbf{0.09}  &             0.021 & $-$4 & $-$0.02 & $-$0.02 & $-$0.03 & $-$0.02 &    --   &    --   &    0.021 & $-$0.03 & 2.0 & 0 \\
$\ldots$ B09              & $-$18.62 & \textbf{$-$0.13} &            0.003 & $-$2 &    0.00 &    0.01 &    0.00 &    0.00 & $-$0.01 & $-$0.01 & \textbf{0.029} & 0.01 & 1.1  & 0 \\
\hline\noalign{\smallskip}
\multicolumn{15}{l}{\textsl{Plasma modeling (section~\ref{sect:plasmamodel}):}} \\
Voigt profile              &  $-$8.28 &           0.01 &          $-$0.003 & $-$4 & $-$0.01 & $-$0.01 &    0.00 &    0.00 &    0.01 &    0.00 & $-$0.003 &    0.01 & $-$1.2 & 1 \\
$g_{\rm acc}$              &  $-$0.54 &        $-$0.01 &          $-$0.005 &    0 &    0.00 &    0.00 &    0.00 &    0.00 &    0.00 &    0.00 &    0.006 &    0.00 & $-$0.1 & 0 \\
$n_{\rm max}$              &    61.46 &        $-$0.01 &             0.006 & $-$1 &    0.02 &    0.04 &    0.02 &    0.01 & $-$0.01 & $-$0.03 &    0.023 &    0.00 &    1.0 & 0 \\
\hline\noalign{\smallskip}
\multicolumn{15}{l}{\textsl{Astrophysical modeling (section~\ref{sect:astromodel}):}} \\
$T_{\rm ion}$ free         & $-$0.02  &           0.00 &             0.000 & $-$1 &    0.00 &    0.00 &    0.00 &    0.00 &    0.00 & $-$0.01 &    0.000 &    0.00 & $-$0.1 & 0 \\
NEI effects\\
$\ldots$ $RT$  free        & $-$3.26  &        $-$0.01 &             0.026 & $-$1 & $-$0.02 & $-$0.02 & $-$0.01 & $-$0.01 & $-$0.01 & $-$0.02 &    0.001 & $-$0.01 &    0.7 & 0 \\
$\ldots$ Ionizing          & $-$5.46  &        $-$0.02 &             0.025 &    0 &    0.01 & $-$0.01 & $-$0.06 & $-$0.06 & $-$0.02 & $-$0.04 &    0.000 & $-$0.01 &    0.8 & 0 \\
$\ldots$ Recombining       & $-$9.19  &           0.02 &          $-$0.036 &    2 & $-$0.02 & $-$0.02 & $-$0.01 &    0.00 &    0.03 &    0.02 &    0.000 &    0.01 & $-$1.5 & 0 \\
$\sigma_{T}$ free      & $-$60.90 & \textbf{0.13}  & \textbf{$-$0.139} &    2 & $-$0.10 & \textbf{$-$0.10} & $-$0.04 & 0.01 & 0.08 & 0.10 & 0.024 & 0.03 & $-$2.3 & 0 \\
He abund.                  & $-$0.07  & \textbf{$-$0.08} &        $-$0.001 &    0 &    0.02 &    0.03 &    0.02 &    0.02 &    0.02 &    0.01 & \textbf{0.025} & 0.02 & $-$0.6 & 0 \\
\hline\noalign{\smallskip}
\multicolumn{15}{l}{\textsl{Spectral components (section~\ref{sect:components}):}} \\
No RS                      & 341.02   &  \textbf{0.05} &   $-$0.015 & \textbf{13} & $-$0.05 & $-$0.04 & $-$0.03 & $-$0.02 &    0.04 &    0.01 & \textbf{$-$0.094} & 0.01 & $\equiv$0 & 4 \\
Hot comp. free             &  $-$1.40 &           0.00 &             0.000 &    2 &    0.00 &    0.00 &    0.00 &    0.00 &    0.00 &    0.00 &    0.003 &    0.00 & 1.3 & 0 \\
CX                         & $-$13.34 &           0.00 &             0.018 & $-$3 & $-$0.02 & $-$0.01 &    0.00 & $-$0.01 & $-$0.01 & $-$0.02 & \textbf{$-$0.042} & 0.00 & $-$1.4 & $-$1 \\
AGN \\
$\ldots$ No AGN            &   624.54 &  \textbf{0.68} &    \textbf{0.523} &    4 & $-$0.01 & $-$0.05 & $-$0.09 & \textbf{$-$0.14} & $-$0.15 & $-$0.12 & \textbf{$-$0.206} & \textbf{$-$0.16} & \textbf{12.8}  & 3 \\
$\ldots$ New AGN           &     8.42 &  \textbf{0.18} &             0.028 &    0 & $-$0.03 & $-$0.03 & $-$0.03 & $-$0.04 & $-$0.04 & $-$0.04 & \textbf{$-$0.041} & $-$0.03 & 1.3  & 0 \\
\hline\noalign{\smallskip}
\multicolumn{15}{l}{\textsl{Fitting techniques (section~\ref{sect:fittingtech}):}} \\
$\chi^2$                   &    54.69 &        $-$0.01 &          $-$0.045 & $-$1 & $-$0.03 & $-$0.01 &    0.00 & $-$0.01 &    0.03 &    0.01 &    0.007 &    0.02 & $-$0.6 & 0 \\
$\chi^2$, no binning       &   ---    &        $-$0.01 & \textbf{$-$0.206} & $-$1 & $-$0.12 & $-$0.07 & $-$0.03 & $-$0.01 &    0.09 &    0.14 & \textbf{0.027} & 0.02 & $-$3.1 & 0 \\
\hline\noalign{\smallskip}
\multicolumn{15}{l}{\textsl{Instrumental effects (appendix~\ref{app:instrumental}):}} \\
No vel. cor.               &    61.70 &           0.00 &          0 & \textbf{13} &    0.00 &    0.00 &   0.00  &    0.00 & $-$0.02 & $-$0.02 &    0.001 &    0.01 & 1.0 & \textbf{$-$23} \\
Small RMF                  &  $-$4.42 &           0.01 &          $-$0.023 &    0 & $-$0.01 & $-$0.02 & $-$0.01 & $-$0.01 & $-$0.01 & $-$0.02 & $-$0.003 &    0.00 & $-$0.2 & 0 \\
XL RMF                     &    12.36 &        $-$0.02 &             0.035 &    0 &    0.00 &    0.03 &    0.02 &    0.02 &    0.02 &    0.01 &    0.010 &    0.00 & 0.1 & 0 \\
No NXB                     &     8.78 &           0.00 &             0.017 &    0 &    0.00 &    0.00 &    0.00 &    0.00 & $-$0.01 & $-$0.01 & $-$0.003 & $-$0.01 & 0.3 & 0 \\
ARF \\
$\ldots$ PS                &    29.54 &           0.02 & \textbf{$-$0.052} &    0 & $-$0.04 & $-$0.02 &    0.00 &    0.00 &    0.01 &    0.04 &    0.003 &    0.00 & $-$0.7 & 0 \\
$\ldots$ No cor.           &    38.48 &  \textbf{0.05} & \textbf{$-$0.076} &    1 & $-$0.03 & $-$0.03 & $-$0.03 & $-$0.03 &    0.02 &    0.05 & $-$0.006 & $-$0.03 & $-$0.6 & 2 \\
$\ldots$ Ground cor.       &   190.52 & \textbf{$-$0.16} & \textbf{$-$0.123} &  0 &    0.03 &    0.00 &    0.02 &    0.06 & $-$0.04 &    0.02 &    0.017 &    0.04 & $-$1.8 & $-$1 \\
$\ldots$ Crab cor.         &    13.36 & \textbf{$-$0.11} &    \textbf{0.066} &  1 &    0.02 &    0.01 &    0.00 &    0.02 &    0.05 &    0.08 & \textbf{0.031} & 0.03 & 0.0 & 0 \\
$\ldots$ New arfgen        &  $-$1.55 &    \textbf{0.78} &           0.004 &    0 &    0.00 &    0.00 &    0.00 &    0.00 &    0.00 &    0.00 &    0.000 &    0.00 &    0.1 & 0 \\
No gain cor.               &   626.73 &           0.01 &             0.003 &    4 & $-$0.13 & $-$0.06 & $-$0.02 & $-$0.01 & $-$0.01 & $-$0.01 & $-$0.008 &    0.00 & $-$0.5 & \textbf{14} \\
\hline\noalign{\smallskip}
\multicolumn{15}{l}{\textsl{Improved model (section~\ref{sect:improve}):}} \\
 & $-$146.77 & -- & -- & -- & $-$14 & \textbf{$-$11} & $-$1 & 5 & 8 & 6 & -- & 5 & \textbf{$-$8.3} & 0 \\
\hline
\end{tabular}
\normalsize
\begin{tabnote}
\footnotemark[$*$] Emission measure $Y$, temperature $kT$, LOS velocity dispersion $\sigma_v$, column density of hot-gas absorption $N_{\rm H,hot}$, and redshift $cz$.\\
\footnotemark[$\dagger$] Elemental abundance relative to the proto-solar values of \citet{lodders2009}. \\
\footnotemark[$\ddagger$] Expected value for the baseline model is 4876. \\
\end{tabnote}
\end{table*}
\setlength{\tabcolsep}{6pt}

The best-fit parameters of our model are given in table~\ref{tab:parameters}. It
is beyond the scope of this paper to discuss the astrophysical interpretation of
the temperature, abundances, and resonance scattering; these are discussed in
much greater detail by T, Z, and RS~papers, respectively.

In the following sections, that form the core of our paper, we investigate in
more detail the systematic effects that affect the best-fit parameters of this
baseline model. We do so by showing in table~\ref{tab:parameters} the difference in
best-fit C-statistic and the best-fit model parameters, for different
assumptions in our modeling. In several cases we also show the relative
difference in the predicted model spectra.

We consider the following systematic effects: the plasma code that is
used (section~\ref{sect:plasmacode}), the atomic database in the background (section~\ref{sect:atomicdata}), different choices for details of the plasma
modeling (section~\ref{sect:plasmamodel}), astrophysical modeling effects
(section~\ref{sect:astromodel}), the role of other spectral components apart from the main
hot plasma (section~\ref{sect:components}), and spectral fitting techniques (section~\ref{sect:fittingtech}).
Those due to instrumental calibration aspects are separately examined in appendix~\ref{app:instrumental}.

\begin{figure*}[!htbp]
\includegraphics[width=17cm]{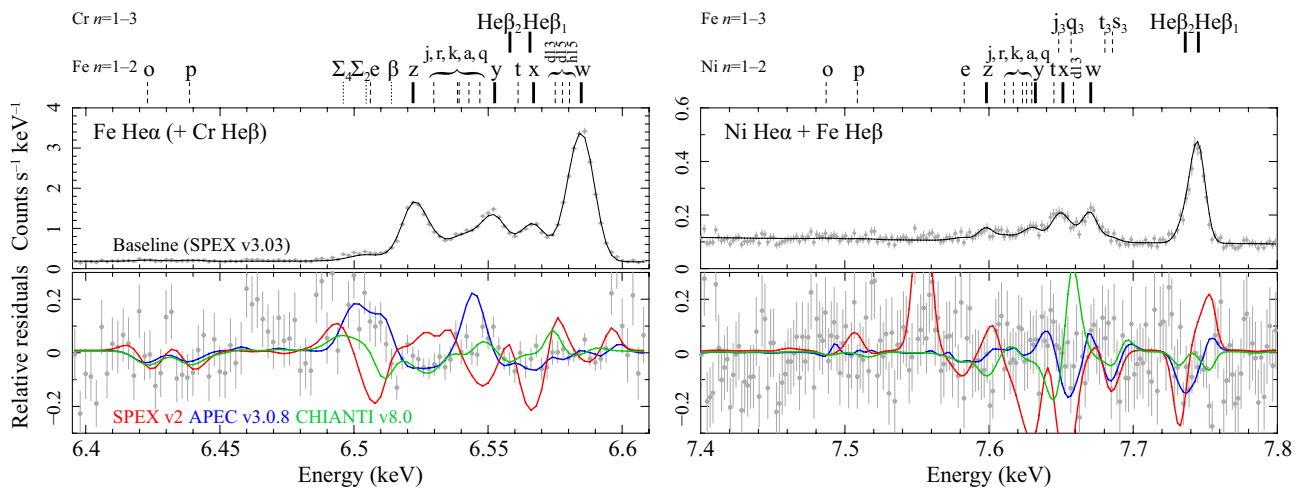}
\caption{The Hitomi SXS spectrum in the Fe (left) and Ni (right) He$\alpha$ bands with the best-fit baseline model in the upper panels,
and the residuals in the lower panels. Also shown in the lower panels are the relative difference between the baseline model
and the best-fit models with various other plasma codes (SPEX v2 in red, APEC v3.0.8 in blue, and CHIANTHI v8.0 in green).
Solid bars above the upper panels show energies of the main He$\alpha$ and He$\beta$ lines, while dashed and dotted ones are 
respectively of the Li-like and Be-like satellite lines. 
For the full-band results and line notations see appendices~\ref{app:speccomp} and \ref{app:linelist}, respectively.
}
\label{fig:spec0}
\end{figure*}

\section{Systematic factors affecting the derived source parameters:
plasma code\label{sect:plasmacode}}

We consider in this paper apart from SPEX version 3.03
(the baseline plasma model) also the old SPEX version 2/\textit{Mekal} plasma model,
the latest SPEX version before the launch of Hitomi (hereafter, the pre-launch version: SPEX version 3.00),
as well as the pre-launch and the latest APEC/AtomDB versions 3.0.2 and 3.0.8
\citep{smith2001, foster2012}, respectively, CHIANTI version 8.0 \citep{dere1997, delzanna2015},
and Cloudy version 13.04 \citep{ferland2013} plasma models.
The best-fit models with these codes highlighting the Fe and Ni He$\alpha$ bands are compared in figure~\ref{fig:spec0}.
The full-band results as well as the relevant atomic data are compared between these codes in
appendices~\ref{app:speccomp} and \ref{app:linelist} (see also section~\ref{sect:mekal}).

\subsection{SPEX versions 3.00 and 3.03\label{sect:spexv3.03}}

Version 3.00 of SPEX was released on January 29, 2016 as the pre-launch version for Hitomi data analysis.
In SPEX version~2, line powers were calculated using the method of \citet{mewe1985},
i.e., using a temperature-dependent parameterization of the line fluxes with empirical density corrections.
This version 3.00 contains fully updated atomic data for the most highly ionized ions,
solving directly the balance equations for the ion energy level populations
incorporating effects like density and radiation field, and uses these level populations to calculate the line power.

Triggered by the early work on the Hitomi SXS data of the Perseus cluster \citep{hitomi2016},
and the follow-up work as presented in this paper, several updates to version 3.00 were made leading to
SPEX version 3.03, released in November 2016, that is used for the present analysis.
Below we list the most important updates for the present work relative to version 3.00.
\begin{enumerate}
\item For Li-like ions, inner-shell transitions were extended from maximum principal quantum number $n=6$ to $n=15$ using FAC calculations.

\item A numerical issue with Be-like ions related to metastable levels was resolved allowing the full use of the new line calculations for these ions.

\item Inner-shell energy levels, Auger rates, and radiative transitions for O-like Fe\emissiontype{XIX} to Be-like Fe\emissiontype{XXIII} were added using \citet{palmeri2003}.

\item A bug in the calculation of trielectronic recombination for Li-like ions was also removed;
in the dielectronic capture from the He-like 1s2s level to Li-like 2s2p$^2$ levels
the relative population of the 1s2s level was ignored leading to a too high population of
these Li-like levels and subsequently to too strong stabilizing radiative transitions from these levels,
and not in agreement with the Hitomi SXS data.

\item The proper branching ratios for excitation and inner-shell ionization to excited levels
that can auto-ionize are now taken into account, leading to improvements for some satellite lines.
\end{enumerate}

To demonstrate the post-launch updates, we present the results of the Hitomi SXS spectral fitting
with both versions 3.00 and 3.03 in table~\ref{tab:parameters}.
The best-fit C-statistic value increases by 2372 from version 3.03 to 3.00,
and the latter gives a 7\% higher temperature, 8\% higher turbulent velocity,
and 30\% lower Fe abundance than the former one. The other abundances also have 3\% to 37\% deviations.
The effective column density of resonance scattering $N_{\rm H,hot}$ becomes zero with version 3.00.

\subsection{Using SPEX version 2 (the Mekal code)\label{sect:mekal}}

The old \textit{Mekal} code, or SPEX version 2 \citep{mewe1995}, contained significantly fewer lines
and chemical elements than the present version of SPEX. In addition, the atomic data
(e.g., line energies) have been improved in the present SPEX
version compared to the old \textit{Mekal} model. This is evident from
table~\ref{tab:parameters}, showing that the best-fit C-statistic value increases by
1125 if we replace the new code by the old code. A detailed comparison
(figures~\ref{fig:spec1}--\ref{fig:spec3} in appendix~\ref{app:speccomp}) shows that there are many differences.
For instance, contrary to the old model, the new model includes Cr and Mn lines (in the
5--6~keV range). Also, updates in the line energies are visible as a sharp
negative residual close to a sharp positive residual.

The old code yields almost the same temperature as the new code, but there are
significant changes in the derived turbulent velocity and the abundances. Small
wavelength errors can be compensated for by adjusting the line
broadening. Abundances are off by 2--4$\sigma$ or up to 5--15\% of the values
obtained from the baseline model.

This is only one example of a comparison between different models.
In appendix~\ref{app:speccomp} (figures~\ref{fig:spec1}--\ref{fig:spec3}),
we show the full Hitomi SXS spectrum in 1.9--9.5~keV with our best-fit baseline model
in the upper panels,
and the residuals in the lower panels. In these lower panels we also show the relative
difference between the baseline model and the best-fit models obtained with
various other plasma codes.

The differences between these models can be divided into two
classes: wavelength differences (leading to a positive residual next to a
negative residual e.g., the Ca\emissiontype{XIX} He$\beta$ line near 4.51~keV has
a different wavelength in the \textit{Mekal} code compared to the baseline
model), or flux differences (leading to a strict positive or negative residual
in the relative residuals e.g., the S\emissiontype{XV} forbidden line near 2.38~keV
is stronger in the \textit{Mekal} model compared to the baseline model).

In appendix~\ref{app:linelist} (tables~\ref{tab:wavelength} and \ref{tab:satellite}),
we list the line energies of the strongest lines in the spectrum.
For comparison, the energies in SPEX are shown together with those
in the APEC version 3.0.8 and CHIANTI version 8.0 codes. All the Lyman- and Helium-series
transitions with model line emissivities $\geq$10$^{-26}$~photon~m$^3$~s$^{-1}$ are listed,
and for satellite lines of He-like, Li-like, and Be-like ions,
the threshold is set to 10$^{-25}$~photon~m$^3$~s$^{-1}$. In addition
we show the Einstein coefficients and emissivities used in the three atomic codes.

\subsection{APEC\label{sect:apec}}

APEC runs were conducted for both the pre-launch version, AtomDB version 3.0.2, and the latest version, AtomDB version 3.0.8. Since the launch of Hitomi, several updates have been made to the database to reflect the needs of the Hitomi data. These updates were not made to ``fit'' the Hitomi SXS data, but instead to reflect the priorities that analysis revealed. These changes were:
\begin{enumerate}
\item The ionization and recombination rate calculation was switched from an interpolatable grid to a fit function, which has a few percent effect on several ion populations depending on the temperatures/ion involved.
\item Wavelengths for higher $n$ transitions of the H- and He-like ions were changed to match Ritz values from the NIST Atomic Spectra Database.
\item Wavelengths for valence shell transitions of Li-like ions were changed to match Ritz values from NIST.
\item Fluorescence yields and wavelengths of inner shell lines were updated to the data of \citet{palmeri2003, palmeri2003a, palmeri2008, palmeri2010, palmeri2012, mendoza2004}.
\item Collisional excitation rates for He-like Fe were changed from an unpublished data set to that of \citet{whiteford2001}.
\item Collisional excitation rates for H-like ions from Al to Ni were changed from FAC calculations to those of \citet{li2015}.
\end{enumerate}

The spectral calculation is done with the BVVAPEC model in Xspec version 12.9.1 \citep{arnaud1996},
while the rebinning and fitting are carried out with SPEX version 3.03.00.
The abundance standard \citep{lodders2009} is applied to the APEC calculations.
This allows a direct comparison between APEC and SPEX.
The ionization balance calculation in APEC, on the other hand,
is based on \citet{bryans2009}, while \citet{urdampilleta2017} is used in SPEX.
This difference is separately discussed in section~\ref{sect:ibal}.

The run with the pre-launch APEC version 3.0.2 gives a best-fit C-statistic which is larger
than the baseline value by 670.

As shown in figure~\ref{fig:spec0} and appendix~\ref{app:speccomp} (figures~\ref{fig:spec1}--\ref{fig:spec3}),
the relative difference between SPEX and APEC is usually within 10\%, except for a few lines,
including Cr\emissiontype{XXIII} He$\alpha$, Mn\emissiontype{XXIV} He$\alpha$,
Fe\emissiontype{XXIV} satellite lines at 6.42~keV, 6.44~keV, 8.03~keV, and 8.04~keV,
Ni\emissiontype{XXVII} He$\alpha$ blended with Ni\emissiontype{XXVI} and Fe\emissiontype{XXIV} satellite lines,
and Fe\emissiontype{XXV} He$\beta$ to He$\eta$ lines. Many differences might be related
to the rates used in level population calculation, e.g., collisional excitation and
spontaneous emission rates (see section~\ref{sect:atomicdata} for details). The line energy
data in APEC version 3.0.8 are in general good agreement with SPEX version 3.03
(see table~\ref{tab:wavelength} in appendix~\ref{app:linelist} for details).

As listed in table~\ref{tab:parameters}, the APEC code gives a similar best-fit
temperature as the SPEX baseline model. The metal abundances obtained with APEC
are lower by 5--10\% for Si, S, Ar, Ca, and Ni than
the best-fit baseline values, while the Cr abundances obtained with the two codes
agree within error bars. The largest difference is with the Fe abundance,
which is 16\% lower in the latest APEC/AtomDB (version 3.0.8) than SPEX.
The best-fit turbulent velocity in $\sigma_v$ (LOS dispersion)
derived with the latest APEC code is 16~km\,s$^{-1}$ lower than the SPEX result.

\subsection{CHIANTI\label{sect:chianti}}

Another atomic code/database widely used in the UV and X-ray spectroscopy for optically thin,
collisionally dominated plasma is the CHIANTI code. Compared to the APEC and SPEX codes, CHIANTI is more focused on modeling
the spectra from relatively cooler plasma in the solar and stellar atmosphere, while in this work, we are testing it in the conditions of hot ICM emission.
The latest version 8.0 \citep{dere1997, delzanna2015} is used.
The current CHIANTI database includes all the relevant H-like and He-like ions except for Cr and Mn, which means that these abundances cannot be estimated.
We calculate the collisional ionization equilibrium spectrum using an IDL-version
\textsl{isothermal} model, setting the ionization balance to \citet{bryans2009}, and
change the solar abundance table to \citet{lodders2009} proto-solar values. To perform the fit to the data,
the IDL calculation is implemented as an input to the \textsl{user} model in SPEX, and the fitting engine of SPEX
repeatedly triggers the IDL run until a best-fit is reached. Since the CHIANTI code does not provide line broadening information,
we apply a multiplicative SPEX Gaussian broadening model {\sl vgau} to the CHIANTI model. This is only a first-order approximation,
since the thermal broadening should vary with the atomic number. A detailed comparison on the best-fit spectra shown in
appendix~\ref{app:speccomp} (figures~\ref{fig:spec1}--\ref{fig:spec3}) reveals several differences in emission features from
the baseline model, at levels ranging from a few \% up to about 20\%. Most of these differences are traced back to
the different input atomic data, which can be found in appendix~\ref{app:linelist} (tables~\ref{tab:wavelength}--\ref{tab:satellite}).

The C-statistic value increases by 327 when fitting with the CHIANTI code. The best-fit temperature, emission measure, turbulent velocity,
and the Fe abundance are roughly consistent with the baseline results, while the remaining abundances differ by 3--19\%. The required column density for resonance scattering is reduced by 10\% with the CHIANTI model.

\subsection{Cloudy\label{sect:cloudy}}

The Cloudy code has been developed as a tool to calculate photoionized plasmas and it is principally used for this application.
It does, however, have a module for calculating CIE plasma spectra, so we have therefore fitted the Perseus spectrum with the
coronal equilibrium model of
Cloudy version 13.04. The abundance standard is set to \citet{lodders2009}. Since
the Cloudy code does not provide the thermal and turbulent broadening, we again apply a
multiplicative SPEX Gaussian broadening model {\sl vgau} to the Cloudy calculations.
As shown in table~\ref{tab:parameters}, the fit with Cloudy yields a
large C-statistic. The most significant residuals appear at the Fe\emissiontype{XXV}
He-series and Fe\emissiontype{XXVI} Ly$\alpha$ lines. The best-fit temperature agrees with
the results of the other codes, but the abundance values differ strongly from those derived from the other codes. We again note that modeling of collisional plasmas is not Cloudy's main purpose.

\section{Systematic factors affecting the derived source parameters: atomic
data\label{sect:atomicdata}}

As shown in table~\ref{tab:parameters}, the atomic code uncertainty contributes
the main uncertainty of many parameters, such as the Si, S, Ar, Ca, Mn, Fe, and Ni
abundances, the hot absorption, and the turbulent velocity. The code uncertainty
mainly comes from the input atomic data, for instance, the ionization balance, collision
excitation/de-excitation rates, recombination rates, and transition probabilities.
In this section, we explore and describe the discrepancies between
the current atomic data used in each code, and estimate the propagated
errors on the fitted parameters.

\begin{figure}[!htbp]
\includegraphics[width=8cm]{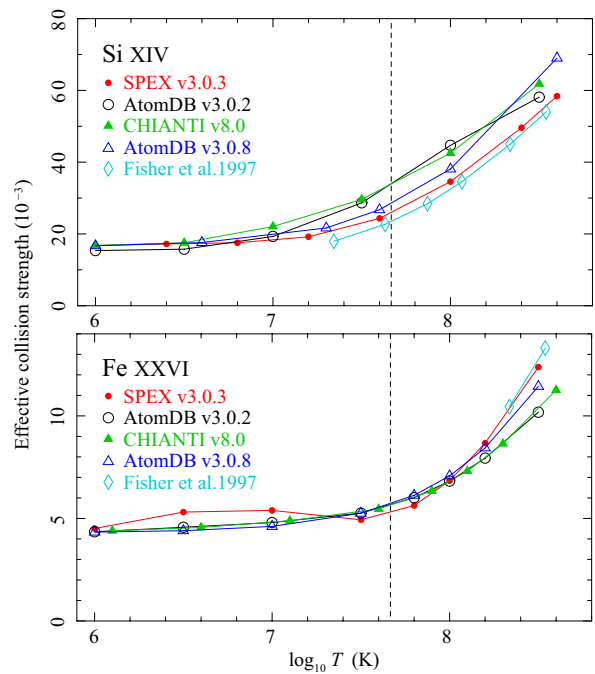}
\caption{Comparisons of effective collision strength as a function of balance temperature. The Ly$\alpha$1 and Ly$\alpha$2
transitions are combined. The vertical dashed lines mark a temperature of 4~keV. }
\label{fig:lya}
\end{figure}

\begin{table*}[!htbp]
\caption{Electron effective collision strengths ($10^{-3}$) of the Ly$\alpha$ transitions for a CIE plasma at 4-keV temperature.}
\label{tab:lya}
\centerline{
\begin{tabular}{lcccccc}
\hline
Ion                     & \multicolumn{5}{c}{Ly$\alpha_1$ transition: 1s ($^2$S$_{1/2}$) -- 2p ($^2$P$_{3/2}$)} & Diff.\footnotemark[$*$] \\ \cline{2-6}
                        & SPEX v3.03 & AtomDB v3.0.8 & AtomDB v3.0.2 & CHIANTI v8.0 &   FAC &       \\
\hline
Si\emissiontype{XIV}    &      17.11 &         18.97 &         22.12 &        22.14 & 18.80 &  23\% \\
S\emissiontype{XVI}     &      12.31 &         13.30 &         15.32 &        15.39 & 13.10 &  20\% \\
Ar\emissiontype{XVIII}  &       9.29 &          9.74 &         11.07 &         8.08 &  9.59 &  27\% \\
Ca\emissiontype{XX}     &       7.25 &          7.40 &          8.25 &         8.08 &  7.27 &  14\% \\
Cr\emissiontype{XXIV}   &       4.61 &          4.68 &          5.00 &          --- &  4.56 &  11\% \\
Mn\emissiontype{XXV}    &       4.17 &          4.24 &          4.47 &          --- &  4.12 &  10\% \\
Fe\emissiontype{XXVI}   &       3.51 &          3.85 &          3.76 &         3.71 &  3.74 &   9\% \\
Ni\emissiontype{XXVIII} &       3.18 &          3.21 &          3.27 &         3.28 &  3.12 &   7\% \\
\hline
Ion                     & \multicolumn{5}{c}{Ly$\alpha_2$ transition: 1s ($^2$S$_{1/2}$) -- 2p ($^2$P$_{1/2}$)} & Diff.\footnotemark[$*$] \\ \cline{2-6}
                        & SPEX v3.03 & AtomDB v3.0.8 & AtomDB v3.0.2 & CHIANTI v8.0 &   FAC &       \\
\hline
Si\emissiontype{XIV}    &       8.55 &          9.54 &         11.15 &        11.05 &  9.48 &  23\% \\
S\emissiontype{XVI}     &       6.15 &          6.68 &          7.75 &         7.68 &  6.63 &  21\% \\
Ar\emissiontype{XVIII}  &       4.64 &          4.92 &          5.62 &         4.57 &  4.86 &  19\% \\
Ca\emissiontype{XX}     &       3.62 &          3.74 &          4.20 &         4.03 &  3.71 &  16\% \\
Cr\emissiontype{XXIV}   &       2.30 &          2.38 &          2.57 &          --- &  2.34 &  13\% \\
Mn\emissiontype{XXV}    &       2.09 &          2.16 &          2.31 &          --- &  2.13 &  13\% \\
Fe\emissiontype{XXVI}   &       1.66 &          1.97 &          1.90 &         1.89 &  1.93 &  16\% \\
Ni\emissiontype{XXVIII} &       1.59 &          1.65 &          1.70 &         1.64 &  1.62 &  10\% \\
\hline
\end{tabular}
}
\begin{tabnote}
\footnotemark[$*$] Relative differences between the codes defined as (maximum$-$minimum)$/$maximum.\\
\end{tabnote}
\end{table*}

\subsection{Collisional excitation \label{sect:lineer}}

\subsubsection{H-like ions\label{sect:lineerh}}

In this section, we address the systematic uncertainties on the collisional excitation rates for H-like ions
from the ground to the 2p levels. The radiative relaxation from the 2p levels back to the ground produces
the Ly$\alpha$ lines.
As shown in table~\ref{tab:lya}, the effective collision strengths of Si\emissiontype{XIV} and Fe\emissiontype{XXVI}
for a 4-keV plasma often differ by 10--30\% among
atomic codes, which contributes an important uncertainty in the abundance measurement (table~\ref{tab:parameters}).
The collision rates used in AtomDB version 3.0.2 and CHIANTI version 8.0 are systematically larger than those in SPEX version 3.03
and AtomDB version 3.0.8, while the latter two are roughly consistent with the calculations
by the Flexible Atomic Code \citep[FAC,][]{gu2008},  version 1.1.1. FAC can calculate both atomic structure and scattering data,
and the relativistic effects are fully taken into account by the Dirac-Coulomb Hamiltonian.
By solving the configuration-interaction wave functions in the Dirac-Fock-Slater central-field potential,
it evaluates the radiative transitions and auto-ionization rates for the input atomic levels,
and computes the effective collisional strengths using a distorted-wave approximation.
The FAC values shown are based on calculations with a default grid that contains 6 grid points.
As check a calculation with a grid of 11 points has also been carried out.
The values of the 11 point grid are about 5\% lower than the values of the default grid calculation.
The consistency between FAC and AtomDB version 3.0.8 is expected,
since the AtomDB values are essentially taken from a FAC calculation by \citet{li2015}.

The differences in the effective collision strengths depend on the electron temperature. In
figure~\ref{fig:lya}, we compare five sets of calculations for Si\emissiontype{XIV}
and Fe\emissiontype{XXVI} Ly$\alpha$ transitions. For Si\emissiontype{XIV},
SPEX uses a R-matrix calculation by \citet{aggarwal1992}, which is roughly consistent with
the AtomDB and CHIANTI values within 8\% at 10$^6$~K, but becomes lower by 30\%
at 10$^{7.7}$~K than the CHIANTI data. This means that even for the simplest H-like ions,
the atomic data for the collision process are not sufficiently converged to match the accuracy of
the current observations. Since the Si abundance is mostly determined by
the Si\emissiontype{XIV} Ly$\alpha$ for the Hitomi SXS data, the 30\% uncertainty
in the collision strength calculation indicates a roughly similar error in the abundance measurement.

For Fe\emissiontype{XXVI}, we compare two representative calculations using a R-matrix method,
\citet{ballance2002} (implemented in CHIANTI version 8.0 and AtomDB version 3.0.2) and
\citet{kisielius1996} (used in SPEX version 3.03), and the FAC calculation in AtomDB version 3.0.8.
The three results roughly agree with each other at 10$^6$~K, while the calculations of \citet{kisielius1996}
is higher than the other two up to 10$^7$~K, and decreases rapidly beyond this temperature,
relative to the others. At the high temperature end (3$\times$10$^8$~K), the difference between the
\citet{ballance2002} and \citet{kisielius1996} values is about 30\%
for the 1s ($^{2}$S$_{1/2}$) -- 2p ($^{2}$P$_{3/2}$) Ly$\alpha_1$ transition.
According to \citet{ballance2002}, the differences at low and high energies are mainly caused by
the treatment of radiation damping and the high-energy approximation, respectively.
This would contribute a minor part of the uncertainty on the Fe abundance measured with the Hitomi SXS data;
the main uncertainty comes from the He$\alpha$ transitions (section~\ref{sect:lineerhe}).

\subsubsection{He-like ions \label{sect:lineerhe}}

\begin{figure*}[!htbp]
\includegraphics[width=17cm]{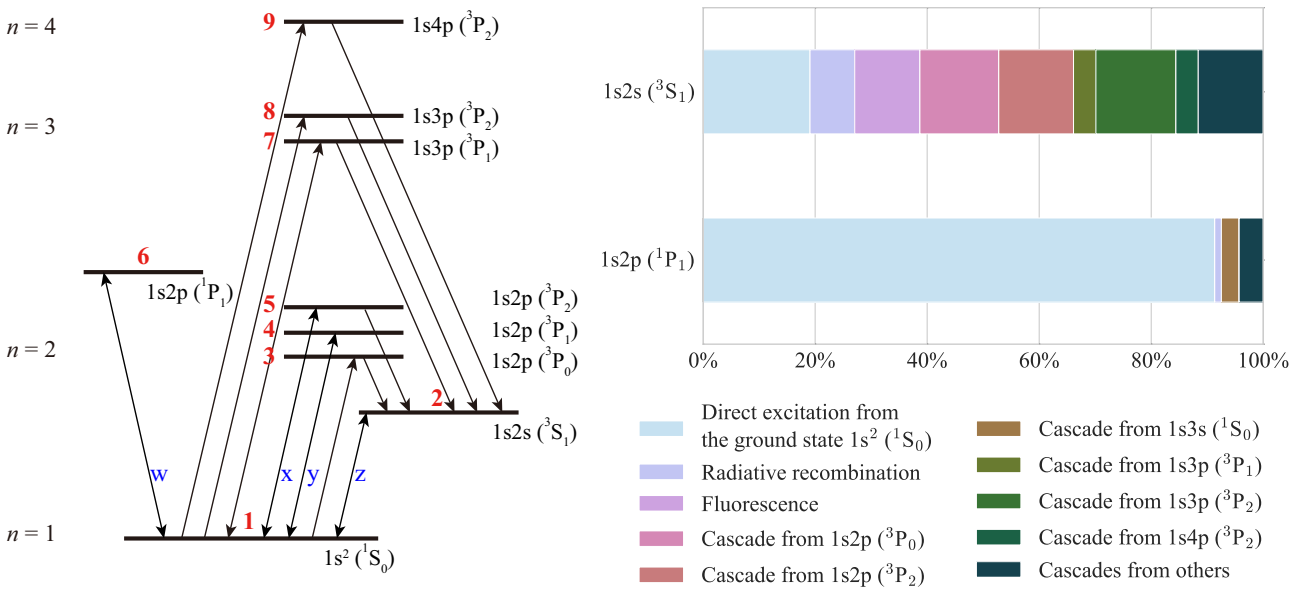}
\caption{(left) Grotrian diagram of the dominant Fe\emissiontype{XXV} levels to the triplet formation. Red numbers are used in table~\ref{tab:trip}.
Blue labels show transitions corresponding to the individual lines of the triplet.
The 1s2s ($^1$S$_{0}$) level is not shown here, as it does not contribute to line production but to continuum emission via the two-photon decay \citep[e.g.,][]{nussbaumer1984}.
(right) Relative contributions to the formations of 1s2s ($^3$S$_{1}$) and 1s2p ($^1$P$_{1}$) levels (upper levels of z and w, respectively).}
\label{fig:triplet}
\end{figure*}

We now turn to the He-like Fe-K multiplet as a test case to assess the flux errors on model
lines by the input atomic data. First we define the range of related atomic levels and data in
figure~\ref{fig:triplet} and table~\ref{tab:trip}. The most dominant populating process
for the upper levels of the resonance and intercombination transitions is electron-impact excitation
from the ground state, and the main loss process is radiative transition back to the
ground state. The upper level of the x line (2p $^3$P$_2$) has a 18\% chance to form
a two-step decay via an intermediate level.

Meanwhile, for a 4-keV plasma, the upper level of the Fe\emissiontype{XXV} forbidden transition (z) is
populated almost equally by: excitation from the ground state; cascades 
from the 2p ($^3$P$_{0}$), 2p ($^3$P$_{2}$), and 3p ($^3$P$_{2}$) levels; 
and radiative recombination from the continuum state. In addition,
inner-shell ionization of Fe\emissiontype{XXIV} drives 8\%, and radiative
transitions from the 3p ($^3$P$_{1}$) and 4p ($^3$P$_{2}$) levels both provide 4\% of the
population. The metastable level can decay to the ground only via radiative
transitions.

\begin{table*}[!htbp]
\caption{Fe\emissiontype{XXV} He$\alpha$ multiplet formation for a CIE plasma at 4-keV temperature.}
\label{tab:trip}
\begin{tabular}{lccccccc}
\hline
Transition\footnotemark[$*$] & Rel. contrib.\footnotemark[$\dagger$] & \multicolumn{5}{c}{Electron effective collision strength (10$^{-3}$)} & Diff.\footnotemark[$\ddagger$] \\ \cline{3-7}
                             &                                       & SPEX v3.03 & AtomDB v3.0.8 & AtomDB v3.0.2 & CHIANTI v8.0 & Open-ADAS &  \\
\hline
1 $\rightarrow$ 2 &          19\% &      0.268 &         0.295 &         0.410 &      0.425 &     0.246 &       42\% \\
1 $\rightarrow$ 3 &          69\% &      0.138 &         0.143 &         0.144 &      0.146 &     0.135 &        8\% \\
1 $\rightarrow$ 4 &          77\% &      0.715 &         0.721 &         0.728 &      0.721 &     0.868 &       18\% \\
1 $\rightarrow$ 5 &          67\% &      0.692 &         0.703 &         0.714 &      0.740 &     0.695 &        6\% \\
1 $\rightarrow$ 6 &          91\% &      4.047 &         4.026 &         4.051 &      4.004 &     4.316 &        7\% \\
1 $\rightarrow$ 7 &          71\% &      0.161 &         0.163 &         0.162 &      0.162 &     0.166 &        3\% \\
1 $\rightarrow$ 8 &          63\% &      0.173 &         0.178 &         0.180 &      0.182 &     0.165 &        9\% \\
1 $\rightarrow$ 9 &          62\% &      0.071 &         0.069 &         0.071 &      0.072 &     0.073 &        5\% \\
\hline
Transition\footnotemark[$*$] & Rel. contrib.\footnotemark[$\dagger$] & \multicolumn{5}{c}{Transition probability (s$^{-1}$)}                  & Diff.\footnotemark[$\ddagger$] \\ \cline{3-7}
                             &                                       & SPEX v3.03 & AtomDB v3.0.8 & AtomDB v3.0.2 & CHIANTI v8.0 & FAC &  \\
\hline
2 $\rightarrow$ 1 (z) &         100\% &   2.080$\times 10^{8}$ &   1.930$\times 10^{8}$ & left &  2.080$\times 10^{8}$ &   1.997$\times 10^{8}$ &       ~7\% \\
4 $\rightarrow$ 1 (y) &         100\% &  4.260$\times 10^{13}$ &  3.720$\times 10^{13}$ & left & 4.350$\times 10^{13}$ &  4.196$\times 10^{13}$ &       14\% \\
5 $\rightarrow$ 1 (x) &          82\% &   6.550$\times 10^{9}$ &   6.578$\times 10^{9}$ & 6.519$\times 10^{9}$ &  6.480$\times 10^{9}$ &   6.568$\times 10^{9}$ &       ~1\% \\
6 $\rightarrow$ 1 (w) &         100\% &  4.565$\times 10^{14}$ &  4.670$\times 10^{14}$ & left & 4.610$\times 10^{14}$ &  4.679$\times 10^{14}$ &       ~2\% \\
7 $\rightarrow$ 1     &          63\% &  1.524$\times 10^{13}$ &  1.060$\times 10^{13}$ & left & 1.126$\times 10^{13}$ &  1.248$\times 10^{13}$ &       30\% \\
3 $\rightarrow$ 2     &         100\% &   3.820$\times 10^{8}$ &   2.770$\times 10^{8}$ & left &  3.740$\times 10^{8}$ &   3.743$\times 10^{8}$ &       27\% \\
5 $\rightarrow$ 2     &          18\% &   1.470$\times 10^{9}$ &   1.420$\times 10^{9}$ & left &  1.420$\times 10^{9}$ &   1.466$\times 10^{9}$ &       ~3\% \\
7 $\rightarrow$ 2     &          34\% &  8.078$\times 10^{12}$ &  7.990$\times 10^{12}$ & left & 7.861$\times 10^{12}$ &  8.057$\times 10^{12}$ &       ~3\% \\
8 $\rightarrow$ 2     &         100\% &  8.932$\times 10^{12}$ &  8.550$\times 10^{12}$ & left & 8.682$\times 10^{12}$ &  8.660$\times 10^{12}$ &       ~4\% \\
9 $\rightarrow$ 2     &          74\% &  3.957$\times 10^{12}$ &  3.550$\times 10^{12}$ & left & 3.642$\times 10^{12}$ &  3.769$\times 10^{12}$ &       10\% \\
\hline
\end{tabular}
\begin{tabnote}
\footnotemark[$*$] Energy-level IDs correspond to the energy levels as denoted in figure~\ref{fig:triplet}.\\
\footnotemark[$\dagger$] Relative contributions to the total gain or loss term of the level derived with SPEX v3.03.\\
\footnotemark[$\ddagger$] Relative differences between the codes defined as (maximum$-$minimum)$/$maximum.\\
\end{tabnote}
\end{table*}

\begin{figure}[!htbp]
\includegraphics[width=8cm]{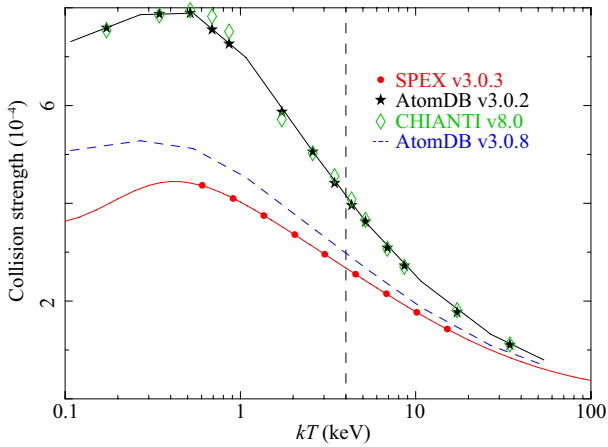}
\caption{Comparison of the effective collision strength from the ground to 1s.2s ($^{3}$S$_1$) state.
The vertical line shows a temperature of 4~keV.}
\label{fig:fe25col}
\end{figure}

\begin{figure}[!htbp]
\includegraphics[width=8cm]{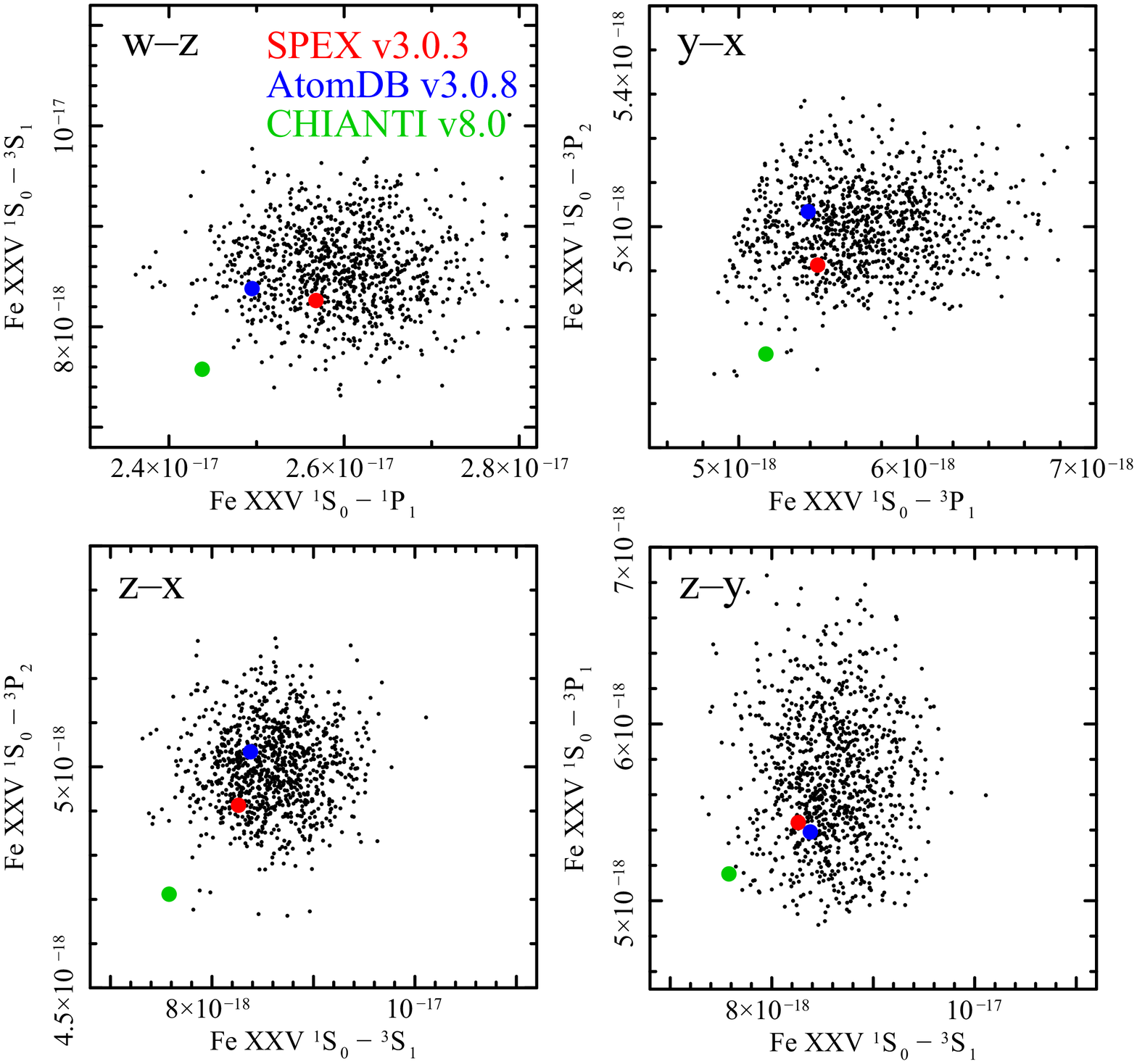}
\caption{Monte-Carlo simulations on the four Fe\emissiontype{XXV} He$\alpha$ lines.
The SPEX, AtomDB, and CHIANTI values are shown in red, blue, and green, respectively.
The line intensities are given in the unit of photon cm$^{3}$ s$^{-1}$. }
\label{fig:triplet_sim}
\end{figure}

We compare the atomic data extracted from the SPEX, AtomDB, and CHIANTI
databases, as well as the collision data from the Open-ADAS database\footnote{ADF04, produced by Alessandra Giunta, 14 Sep 2012. See $\langle$http://open.adas.ac.uk$\rangle$.},
and the radiative transition data from the FAC calculation. The effective collision strengths
used in SPEX, AtomDB version 3.0.2 and CHIANTI version 8.0, and AtomDB version 3.0.8 are taken from the published data in
\citet[][distorted wave]{zhang1987}, \citet[][R-matrix]{whiteford2005}, and \citet[][R-matrix]{whiteford2001}, respectively.
The data from Open-ADAS is calculated with the distorted-wave approximation. We do not show
the FAC results on the collisional excitation, since it does not provide
explicitly the contributions from resonance excitation channels, which are incorporated in the other calculations.

As shown in table~\ref{tab:trip}, the collision data converge relatively well ($<$18\%) on the ground
to $^1$P and $^3$P level transitions, but differ by up to 42\% on the ground to $^3$S transition.
As shown in figure~\ref{fig:fe25col},
the effective collision strengths used in CHIANTI version 8.0 / AtomdB version 3.0.2 are systematically
larger than that in the SPEX version 3.03, by a factor of two at 1~keV, and about 40\% at 10~keV.
The values in AtomDB version 3.0.8 lie in the middle, about 10\% higher than the SPEX values at 4~keV.
It appears that the R-matrix calculations (AtomDB and CHIANTI) are systematically higher,
by 10--40\%, than the distorted-wave calculations (SPEX and Open-ADAS).
Since the forbidden transition from $^3$S to the ground gives a line intensity only second
to the resonance line for a 4-keV plasma, while the latter is subject to resonance scattering (section~\ref{sect:selfab}),
the uncertainty of the $^3$S excitation should contribute a significant portion of the total error of the Fe abundance.
The radiative transition data used in different codes agree within a 15\% level for the He-like triplet lines.
The transition rates from higher levels (e.g., $n=$3 and 4) to the ground can have a larger uncertainties up to 30\%,
which will be discussed in more detail in section~\ref{sect:transition}.

Assuming that the
deviations between different data gives a rough measure of the atomic process
uncertainties, we carry out a Monte-Carlo simulation to quantify the atomic uncertainties
on the He-like triplet line ratios. We generate 1000 sets of collisional excitation rates by randomizing
based on the five sets of collision data in table~\ref{tab:trip}. The same is done for the transition probability.
Then we run the SPEX calculation repeatedly, each time with one set of randomized collision and radiative data,
to determine the flux error on each individual line. There are two potential caveats: first,
the Monte-Carlo method assumes that all the rate errors are
independent, which is not always true for the atomic calculations; second, the differences between SPEX
and other codes on the atomic data of the recombination processes and the fluorescent yields, as well as on the atomic
structure such as the maximum principal quantum number, are not taken into account in the simulation.
Therefore the error obtained in the simulation should be regarded as a lower limit.

The results of the 1000 simulations are shown in figure~\ref{fig:triplet_sim}.
The simulation predicts that the resonance (w), intercombination (x and y), and
the forbidden (z) lines have uncertainties of $\sim$4\%, 2\% and 8\%, and 6\%,
respectively. The y and z lines have the larger atomic uncertainties than the other two,
probably caused by the relatively large errors of the collision strengths and the complex formation of the $^3$P and $^3$S levels.
The actual AtomDB version 3.0.8 and SPEX version 3.03 line intensities indicate similar uncertainties.
The CHIANTI version 8.0 triplet line fluxes are systematically lower than the simulation results and the other two codes.
This could be caused by the fact that CHIANTI has the lowest maximum principal quantum number,
and hence possibly a lowest radiative decay contribution to the $n=$2 levels, among the three atomic codes.
When multiplying the CHIANTI fluxes by a factor of 1.05, they become well in line with the simulation results.

\subsubsection{Best fit with adjusted line ratios for the x and y lines \label{sect:xymod}}

We have tested the sensitivity of our results on the He-like Fe lines further as
follows. We made the intensity of the x and y lines relative to the forbidden line
a free parameter. Technically, this was achieved by applying two \textsl{line}
components to the x and y line. This model produces the transmission $T(E)$ for in
our case an absorption or emission line as
$T(E)=\exp{[-\tau_0\phi(E)]}$
with $\phi(E)$ the Gaussian optical depth profile.
We have frozen the line energy of this absorption line to the energies of the x and y
lines, respectively, and the width to the width of the emission line
(using the best-fit thermal and turbulent broadening from the baseline model).
Thus, the only two additional free parameters are the nominal optical depths
$\tau_0$ of both lines, positive values indicating lower flux, negative values higher flux.
The best-fit parameters are $\tau_0=$0.035$\pm$0.028 for x and
$\tau_0=-$0.068$\pm$0.025 for y. From this we derive that for the best-fit model the
flux of the x-line should be lower by 3$\pm$3\% and that of y should be higher by 8$\pm$3\%
compared to our SPEX plasma model in order to give the best agreement with the observed spectrum (table~\ref{tab:parameters}).

The atomic uncertainties on the x, y, and z lines are calculated using a Monte-Carlo
simulation in section~\ref{sect:lineerhe}. Based on the simulated data, we further estimate that
the errors on the x and y relative to the forbidden line ratios are 6.2\% and 9.2\%, respectively.
Hence the best-fit modifications to the x and y lines are well in line with the expected
atomic errors.

\subsection{Transition probability \label{sect:transition}}

\begin{figure}[!htbp]
\includegraphics[width=8cm]{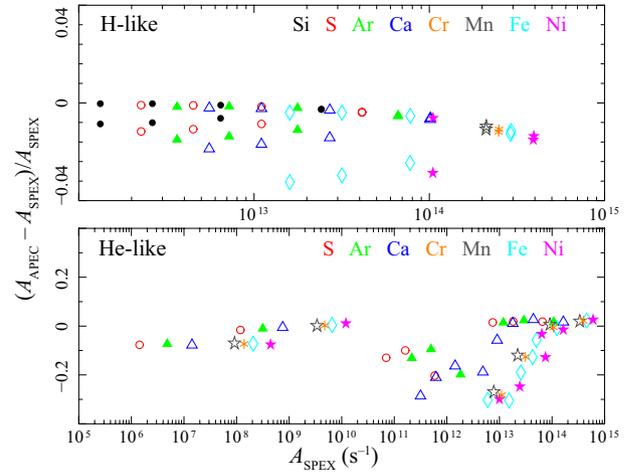}
\caption{The differences of Einstein $A$ values of the strongest transitions (table~\ref{tab:wavelength})
in SPEX v3.03 and AtomDB v3.0.8 for the H-like (upper) and He-like (lower) ions. The
upper states are in the range of $n =$2--5 for H-like and 2--4 for He-like ions.}
\label{fig:tr1}
\end{figure}

\begin{figure}[!htbp]
\includegraphics[width=8cm]{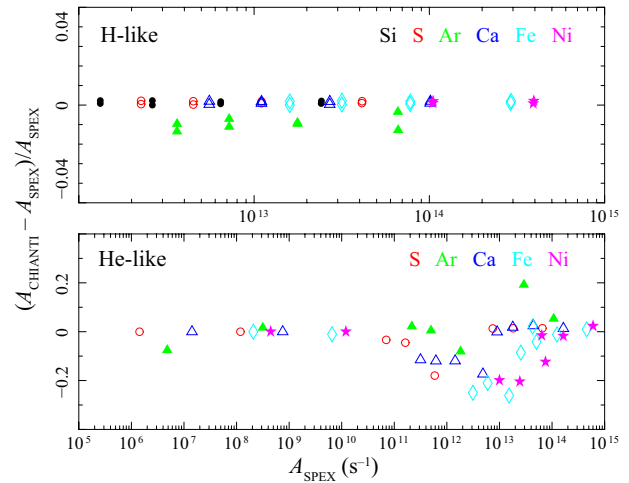}
\caption{Same as figure~\ref{fig:tr1}, but for comparison between SPEX v3.03 and CHIANTI v8.0.}
\label{fig:tr2}
\end{figure}

Besides the radiative transition data for the He-like triplet shown above, here we make a more
systematic comparison of the transition probabilities among the atomic codes. The radiative data
for selected strong lines are shown in appendix~\ref{app:linelist}
(table~\ref{tab:wavelength}). In figures~\ref{fig:tr1} and
\ref{fig:tr2}, we demonstrate that the Einstein $A$ values for H-like ions are consistent within a few
percent among the codes, while for He-like ions, especially for transitions from $n=$3 or more
to the ground, the $A$ values have larger uncertainties up to 30\%. The SPEX $A$ values are systematically
higher than those in AtomDB and CHIANTI. Partly owing to the difference in the transition data,
the He$\beta$, He$\gamma$, and He$\delta$ line
intensities calculated by SPEX are higher than the AtomDB and CHIANTI lines (see details in table~\ref{tab:wavelength}).
These lines contribute a minor role in the abundance measurements.

\subsection{Satellite line emission \label{sect:satellite}}
\begin{figure}[!htbp]
\includegraphics[width=8cm]{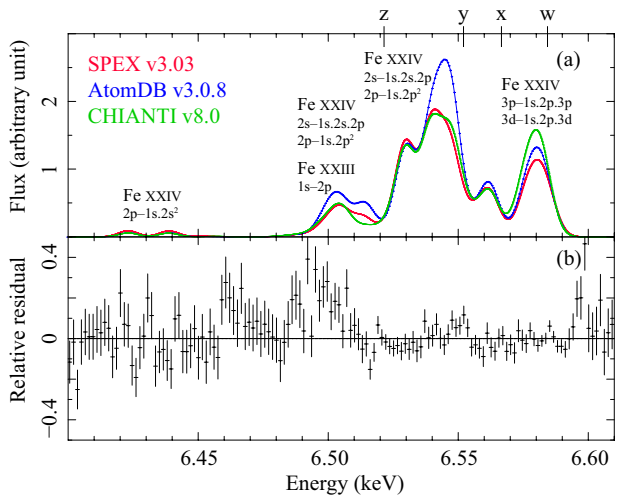}
\caption{(a) Li-like Fe\emissiontype{XXIV} and Be-like Fe\emissiontype{XXIII} lines of a 4-keV CIE plasma in the 6.4--6.61~keV band. The observed line broadening is taken into account. (b) Residual of the Hitomi SXS spectrum to the baseline fit in the same band.}
\label{fig:sat}
\end{figure}

\begin{table*}[!htbp]
\caption{Energies, Auger transition rates, and branching ratios of Fe\emissiontype{XXIV}.}
\label{tab:sat}
\centerline{
\begin{tabular}{@{\extracolsep{4pt}}lcccccc}
\hline
Level                            & Energy\footnotemark[$*$] & \multicolumn{2}{c}{Auger transition rate (s$^{-1}$)} & \multicolumn{2}{c}{Branching ratio for radiative transition} \\ \cline{3-4} \cline{5-6}
                                 &        (keV) &              SPEX v3.03 &              AtomDB v3.0.8 &                    SPEX v3.03 &                AtomDB v3.0.8 \\
\hline
1s.2s$^{2}$ $^{2}$S$_{1/2}$      &      6.60040 & 1.46$\times$10$^{14}$ &  1.43$\times$10$^{14}$ &  0.12 &  0.12 \\
1s.2s.($^3$S).2p $^{4}$P$_{1/2}$ &      6.61369 & 7.79$\times$10$^{10}$ &  1.33$\times$10$^{10}$ &  0.98 &  1.00 \\
1s.2s.($^3$S).2p $^{4}$P$_{3/2}$ &      6.61666 & 7.23$\times$10$^{11}$ &  3.91$\times$10$^{11}$ &  0.96 &  0.98 \\
1s.2s.($^3$S).2p $^{4}$P$_{5/2}$ &      6.62781 & 3.53$\times$10$^{4}$  &  1.97$\times$10$^{9}$  &  1.00 &  0.76 \\
1s.2s.($^1$S).2p $^{2}$P$_{1/2}$ &      6.65348 & 3.89$\times$10$^{13}$ &  4.24$\times$10$^{13}$ &  0.89 &  0.88 \\
1s.2s.($^3$S).2p $^{2}$P$_{3/2}$ &      6.66194 & 5.70$\times$10$^{8}$  &  1.41$\times$10$^{11}$ &  1.00 &  1.00 \\
1s.2p$^2$ $^{4}$P$_{1/2}$        &      6.67097 & 2.41$\times$10$^{11}$ &  3.37$\times$10$^{11}$ &  0.99 &  0.99 \\
1s.2s.($^3$S).2p $^{2}$P$_{1/2}$ &      6.67644 & 8.18$\times$10$^{13}$ &  6.77$\times$10$^{13}$ &  0.69 &  0.73 \\
1s.2s.($^1$S).2p $^{2}$P$_{3/2}$ &      6.67915 & 1.10$\times$10$^{14}$ &  1.07$\times$10$^{14}$ &  0.01 &  0.04 \\
1s.2p$^2$ $^{4}$P$_{3/2}$        &      6.67928 & 8.46$\times$10$^{11}$ &  9.66$\times$10$^{11}$ &  0.92 &  0.92 \\
1s.2p$^2$ $^{4}$P$_{5/2}$        &      6.68498 & 2.27$\times$10$^{13}$ &  2.61$\times$10$^{13}$ &  0.60 &  0.58 \\
1s.2p$^2$ $^{2}$D$_{3/2}$        &      6.70268 & 1.29$\times$10$^{14}$ &  1.25$\times$10$^{14}$ &  0.74 &  0.74 \\
1s.2p$^2$ $^{2}$P$_{1/2}$        &      6.70458 & 9.93$\times$10$^{11}$ &  9.39$\times$10$^{11}$ &  1.00 &  1.00 \\
1s.2p$^2$ $^{2}$D$_{5/2}$        &      6.70902 & 1.39$\times$10$^{14}$ &  1.37$\times$10$^{14}$ &  0.60 &   0.6 \\
1s.2p$^2$ $^{2}$D$_{3/2}$        &      6.72236 & 3.48$\times$10$^{13}$ &  3.28$\times$10$^{13}$ &  0.95 &  0.95 \\
1s.2p$^2$ $^{2}$S$_{1/2}$        &      6.74147 & 2.53$\times$10$^{13}$ &  2.92$\times$10$^{13}$ &  0.91 &   0.9 \\
\hline
\end{tabular}
}
\begin{tabnote}
\footnotemark[$*$] Energy levels above the ground state in SPEX v3.03.
\end{tabnote}
\end{table*}

The line energies, radiative transitions, and emissions of the satellite lines for
a 4-keV CIE plasma are compared in table~\ref{tab:satellite} in appendix~\ref{app:linelist}. The atomic level-dependent Auger transition rates and radiative-to-total branching ratios
are shown in table~\ref{tab:sat}, and the resulting line spectra for Fe are plotted in figure~\ref{fig:sat}. The most
noticeable issue is that APEC version 3.0.8 gives higher Fe\emissiontype{XXIV} fluxes at $\sim$6.5~keV and 6.545~keV than the other two codes, which
is driven by a recent update of APEC by incorporating the dielectronic recombination (DR) rates and branching ratios calculated
in \citet{palmeri2003}. This could partially explain the different Fe abundances with SPEX and APEC as shown in table~\ref{tab:parameters}.

\subsection{Ionization equilibrium concentrations \label{sect:ibal}}

\begin{figure}[!htbp]
\includegraphics[width=8cm]{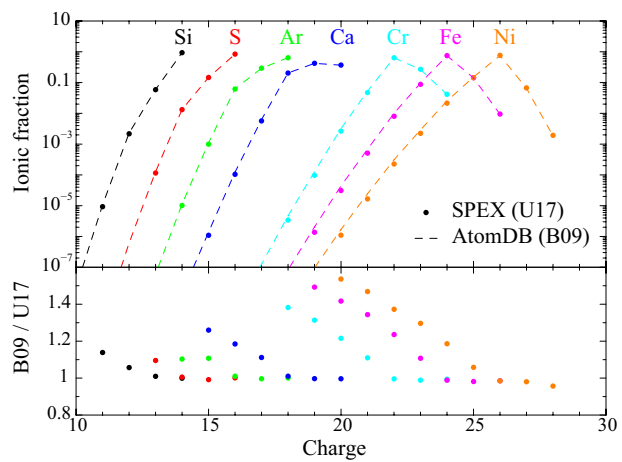}
\caption{(upper) Ionic fraction of a 4-keV CIE plasma in SPEX v3.03 (U17: dots) and APEC v3.0.8 (B09: curves). (lower) Ratios between the two.}
\label{fig:ionbal}
\end{figure}

Figure~\ref{fig:ionbal} shows relative ionic fractions of a 4-keV CIE plasma based
on the SPEX and APEC calculations. In SPEX the ionization balance mode was set to
\citet{urdampilleta2017} which allows to include inner-shell ionization
contributions to the spectrum, while in APEC the balance from \citet{bryans2009} was assumed.
For He-like, H-like, and bare ions of Si--Cu, these two calculations agree with
each other within 5\%. For Li-like ions, they agree within 10\%. For higher
sequences, however, larger differences are seen as the APEC values are
systematically larger, by up to 57\%, than the SPEX values.

Here we assess the uncertainties on ionization concentration by replacing the
baseline \citet[][hereafter U17]{urdampilleta2017} balance with historical ones,
namely \citet[][AR85]{arnaud1985}, \citet[][AR92]{arnaud1992}\footnote{Because \citet{arnaud1992} reported updates only on the Fe ionization concentration, the calculations of \citet{arnaud1985} are utilized for the other elements.},
and \citet[][B09]{bryans2009}. It should be noted that the AR85 and AR92 balances
do not include trace elements,
such as Cr and Mn. As shown in table~\ref{tab:parameters}, the baseline model
with the AR85 and AR92 ionization balances becomes much worse by $\delta C_{\rm stat}$ of about 100,
and the best-fit temperature and abundances changes by 1--3\%. The B09 balance provides an equally good
fit as the U17 one, yielding almost the same parameters except for
the Fe abundance, which increases by 4\%. The $N_{\rm H}$ of the self-absorption component
changes by 6--13\% for different balances. By
comparing the values from the mostly used B09 and U17 balances, the systematic uncertainty on abundances from
ionization concentration is 1--4\%.

\begin{figure}[!htbp]
\includegraphics[width=8cm]{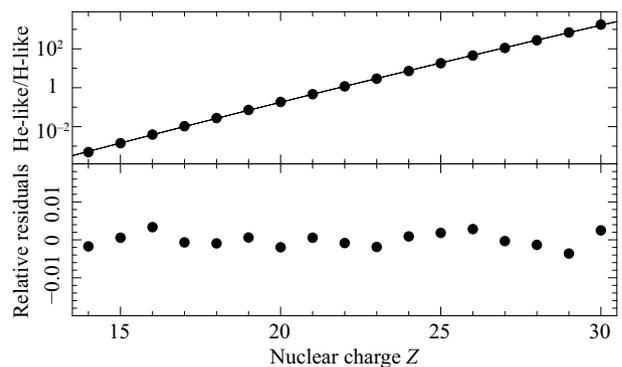}
\caption{(upper) He-like to H-like ion ratio as a function of nuclear charge $Z$ for a 4-keV plasma. The solid line shows a
polynomial fit in logarithm space. (lower) Relative residuals of a polynomial fit.}
\label{fig:he-h-con}
\end{figure}

A related issue is the uncertainty on the He-like to H-like ion ratios. As shown in
appendix~\ref{app:speccomp} (figures~\ref{fig:spec1}--\ref{fig:spec3}), the He- and Ly-series are the
dominant line feature of the Perseus spectrum, and their ratios largely determine the temperature
measurement. Here we examine the He-like to H-like ion ratios as a function of nuclear charge $Z$,
which is expected in theory to be a perfectly smooth function. The calculation is based on SPEX version 3.03.
As shown in figure~\ref{fig:he-h-con}, the He-like to H-like ion ratio indeed appears as a nearly linear function
in logarithmic space, and the scatter is within 0.5\%.

\section{Systematic factors affecting the derived source parameters: plasma
modeling\label{sect:plasmamodel}}

Although it is in principle straightforward to calculate a spectrum from the atomic data, practically these calculations
are based on a range of approximations, and usually include only limited physical processes ---
treatment of specific physical processes is limited or missing entirely. This section
explores these technical issues in the plasma modeling and discusses their impacts on the fitted parameters.

\subsection{Voigt profiles\label{sect:voigt}}

\begin{figure}[!htbp]
\includegraphics[width=8cm]{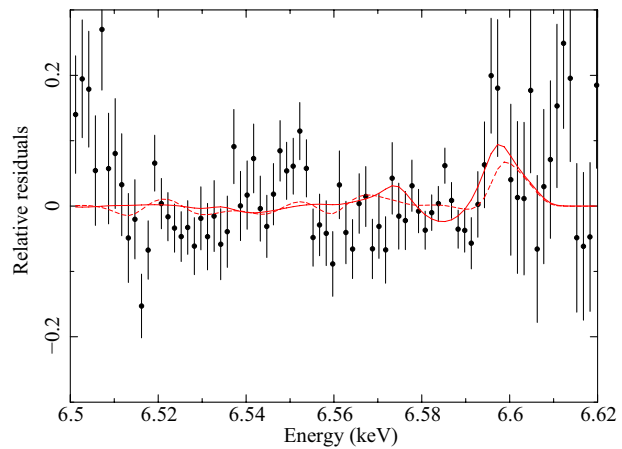}
\caption{Residuals of the baseline model near the He-like 1s-2p transitions.
The dotted curve shows the predicted model change when instead of Doppler
profiles Voigt profiles are chosen, but without altering the parameters
of the baseline model. The solid curve shows the same but after refitting
the spectrum.}
\label{fig:voigt}
\end{figure}

In our baseline model we have approximated the line profiles using Doppler
profiles (Gaussians). This gives a significant increase in speed in obtaining our spectral
fits. However, the true profiles are Voigt profiles. We have tested the
sensitivity of our results to these intrinsic line profile assumptions.
The Lorentzian widths of the Voigt profiles are fixed to the natural widths in SPEX version 3.03.
Figure~\ref{fig:voigt} shows our results. The changes are substantial (5--10\%)
near the Fe\emissiontype{XXV} resonance line observed at 6.60~keV. In all other parts
of the spectrum the changes are smaller, due to the fact that the lines are
weaker.

\subsection{Continuum contributions from heavy elements\label{sect:ac}}

\begin{figure}[!htbp]
\includegraphics[width=8cm]{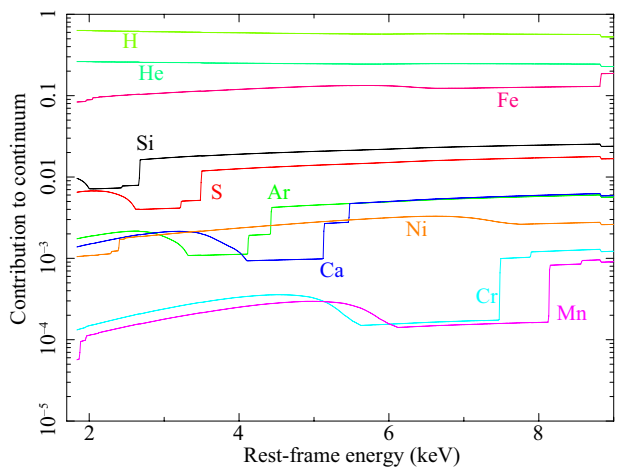}
\caption{Relative contribution of the different elements to the continuum
emission of the main thermal component in the Perseus cluster.}
\label{fig:aburat}
\end{figure}

Not only abundant elements like H, He, O, and Fe contribute significantly to the
continuum emission, but also the contributions of less abundant elements like Cr
or Mn are detectable. We discovered this by accident when we tested our baseline
model with the old version of SPEX (version 2). In that old version only the 15 most
abundant elements with nuclear charge less than 30 were taken into account in
the line emission, yet the model could produce some very crude constraints on
the Cr and Mn abundance, while the line emission of both elements was not
accounted for by the model.

What is the explanation for this? In figure~\ref{fig:aburat} we show the relative
contribution of each element to the continuum emission (including here also the
AGN continuum). About 90\% of the emission is due to H and He, about 10\% is due
to Fe, and all other elements contribute less than a few percent at most. In
particular for the elements between Si and Mn clearly the smooth two-photon
emission bumps and the free-bound edges are visible.

The present spectrum has 487621 counts with a nominal uncertainty of 698 counts.
Cr and Mn contribute 78 and 104 counts to the continuum, respectively. Therefore
their contribution is small but if the abundances would have been off by a
factor of 10, their continuum contribution with their specific structure as
shown in figure~\ref{fig:aburat} would have allowed to constrain their abundances.

In SPEX all contributions to
the radiative recombination (free-bound) continuum smaller than a threshold are
omitted for computational efficiency (the free-bound continuum calculation takes
most of the computing time for high-resolution spectra because of the large
number of energy bins and atomic shells that need to be calculated). The
threshold is controlled by the parameter "gacc" that can be set by the user. Its
default value is 10$^{-3}$, but for figure~\ref{fig:aburat} we have put it to
10$^{-7}$.

This same value is used in the entry $g_{\rm{acc}}$ listed in
table~\ref{tab:parameters}. It can be seen that changing this parameter has only
a very minor effect on the fit (improvement of C-statistic only 0.41), but the
additional computational burden is heavy.

\subsection{Maximum principal quantum number $n$ in the calculations\label{sect:maxn}}

Collisional excitation by thermal electrons mostly populates the inner shells of
the atomic structure. Although the emission lines from outer shells are usually
rather weak, some of them become visible in the Hitomi SXS spectrum
(appendix~\ref{app:linelist}; table~\ref{tab:wavelength}).
Here we test the impact on the obtained spectral
parameters by limiting the maximum principal quantum number $n$ in the
calculation. As shown in table~\ref{tab:parameters}, when excluding the outer
shells with $n > 5$, the fitting with the baseline model gets poorer by
$\delta C_{\rm stat}\approx$61, and the best-fit metal abundances become slightly larger by a few
percent. This is because the outer shell population will also contribute to
the inner shell (e.g., Ly$\alpha$, Ly$\beta$) transitions by radiative
cascading. As shown in table~\ref{tab:wavelength}, the Hitomi SXS data require the
plasma code to calculate to at least $n=10$ for the Fe\emissiontype{XXV} lines.

\subsection{Hyperfine mediated transitions \label{sect:hmt}}

The isotopic composition of Fe contains approximately 2\% of $^{57}$Fe, which
has non-zero nuclear spin and thus might be expected to exhibit a
hyperfine-mediated transition from 1s2p $^{3}$P$_0$ to ground, resulting in a
weak third intercombination line. The transition rate has been
calculated by \citet{johnson1997}, who find that it is about
6\% of the transition rate to the 1s2s $^3$S$_1$ state, so that the strength of
the 1s2p $^3$P$_0$ transition to ground is negligible for Fe. The low
branching ratio to ground can be attributed to the relatively weak
magnetic moment of $^{57}$Fe. We caution that all odd $Z$ elements have
non-zero nuclear magnetic moments, and for most of those ions in the Fe
group, the hyperfine mediated decay channel to ground is actually dominant.

\section{Systematic factors affecting the derived source parameters:
astrophysical model\label{sect:astromodel}}

The atomic data and plasma code are eventually integrated into the spectral models.
To verify the spectral modeling with the Hitomi SXS data, it is important to test it
in a proper astrophysical context. In this session, we will incorporate several astrophysical
effects, such as non-equilibrium and multi-temperature,
examine their spectral features with the data, and calculate the related uncertainties on the fitted parameters.
The physical implication of these effects will be discussed in other Hitomi Collaboration papers.

\subsection{Ion temperature versus turbulence \label{sect:iont}}

The basic assumption made in our earlier paper on turbulence \citep{hitomi2016}
is that the ion temperature of the cluster gas equals the electron temperature.
Given the relatively high density in the core of the Perseus cluster \citep[$\sim$0.05~cm$^{-3}$:][]{zhuravleva2014}
compared to the outskirts \citep[$\sim$10$^{-4}$~cm$^{-3}$:][]{urban2014},
this assumption may be justified, but in other circumstances it may be different.

In order to test this, we have decoupled the ion temperature from the electron
temperature in our model and refitted the spectrum. We get an insignificant
improvement of our fit ($\delta C_{\rm stat} = $ $-$0.02) with the best-fit values
of the ion temperature of $kT_{\rm ion} = $~4.1~($-$2.3, $+$3.2)~keV and turbulent velocity
$\sigma_v = $~156~($-$21, $+$13)~km\,s$^{-1}$. However, there is a strong anti-correlation between
both parameters. Without constraints on the ion temperature,
$\sigma_v$ can be anywhere between 134 and 168~km\,s$^{-1}$.
The best-fit values of these parameters depend on details of
the spectral analysis method, although the differences are smaller than the
statistical errors. Such systematic effects are separately discussed in V~paper.
Note that for a fixed ion temperature, the uncertainty on the turbulent velocity is much
smaller, i.e., only 3~km\,s$^{-1}$. We show the (minor) effects of a free ion
temperature on the other parameters in table~\ref{tab:parameters}.

\subsection{Deviations from collisional ionization equilibrium \label{sect:nei}}

The core of the Perseus cluster is a very dynamical environment, with a
relatively high density and an active galactic nucleus at its center. Therefore,
in principle one might expect non-equilibrium ionization effects to play a role.
We have tested this as follows.

The most simple test is to decouple the temperature used for the ionization
balance calculations, $T_{\rm bal}$, from the (electron) temperature $T_{\rm
spec}$ used for the evaluation of the emitted spectrum for the set of ionic
abundances obtained using $T_{\rm bal}$. This can be achieved within the SPEX
package by making the parameter $RT \equiv T_{\rm bal}/T_{\rm spec}$ a free
parameter. We obtain a best fit for $RT=$0.980$\pm$0.011, i.e., close to unity,
with only a modest improvement in C-statistic of 3.26.

Alternatively, we can replace the basic CIE model by a genuine non-equilibrium
ionization (NEI) model in SPEX. This model can mimic a plasma that suddenly
changes its electron temperature from a value $T_1$ to a value $T_2$. The
spectrum is then evaluated after a time $t$, related to the measured relaxation
timescale $U$ by $U=\int n_{\rm e}{\rm d}t$, the electron density
integrated over time from the instant that the temperature suddenly changes.

The first case we consider is an ionizing plasma $T_1<T_2$ (labeled ``Ionizing'' in table~\ref{tab:parameters}),
which has $T_1=$1.5$\pm$0.4~keV, $T_2=$3.994$\pm$0.021~keV, and $U=($1.4$\pm$0.3$)\times$10$^{18}$~m$^{-3}$~s.
The ionizing plasma model improves the C-statistic by 5.46.

Further, we tested a recombining model by inverting the role of $T_1$ and $T_2$ (model labeled ``Recombining'').
Leaving $T_1$ free it appears that it gets to a very high value.
Therefore we choose to fix $T_1$ to a high value (100~keV) so we start essentially with a fully ionized plasma.
We obtain $T_2=$3.933$\pm$0.020~keV, $U=($2.5$\pm$0.2$)\times$10$^{18}$~m$^{-3}$~s, and an improvement in C-statistic of 9.19.

The above may suggest that there are some significant although minor
non-equilibrium effects. However, we cannot claim such effects here. First,
nominally our fits are very close to equilibrium ($RT\approx1$ or $U\approx$10$^{19}$~m$^{-3}$~s).
The best-fit value for $RT$ may differ from unity at the 1.9-$\sigma$ confidence level,
but the absolute difference is only 2.0\%. It is likely true that the systematic uncertainties on
the ionization and recombination rates are large enough to account for such a
small deviation from equilibrium. For example, when we increase all ionization rates for
iron ions arbitrarily by 5\%, the peak concentration of Fe\emissiontype{XXV} for the
baseline model would increase from 0.747 to 0.750; a lowering of the
temperature by 1\% would have the same effect on the Fe\emissiontype{XXV}
concentration.

Another issue is that introducing multi-temperature structure
(section~\ref{sect:multit}) gives much larger improvements to the fit. Clearly,
the Perseus core region contains multiple-temperature components, and at such a
level that weak non-equilibrium effects cannot be separated from it.

\subsection{Effects of the spatial structure of the Perseus cluster \label{sect:gdem}}

Up to now, we have treated the Perseus spectrum with relatively simple spectral
models. In reality, Perseus shows temperature and abundance gradients. How do
they affect our analysis? We investigate this through simulation. Our goal here is
to estimate the systematic uncertainties on the derived parameters
resulting from neglecting the spatial structure of Perseus.

We proceed as follows. We have taken the radial temperature and density profile
derived from deprojected Chandra spectra as given by \citet[][extended data
figure~1]{zhuravleva2014}. For the radial abundance profile we have adopted the
average profile for a large sample of clusters based on XMM-Newton data
\citep{mernier2016b}. We have not chosen their profile derived from the Perseus
data alone, because that is noisier than the average profile for the full set of
clusters. \citet{mernier2016b} show that in general the radial abundance profiles
of individual clusters agree well with this average profile.

\begin{figure}[!htbp]
\includegraphics[width=8cm]{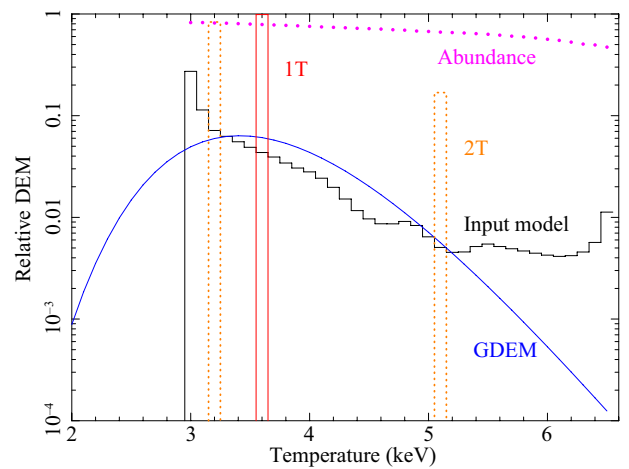}
\caption{The predicted DEM profile of the Perseus cluster within
the FOV of the Hitomi SXS (black histogram) and the corresponding
average abundance profile (relative to the proto-solar values of \citealt{lodders2009}; magenta dots).
The best-fit isothermal (1T) and two-temperature (2T) models to this
DEM are shown with the red and orange histograms, respectively,
and the best-fit GDEM model with the blue curve.}
\label{fig:zdemsim}
\end{figure}

We have then integrated these 3D profiles over the line of sight through the
projected FOV of the Hitomi SXS for our present observations. We accounted for
the different pointing position for Obs\,2$+$3 compared to Obs\,4 by
weighing with the relative exposure times. This way we have obtained the
differential emission measure distribution (DEM) within the FOV of the Hitomi SXS.
We have binned it in 0.1-keV wide temperature bins.
The total emission measure is 1.003$\times$10$^{73}$~m$^{-3}$.
Figure~\ref{fig:zdemsim} shows this distribution (normalized to integral unity) as well
as the average abundance for each temperature bin.

We see that the DEM is strongly peaked towards 3~keV, and decreases rapidly
towards higher temperatures. This peak corresponds to the coldest gas in the
center of the cluster using the \citet{zhuravleva2014} parameterization (we have
assumed that the temperature remains constant for radii smaller than 10~kpc).
The DEM then flattens near 5~keV and turns up again above 6.4~keV. This
corresponds to the peak in the radial temperature distribution around 250~kpc.
The abundance drops almost continuously from 0.82 in the center (at 3~keV) to
0.47 at 6.5~keV.

Thus, we are faced with an extremely skewed DEM distribution over a range of
only a factor of two difference in temperature, combined with a monotonous
declining abundance pattern that also differs by a factor of two from low to
high temperatures. How does this affect our modeling?

We have taken our baseline model, and replaced the main $\sim$4-keV emission
component with the 36 temperature components shown in figure~\ref{fig:zdemsim}.
The abundances for the different
temperature components are the ones shown in figure~\ref{fig:zdemsim}. For
simplicity we assume that all elements have the same abundance. All other
spectral components (absorption, AGN contribution, etc.) are
taken to be exactly the same as in our best-fit baseline model. We then
simulated this spectrum with the same exposure time as the observed Hitomi SXS
spectrum, and fitted this simulated spectrum the same way as our baseline model.

In order to avoid the overhead of having to simulate many different cases, we
have turned the random noise in our simulations off. In this way we get with a
single simulation the best-fit parameters (and their uncertainties where
needed). A perfect fit would then yield a formal C-statistic of 0.

\begin{table}[!htbp]
\caption{Best-fit parameters of the simulated Perseus spectrum.}
\label{tab:simspec}
\centerline{
\begin{tabular}{lcccc}
\hline
Parameter  & 1T   & GDEM   & 2T & input \\
\hline
$C_{\rm stat}$             & 36.37 & 3.27  & 2.64  & 0     \\
$Y_1$ ($10^{72}$~m$^{-3}$) &  9.89 & 10.11 & 8.36  & 10.03 \\
$kT_1$ (keV)               & 3.622 & 3.529 & 3.292 & 3.624 \\
$\sigma_T$\footnotemark[$*$] & --- & 0.112 & ---   & ---    \\
$Y_2$ ($10^{72}$~m$^{-3}$) & ---   & ---   & 1.69  & ---   \\
$kT_2$ (keV)               & ---   & ---   & 5.12  & ---   \\
Si                         & 0.853 & 0.787 & 0.803 & 0.778 \\
S                          & 0.845 & 0.787 & 0.797 & 0.778 \\
Ar                         & 0.810 & 0.786 & 0.784 & 0.778 \\
Ca                         & 0.778 & 0.784 & 0.778 & 0.778 \\
Cr                         & 0.716 & 0.763 & 0.768 & 0.778 \\
Mn                         & 0.697 & 0.751 & 0.777 & 0.778 \\
Fe                         & 0.725 & 0.747 & 0.758 & 0.778 \\
Ni                         & 0.747 & 0.763 & 0.769 & 0.778 \\
\hline
\end{tabular}
}
\begin{tabnote}
\footnotemark[$*$] A logarithmic temperature scale $\sigma_{T}$ of the GDEM model. \\
\end{tabnote}
\end{table}

We first fit this simulated spectrum with our baseline model, where the thermal
emission is modeled as a single-temperature component (labeled as 1T).
The best-fit reaches a C-statistic value of 36.37, i.e.,
the isothermal approximation is poorer by 36.37 compared with the true underlying spectrum.
This fit (table~\ref{tab:simspec})
shows some clear biases. First, the abundance of Si and S, with lines at the
low-energy end of the spectrum, are too high by about 10\% compared to the input
model (the input model does not have a single abundance, but we list the
emission-measure weighted abundance for the input model in
table~\ref{tab:simspec}). On the other hand, the Fe and Ni abundances are too low
by 4\%. As a result, the Si/Fe ratio is even off by 15\%. This bias can be
understood from the different temperature dependence of the Si/S lines compared
to the Fe/Ni lines. Our model forces these lines to be formed at the same
temperature, and the only way to get the line fluxes more or less right is to
adjust the abundances.

Interestingly, the Cr and Mn abundances are even lower, by 8--10\%. This is due to
the fact that the 1T model in the simulation under-predicts the true continuum
near the dominant Cr and Mn lines by about 0.3\%. As a result, the total
simulated flux near these lines can be recovered only by reducing the
abundances.

The temperature for this simulated 1T model (3.62~keV) is slightly lower than
the temperature for the baseline model (4.05~keV). There may be various reasons
for this. First, our spherically symmetric model for the Perseus cluster that we
used may be too simplistic. For example, the Chandra intensity map of the
Perseus cluster \citep[][figure~1]{zhuravleva2014} shows non-azimuthal fluctuations
up to about 50\% due to various structures within the Perseus core. Also, there
are calibration uncertainties; for instance, for 4-keV plasmas,
\citet{schellenberger2015} shows differences between temperatures derived from
Chandra and XMM-Newton that can easily reach 10\%. It is not unfeasible that
similar differences would exist between the Hitomi SXS temperature scale and that of
Chandra. Finally, even with fully deprojected spectra, at the same distance from
the cluster center multiple temperature components may co-exist due to different
cooling or heating histories of different plasma elements
\citep[e.g.][]{kaastra2004}.

We then fit the simulated spectrum with the Gaussian DEM (GDEM) model, where the
DEM is log-normally distributed (the blue curve in figure~\ref{fig:zdemsim}).
This model gives a much better description of the simulated spectrum
(table~\ref{tab:simspec}), with a C-statistic of only 3.27. The corresponding DEM is
quite different from the DEM of our input model (the black histogram in figure~\ref{fig:zdemsim})
but because it has the same total emission measure, average temperature and variance
as the input DEM distribution, the corresponding spectra are very similar.
Note that while the model parameter for the temperature of the GDEM model is
3.53~keV, its emission-measure weighted temperature is 3.59~keV, which is very
close to the emission-measure weighted temperature of the input model (3.62~keV)
or the 1T fit (3.62~keV).
There is still a small bias in the derived abundances, but it is less than 4\%
for all elements.

The last model we fit to this simulated spectrum is a two-temperature component model (2T)
with the abundances of both components tied together. This provides the best-fit
(table~\ref{tab:simspec}) with a C-statistic value of only 2.64 and and abundance bias smaller than 3\%.

\begin{figure}[!htbp]
\includegraphics[width=8cm]{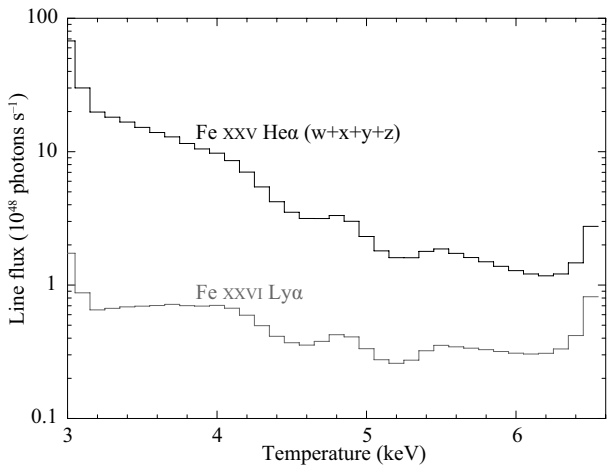}
\caption{The predicted distribution of the line flux of the main $n=$1--2
transitions of Fe\emissiontype{XXV} and Fe\emissiontype{XXVI} within the Hitomi SXS FOV.}
\label{fig:felinesim}
\end{figure}

Finally, we have investigated the properties of the strongest lines in the
spectrum. Defining line fluxes can be done in two ways: either taking the
``pure'' line flux, or also including other weak lines that are blended with the
line of interest at the spectral resolution of the instrument. We have chosen
the latter approach, and included the flux from all lines within $\pm 2$~eV from
the line of interest. Figure~\ref{fig:felinesim} shows the combined line flux of
the four He$\alpha$ transitions w, x, y, and z of Fe\emissiontype{XXV} and the sum of both
Ly$\alpha$ lines of Fe\emissiontype{XXVI}. No resonance scattering has been taken into
account in these calculations. It is seen that the Fe\emissiontype{XXV} emission is more
concentrated towards lower temperatures (the emission-weighted temperature for
this ion is 3.69~keV), while Fe\emissiontype{XXVI} has a much flatter distribution
(average temperature for this ion is 4.39~keV). Also, the ratio of the sum of
the x, y, and z line fluxes to the w line flux changes significantly over this
temperature range: from 0.79 at 3~keV to 0.63 at 6.5~keV.

\subsection{Multi-temperature fitting of the Hitomi SXS data \label{sect:multit}}

\begin{table*}[!htbp]
\caption{Parameters of the multi-temperature fits.}
\footnotesize
\label{tab:multi-t}
\centerline{
\begin{tabular}{@{}lccccccccccccc@{}}
\hline
Model          & $C_{\rm stat}$\footnotemark[$*$]   &$Y$ & $kT$                                  & $\sigma_v$                  & \multicolumn{8}{c}{Abundance (solar)}                                     & $N_{\rm H,hot}$      \\ \cline{6-13}
               &         & ($10^{73}$\,m$^{-3})$         & (keV)                                 & (km\,s$^{-1}$)              & Si & S    & Ar   & Ca   & Cr   & Mn    & Fe                               & Ni   & ($10^{24}$\,m$^{2}$) \\
\hline
Baseline\footnotemark[$\dagger$] & 4926.03 & 3.73        & 3.969                                 & 156                         & 0.91 & 0.94 & 0.83 & 0.88 & 0.70 & 0.74 & 0.827                           & 0.76 & 18.8 \\
GDEM           & 4865.13 & 3.85                          & 3.830, 0.130\footnotemark[$\ddagger$] & 158                         & 0.81 & 0.84 & 0.79 & 0.89 & 0.78 & 0.84 & 0.851                           & 0.79 & 16.5 \\
2CIE$_{\rm A}$ & 4867.31 & 2.62, 1.22\footnotemark[$\S$] & 3.360, 5.140\footnotemark[$\S$]       & 157                         & 0.82 & 0.85 & 0.78 & 0.88 & 0.78 & 0.84 & 0.851                           & 0.79 & 15.6 \\
2CIE$_{\rm B}$ & 4800.50 & 2.22, 1.53\footnotemark[$\S$] & 3.142, 5.166\footnotemark[$\S$]       & 106, 215\footnotemark[$\S$] & 0.81 & 0.85 & 0.80 & 0.90 & 0.80 & 0.84 & 1.041, 0.708\footnotemark[$\S$] & 0.80 &  9.9 \\
3CIE           & 4790.72 & 2.22, 1.26\footnotemark[$\|$] & 3.578, 5.118\footnotemark[$\|$]       & 112, 234\footnotemark[$\|$] & 0.78 & 0.84 & 0.82 & 0.92 & 0.78 & 0.81 & 0.916, 0.705\footnotemark[$\|$] & 0.79 & 11.9 \\
\hline
\end{tabular}
}
\normalsize
\begin{tabnote}
\footnotemark[$*$] Expected values for the baseline and multi-temperature models are 4876. \\
\footnotemark[$\dagger$] The best-fit parameters of the baseline model adopted from table~\ref{tab:parameters} for comparison.\\
\footnotemark[$\ddagger$] $\sigma_{T}$ of the GDEM model. It is a common logarithmic temperature scale. \\
\footnotemark[$\S$] Parameters of the cool and hot ICM components of the 2CIE modeling. \\
\footnotemark[$\|$] The third component has a temperature of 1.9~keV and best-fit $Y = 0.33 \times$ $10^{73}$\,m$^{-3}$. The values of $\sigma_v$ and Fe abundance are tied to those of the $\sim$3.5-keV component. \\
\end{tabnote}
\end{table*}

As shown in section~\ref{sect:gdem}, the central region of the Perseus cluster contains multiple temperature
components. To evaluate the impact of the multi-temperature structure on the ICM parameters
(e.g., turbulent velocity and abundances) for the real data, we
carry out a multi-temperature fit to the Hitomi SXS spectrum. It is known
that there is often more than one solution to fit a multi-temperature structure,
since models with different combinations of temperatures and abundances might essentially
yield a similar spectrum. Exploring these solutions is the focus of T~paper.
In this paper, we present three basic approximations for
the temperature structure, and test them using the Hitomi SXS data.

First we assume that the temperature distribution follows a GDEM form. As shown in
section~\ref{sect:gdem}, the GDEM model provides a proper approximation to the radial temperature
profile of the Perseus cluster as derived from Chandra data. In the fit, we adopt
the peak temperature, Gaussian width of the DEM, abundances, and turbulent velocity as free
parameters, and the remaining components (AGN and resonance scattering)
are modeled in the same way as in the baseline model (section~\ref{sect:baseline}). The
effective-area correction factor (appendix~\ref{app:fudge}) is also left free, as the continuum of
the GDEM model is slightly different from the single-temperature baseline model. The results
of the GDEM fits are shown in table~\ref{tab:multi-t}. The C-statistic improves by
61 compared to the baseline fit. The best-fit central temperature $T$ is 3.83$\pm$0.05~keV,
and the Gaussian width $\sigma_{T} =$~0.13$\pm$0.01, which indicates a
significant deviation from isothermality. Note that $\sigma_{T}$ is defined in units
of $\log_{10}\,(T)$, hence the value of $\sigma_{T}$ corresponds roughly to 35\% of
$T$ or 1.3~keV. The GDEM fitting gives lower Si, S, and Ar abundances, a similar Ca abundance, and
slightly higher Cr, Mn, Fe, and Ni abundances than the single-temperature run. The abundance
changes agree well with the prediction in table~\ref{tab:simspec},
indicating that the GDEM results are more close to the real values than the baseline results.
The turbulent velocity remains intact in the new fit.

As a second approach, we apply a model with two discrete temperatures. First we assume that the two temperature
components have the same set of abundances and turbulent velocity, as well as the same foreground
absorption with a column density of 1.38$\times$10$^{21}$ cm$^{-2}$. This setting is called 2CIE$_{\rm A}$. The other spectral components are
inserted in the same way as the baseline model, and the effective area fudge factor is left free in
the fitting. As shown in table~\ref{tab:multi-t}, the
C-statistic improves by 59 compared to the baseline fit. The best-fit two temperatures are 3.36$\pm$0.29~keV
and 5.14$\pm$0.30~keV, and the abundances
and turbulent velocity agree well with those with the GDEM model. The two-temperature fit can be further improved by allowing the Fe abundances and turbulence of the two
components to vary freely. This setting is then called 2CIE$_{\rm B}$. This fitting improves
the C statistic by 126 from the baseline fit.

\begin{figure}[!htbp]
\includegraphics[width=8cm]{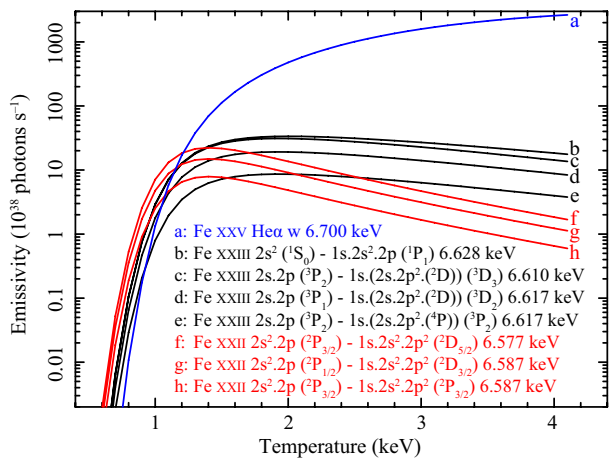}
\caption{Emissivities of B-like Fe\emissiontype{XXII} and Be-like Fe\emissiontype{XXIII} lines at around rest-frame energies of 6.58 and 6.61~keV, respectively, as a function of temperature. The Fe\emissiontype{XXV} He$\alpha$ w line is plotted as a reference.}
\label{fig:fe23}
\end{figure}

\begin{figure}[!htbp]
\includegraphics[width=8cm]{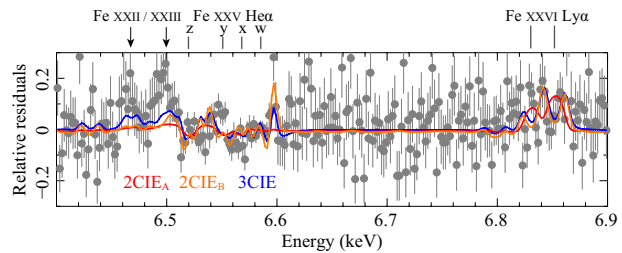}
\caption{Residuals of the baseline fitting in the Fe band.
The red, orange, and blue curves show the changes by the 2CIE$_{\rm A}$, 2CIE$_{\rm B}$, and 3CIE models, respectively. The black arrows indicate the redshifted energies of Fe\emissiontype{XXII} and Fe\emissiontype{XXIII} DR line complex.}
\label{fig:2tcom}
\end{figure}

The fitting result (figure~\ref{fig:spec2} in appendix~\ref{app:speccomp}) shows positive residuals,
about 10--20\% of the continuum level, at $\sim$6.47~keV and 6.50~keV in the baseline fit.
These residuals remain intact in the fittings with the GDEM or the two-temperature models.
Assuming that these features are emitted at the rest-frame of the Perseus cluster,
they coincide with the Fe\emissiontype{XXII} and Fe\emissiontype{XXIII} DR lines, at 6.58~keV and 6.61~keV, respectively.
As shown in figure~\ref{fig:fe23}, these Fe\emissiontype{XXII} and Fe\emissiontype{XXIII} lines are
important in low temperature (1--2~keV) plasma. Indeed such a component was recently reported to be
associated with the H$\alpha$-bright filaments in the Perseus cluster \citep{walker2015}.
Hence we extend the 2CIE$_{\rm B}$ model by adding a third CIE component. It becomes the 3CIE model
shown in table~\ref{tab:multi-t}. Since the third component cannot be determined well with the Hitomi SXS spectrum,
we tie all of its parameters, except for the temperature and emission measure, to those of the $\sim$3.5-keV component.
The best-fit temperature and emission measure are 1.9~keV and 0.33$\times$ 10$^{73}$~m$^{-3}$.
The 3CIE model improves the C-statistic by 135 from the baseline fit, although the improvement from
the 2CIE$_{\rm B}$ model is not significant ($\delta C_{\rm stat}=$10). As shown in figure~\ref{fig:2tcom},
compared to the two-temperature fits, the 3CIE fit provides a better description to the Fe\emissiontype{XXII} and
Fe\emissiontype{XXIII} complex at 6.47--6.50~keV.

\subsection{Helium abundance \label{sect:heab}}

Helium is an interesting element. It does not have line transitions in the
X-ray band, yet its continuum contribution relative to hydrogen varies by
$\sim$5\% over the Hitomi SXS band, and therefore our results are affected by the
adopted He abundance. It has been discussed that the He abundance in cluster
cores may be enhanced by a factor of two or more due to sedimentation
\citep{fabian1977,gilfanov1984,qin2000,chuzhoy2003,ettori2006,medvedev2014,berlok2016}.
However, the magnitude of the effect is hard to predict due to the role of the magnetic topology, plasma instabilities, gas mixing by mergers and turbulence, and the formation of a cool core.

We have tested the effects of an enhanced He abundance on our baseline
model by enhancing the He abundance to 1.1 times its original value. The effects
are shown in table~\ref{tab:parameters}. The main effect is an enhancement of
the abundances of all metals by 0.02--0.03.

\section{Systematic factors affecting the derived source parameters: spectral
components\label{sect:components}}

Besides the (near-)thermal emission from the ICM, the Hitomi SXS spectrum might contain
additional spectral components, such as the resonance scattering and the charge
exchange between hot and cold matter. Are these components properly modeled in the
current atomic codes? We investigate the additional spectral components and
calculate the induced uncertainties on the derived properties of the main thermal component.

\subsection{Self-absorption by hot gas\label{sect:selfab}}

\begin{table}[!htbp]
\caption{Strongest absorption lines in the Perseus spectrum\footnotemark[$*$].}
\label{tab:abslines}
\centerline{
\begin{tabular}{lcccc}
\hline
Line                                & $E$    & $\tau_0$ & $f$ & $A$ \\
                                    & (keV)  &          &    & ($10^{12}$~s$^{-1}$) \\ 
\hline
O\emissiontype{VIII}   Ly$\alpha_2$ & 0.6535 & 0.017 & 0.139 & 2.6 \\
O\emissiontype{VIII}   Ly$\alpha_1$ & 0.6537 & 0.033 & 0.277 & 2.6 \\
Ne\emissiontype{X}     Ly$\alpha_2$ & 1.0215 & 0.010 & 0.139 & 6.3 \\
Ne\emissiontype{X}     Ly$\alpha_1$ & 1.0220 & 0.020 & 0.277 & 6.3 \\
Mg\emissiontype{XII}   Ly$\alpha_2$ & 1.4717 & 0.008 & 0.139 & 13 \\
Mg\emissiontype{XII}   Ly$\alpha_1$ & 1.4726 & 0.016 & 0.277 & 13 \\
Fe\emissiontype{XXIII} L$\beta$     & 1.1251 & 0.007 & 0.256 & 4.7 \\
Fe\emissiontype{XXIII} L$\beta$     & 1.1290 & 0.011 & 0.410 & 7.6 \\
Fe\emissiontype{XXIV}  L$\beta$     & 1.1627 & 0.036 & 0.126 & 7.4 \\
Fe\emissiontype{XXIV}  L$\beta$     & 1.1674 & 0.070 & 0.243 & 7.2 \\
Fe\emissiontype{XXIV}  L$\gamma$    & 1.5505 & 0.007 & 0.032 & 3.3 \\
Fe\emissiontype{XXIV}  L$\gamma$    & 1.5525 & 0.013 & 0.062 & 3.2 \\
Fe\emissiontype{XXIV}  L$\delta$    & 1.7304 & 0.005 & 0.026 & 1.7 \\
\hline
Si\emissiontype{XIV}   Ly$\alpha_2$ & 2.0043 & 0.018 & 0.139 & 24 \\
Si\emissiontype{XIV}   Ly$\alpha_1$ & 2.0061 & 0.037 & 0.277 & 24 \\
Si\emissiontype{XIV}   Ly$\beta_1$  & 2.3766 & 0.006 & 0.053 & 6.5 \\
S\emissiontype{XV}     He$\alpha$ w & 2.4606 & 0.008 & 0.767 & 67 \\
S\emissiontype{XVI}    Ly$\alpha_2$ & 2.6197 & 0.016 & 0.139 & 41 \\
S\emissiontype{XVI}    Ly$\alpha_1$ & 2.6227 & 0.031 & 0.277 & 41 \\
Ar\emissiontype{XVII}  He$\alpha$ w & 3.1398 & 0.005 & 0.775 & 111 \\
Ar\emissiontype{XVIII} Ly$\alpha_1$ & 3.3230 & 0.010 & 0.277 & 66 \\
Ca\emissiontype{XIX}   He$\alpha$ w & 3.9023 & 0.011 & 0.782 & 172 \\
Ca\emissiontype{XX}    Ly$\alpha_1$ & 4.1075 & 0.008 & 0.277 & 101 \\
Fe\emissiontype{XXIV}  r\footnotemark[$\dagger$] & 6.6533 & 0.008 & 0.157 & 343 \\
Fe\emissiontype{XXIV}  q\footnotemark[$\dagger$] & 6.6619 & 0.025 & 0.489 & 471 \\
Fe\emissiontype{XXV}   He$\alpha$ w & 6.7004 & 0.338 & 0.798 & 518 \\
Fe\emissiontype{XXV}   He$\beta_1$  & 7.8810 & 0.056 & 0.156 & 140 \\
Fe\emissiontype{XXV}   He$\delta_1$ & 8.2955 & 0.020 & 0.058 & 58 \\
Fe\emissiontype{XXV}   He$\gamma_1$ & 8.4874 & 0.009 & 0.028 & 29 \\
Fe\emissiontype{XXVI}  Ly$\alpha_2$ & 6.9517 & 0.011 & 0.139 & 291 \\
Fe\emissiontype{XXVI}  Ly$\alpha_1$ & 6.9732 & 0.021 & 0.277 & 292 \\
Ni\emissiontype{XXVII} He$\alpha$ w & 7.8051 & 0.013 & 0.683 & 602 \\
\hline
\end{tabular}
}
\begin{tabnote}
\footnotemark[$*$] Data based on SPEX v3.03: the rest-frame energy $E$, optical depth $\tau_0$ at line center with the best-fit parameters of the baseline model, oscillator strength $f$, and transition probability $A$. \\
\footnotemark[$\dagger$] DR satellite transitions of 2s -- 1s.2s.2p. See table~\ref{tab:satellite}. \\
\end{tabnote}
\end{table}

As indicated in section~\ref{sect:baseline}, we have included a simple model to
account for the absorption of photons through the cluster gas itself.
In table~\ref{tab:abslines} we show the transitions with strong line absorptions
in the Hitomi SXS band, including the band that would have
been observed if the gate valve would have been opened.
The optical depth $\tau_0$ at line center is derived by assuming the best-fit baseline parameters
of the column density of the hot gas $N_{\rm H, hot}$, abundances,
and velocity dispersion $\sigma_v$ (table~\ref{tab:parameters}).
The transitions with optical depths larger than 0.005 are listed.
We also list the oscillator strength $f$ and the total transition probability $A$ from the upper level of the line
that is used in these calculations (Voigt absorption profiles are being used).

Clearly, the Fe\emissiontype{XXV} resonance line (He$\alpha$ w) has the highest optical depth,
but we see significant contributions from the other lines of the same Rydberg series,
as well as for other ions of Fe and other elements. Also the optical depth of the
Fe\emissiontype{XXIV} lines that block a part of the He-like intercombination line
\citep{mehdipour2015} is up to 2\%, a level that is detectable (for
the intercombination line, the statistical uncertainty of the spectrum  over one
instrumental FWHM of 5~eV is about 3\%).

In our baseline model, we have coupled the turbulent velocity, the Doppler velocity and the
temperature to the corresponding parameters of the dominant thermal emission component.
We have also tested a model where we have decoupled these quantities. We obtain an insignificant
improvement of our model (see table~\ref{tab:parameters}) with a
temperature of 3.8$\pm$0.6~keV for the absorbing gas, a velocity relative to the hot gas of 10$\pm$30~km\,s$^{-1}$,
a LOS turbulent velocity dispersion of 191$\pm$35~km\,s$^{-1}$ and a column density of (20.1$\pm$2.2)$\times$10$^{24}$~m$^{-2}$.
All these parameters are fully consistent with the parameters of the emission component within the
uncertainties of those emission parameters, but obviously we cannot exclude that the properties
of the absorbing gas are --- on average --- within the range indicated by the above uncertainties.

Our model substitutes a simple absorption model for resonance scattering effects. 
It assumes a common hydrogen-equivalent column density for all the transitions listed in \ref{tab:abslines}, 
ignoring the spatial structure of the ICM. 
The model also ignores the re-emission process after absorption, which possibly results in 
lower estimation of optical depths.
A more accurate characterization of resonance scattering requires radiative simulations, 
which will be separately presented in RS~paper.

\subsection{Charge exchange contributions\label{sect:cx}}

\begin{figure}[!htbp]
\includegraphics[width=8cm]{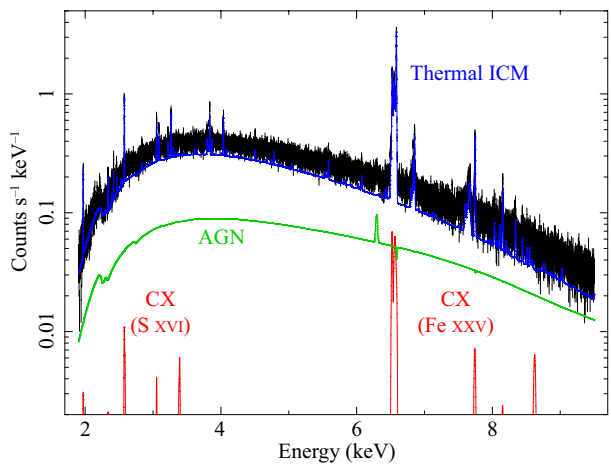}
\caption{The best-fit result with the CX emission. Each X-ray emission component is shown: thermal ICM in blue, AGN in green, and CX in red.}
\label{fig:cxspec}
\end{figure}

\begin{figure}[!htbp]
\includegraphics[width=8cm]{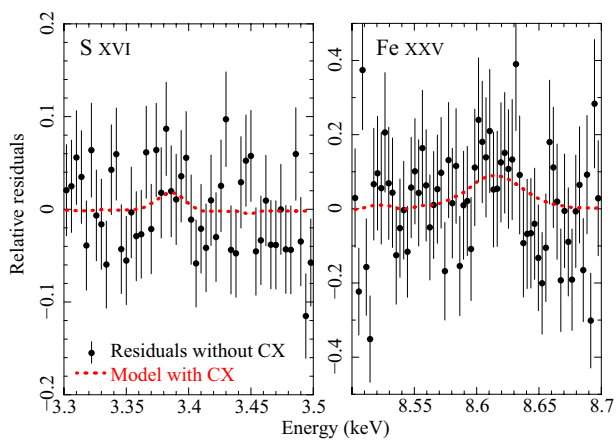}
\caption{Residuals of the baseline fitting in the S\emissiontype{XVI} 3.2--3.6~keV (left) and Fe\emissiontype{XXV} 8.3--8.8~keV (right) bands.
The red curve in each panel shows the model change by including the CX component. }
\label{fig:cxmod}
\end{figure}

Charge exchange (CX) happens when a neutral atom collides with a sufficiently charged ion,
which recombines with the electron(s) captured from the atom. The product ion
often has a highly excited state with large principal quantum number $n$, and
thereafter, the decay of the excited electron(s) will fill the inner-shell vacancies
through line emission. Therefore, the most
characteristic features of the CX emission in X-rays are the transitions from
high-$n$ shells to the ground, which are much stronger than those in the CIE case.
The CX spectrum also exhibits higher Ly$\beta$-to-Ly$\alpha$ and forbidden-to-resonance (z-to-w) ratios,
although these features can be achieved by other atomic processes \citep{gu2016b}.

In this section we examine the CX contributions to the ICM emission
of the Perseus cluster with the Hitomi SXS spectrum.
The CX component adopted here is described in
\citet{gu2016a}. It uses velocity-dependent, $nlS$- (He-like) or $n$- (H-like)
resolved reaction cross sections, based primarily on the multi-channel
Landau-Zener calculations \citep{mullen2016}. The low-energy weight function
\citep[equation~4 of][]{gu2016a} is
applied to the H-like data in which the $l$-distribution cannot be obtained by the
Landau-Zener calculations. For the Li-like and Be-like sequences, it includes velocity- and
$nl$-resolved cross sections which are derived from an empirical scaling relation
presented in \citet{gu2016a}.
In this model, only the atomic component of the cold gas is considered,
although in reality the molecular gas also contributes.
The collision velocity is set to 200~km\,s$^{-1}$ \citep{conselice2001}, and the ionization
temperature and abundances of the CX ions are fixed to the best-fit values of the
ICM thermal component. As shown in table~\ref{tab:parameters}, a new baseline run including
the CX component results in a minor C-statistic improvement ($\delta C_{\rm stat}=$13).

The fitting prefers that the CX lines are more broadened than the CIE lines, with a turbulent velocity $v_{\rm mic}\gtrsim$600~km~s$^{-1}$ (or $\sigma_v\gtrsim$400~km~s$^{-1}$).
Since the actual line profile cannot be determined by the current data, we fix the turbulent velocity of the CX component at $v_{\rm mic}=$ 800~km~s$^{-1}$,
corresponding to $\sigma_{v}$ of 566~km~s$^{-1}$ which is the upper limit of the neutral atomic line width of
the molecular cloud near NGC\,1275 as reported in, e.g., \citet{salome2011}. The large line width might be caused by
a combined effect: it can be partially contributed by the kinematics of the neutral cloud and the ICM,
and partially from the atomic uncertainty of the capture state \citep{gu2016a}, as the CX lines from $n\geq$10 levels are often blended.
Changing the turbulent velocity to a larger value (e.g, 1000~km~s$^{-1}$) has a negligible effect on the fitting.

As shown in figures~\ref{fig:cxspec} and \ref{fig:cxmod}, the CX model predicts that the most promising
high-$n$ transitions are seen in the S\emissiontype{XVI} band, which has been reported
by \citet{hitomi-3.5}, as well as the Fe\emissiontype{XXV} band. The CX lines contribute
to $\sim$1\% of the continuum for S\emissiontype{XVI} at $\sim$3.4 keV, and $\sim$3\%
for Fe\emissiontype{XXV} at $\sim$8.6 keV. To measure the statistical uncertainties, we
replace the CX model with two Gaussian lines at the energies of the S\emissiontype{XVI} and
Fe\emissiontype{XXV} high-$n$ transitions. The Gaussian FWHM is set
free for each line. The S\emissiontype{XVI} and Fe\emissiontype{XXV} CX lines have 1.6-$\sigma$ and
2.4-$\sigma$ significances, respectively. However, it is premature to claim the detection of
CX with the current data, since the uncertainty from the effective area/gain
calibration is large and energy dependent, as discussed by \citet{hitomi-3.5}.
For the remaining ions, the high-$n$ transitions are negligible, either due to the
low abundances, or blending with strong thermal lines.

As shown in table~\ref{tab:parameters}, inclusion of the CX component has minor effects on the ICM temperature, emission measure,
and turbulent velocity. The Fe and Si abundances are reduced by $\approx$ 5\% and 2\%, respectively, and the S, Cr, Mn, and Ni are affected by 1--3\%.
Since the CX emission has a larger forbidden-to-resonance (z-to-w) ratio than the thermal emission, the equivalent $N_{\rm H,hot}$ for the possible
resonance scattering is reduced by about 7\%. The effect on the resonance scattering study will be further discussed in RS~paper.

\subsection{AGN contribution\label{sect:noagn}}

To assess the uncertainty from the AGN flux, here we first consider an extreme condition:
the central AGN is quite dim and its power-law emission is negligible. As shown in table~\ref{tab:parameters},
the non-AGN run gives a much worse fit ($\delta C_{\rm stat}=$ 625)
than the original baseline fit, and the best-fit temperature shifts by 0.5 keV. The abundances are systematically lower by 0.01--0.21~solar.

Next we examine a more realistic case for possible systematic uncertainty related to the detailed AGN modeling.
The AGN spectrum in the baseline model was established in the early study for AGN~paper with the PSF photometry.
The technique is essentially unchanged in the final analysis but the energy band is extended up to 
20~keV with the \texttt{sxsextend} tool.
The broader-band spectrum requires a slightly flatter photon index and a $\approx$20\% lower flux 
in 2--10~keV (see AGN~paper for details).
Another notable update is the RMF type, which has been changed from the large size (also used in our baseline model) 
to the extra-large size to include electron-loss continuum.
As we examine the effect of using different types of RMF separately (section~\ref{sect:sxlrmf}),
we use the new AGN model derived in the same method as in AGN~paper but with the large-size RMF 
for a straightforward comparison to the baseline model.
Therefore, slightly different parameter values from AGN~paper are adopted in our test:
the photon index of 1.85 and the 2--10-keV flux of 2.9$\times$10$^{-14}$~W~m$^{-2}$ (private communication with H.~Noda and Y.~Fukazawa, 2017).

The new AGN model run gives a slightly poorer fit ($\delta C_{\rm stat}=$ 11) than the original baseline model.
The lower AGN flux requires a significant rise of the ICM continuum by 6\%, which results in 3--4\% lower abundances.
The change in the ICM gas temperature becomes insignificant, unlike the no-AGN case.

\section{Systematic factors affecting the derived source parameters:
fitting techniques\label{sect:fittingtech}}

In this section we discuss the effects of applying different fitting techniques on the derived parameters of the baseline model.

\subsection{Comparison of $\chi^2$-statistics versus the default C-statistics\label{sect:chi2}}

\begin{figure}[!htbp]
\includegraphics[width=8cm]{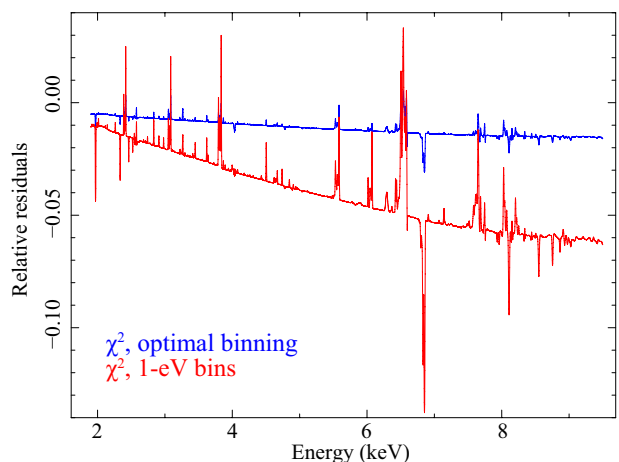}
\caption{Relative differences between a $\chi^2$ fit and a C-statistic fit
(blue curve) for the baseline model. The red curve shows the same but with 1 eV bins
instead of optimal binning.}
\label{fig:chi}
\end{figure}

It is well known that the use of $\chi^2$-statistics in spectral fitting can
give bias in the estimated parameters \citep[e.g.][]{nousek1989,mighell1999}.
The proper way to resolve this is to use C-statistics \citep{cash1979}, and we
have used that for our baseline model. We use the modification of C-statistic as
proposed by Castor (see the Xspec manual\footnote{$\langle$https://heasarc.gsfc.nasa.gov/xanadu/xspec/manual/$\rangle$.}, \citealt{arnaud1996}). This
modification is the standard in the Xspec and SPEX packages. Our present Hitomi
SXS spectrum offers an excellent opportunity to demonstrate the bias that one gets
when using $\chi^2$-statistics. We have taken the baseline model and replaced
the C-statistic with the $\chi^2$ statistic in the spectral fit. The best-fit
model has $\chi^2=$6192 for 5790 degrees of freedom. The value of the
C-statistic that corresponds to this $\chi^2$-optimized fit is 88 higher than
for the baseline model. We show the relative difference between both models in
figure~\ref{fig:chi}.

It is seen that the continuum for the $\chi^2$ fits is about 1\% lower than for
the baseline model, while some of the stronger emission lines have similar
fluxes for both cases. This 1\% bias is caused by the well-known effect that
$\chi^2$ fits tend to give lower fluxes by giving relatively more weight to the
data points that by chance have a flux below the expected value than to the data
points that have a flux above it. Our present spectrum has typically 100 counts
in most continuum bins, and according to \citet{mighell1999} this would give a
bias of about 1 count, in remarkable good agreement with our findings here. Note
that for typically 100 counts per bin, the Poissonian error bars are about 10
counts, hence much larger than the differences between the models. This shows
that biased fits are easily overlooked if plotted at full resolution. Only
rebinning the best-fit drastically (with a factor of at least a hundred or so)
would show the bias.

The bias becomes even stronger if in addition to using $\chi^2$-statistics we
drop the optimal binning and use 1-eV bins (see the red curve in
figure~\ref{fig:chi}). In addition to a lower flux, there is now also a
significant bias in the temperature, leading to a different overall slope of the
spectrum. The bias is even 6\% at the highest energies.

\subsection{Optimal binning versus other binning \label{sect:opbin}}

We have also tested how our results depend on the adopted bin size. When we use
C-statistics, we find no difference at all for the parameters shown in
table~\ref{tab:parameters} when comparing our optimal binning with a uniform
binning of 1~eV. This is easily understood by noting that our optimal binning
already gives a bin size of 1--2 eV for all bins (see appendix~\ref{app:binning}),
and that it is more the order of magnitude of the
bins rather than the precise bin size that matters for the sensitivity of
statistical tests \citep[][see figure~C3]{kaastra2016}.

Note that when $\chi^2$ is being used, binning is important, but as we
demonstrate in section~\ref{sect:chi2}, the use of $\chi^2$-statistics should be
avoided.

\subsection{Local versus global fit \label{sect:localfit}}

Astrophysical spectroscopic analysis in the radio through ultraviolet bands
often relies upon precise measurements of selected, strong emission lines
(e.g., H\emissiontype{I} 21~cm, Fe\emissiontype{II} 1.257~$\mu$m,
and [O\emissiontype{III}] 5007~\AA) whose atomic and diagnostic properties are well-understood.
This can also be done with X-ray spectroscopy \citep{hitomi2016},
but both the physics of X-ray--emitting plasmas and the availability of
high-resolution spectrometers create significant challenges.

The Hitomi SXS spectrum of Perseus presents a clear combination of emission lines
with a continuum, implying it can be completely understood via fits with the sum of
a simple continuum plus a series of Gaussian emission lines, with astrophysical parameters
derived from positions, widths, and flux ratios of the Gaussian parameters.
An advantage of this approach is that it requires a relatively small amount of reliable atomic data,
enabling the use of experimentally-verified and theoretically-understood features.
For example, the ratio of the line intensity of the Ly$\alpha$ line to
its resolved DR satellites depends critically on the electron temperature;
advanced line diagnostics using multiple DR satellite lines even test
whether the underlying plasma is in thermal equilibrium \citep{gabriel1979, kaastra2009}.

Although elegant, and without doubt useful to obtain an approximate description,
this approach will miss details resulting from a self-consistent fit of the full spectrum.
Three key problems occur with X-ray spectral analysis via purely local line fits:
\begin{enumerate}
\item Unlike other spectral bands, the line and continuum emission arise from the same plasma.
Therefore, simplifying the continuum to a spline fit or even bremsstrahlung emission independent
of the line components ensures the resulting analysis will miss features.
The X-ray continuum, even in strict collisional ionization equilibrium, contains significant contributions
from radiative recombination continua and two photon emission \citep[see e.g., figure~8 in][]{kaastra2008}.
While these components can be included in the fit, e.g., the APEC No-Line model used in \citet{plucinsky2017},
separating line from continuum emission makes finding a self-consistent model all but impossible.
\item X-ray spectrometers, even the SXS, have only limited resolution, while the X-ray bandpass
has a plethora of strong lines, making line blending an ongoing problem. Table~\ref{tab:wavelength}
in appendix~\ref{app:linelist}
shows several instances of lines from different elements separated by less than the instrumental
resolution. Worse, the narrow bandpasses of oft-used diagnostic lines such as the He${\alpha}$ complex
includes a multitude of DR satellite lines together with the strong ``triplet'' (actually a quartet) lines.
Many of the lines have multiple excitation channels all of which must be known in order to fit the complex reliably.
This is especially true of the forbidden line (z), as discussed earlier. For the SXS spectrum,
Gaussian lines were used to determine the turbulent motion \citep{hitomi2016},
but a local form of the global fit is required to extract the maximum amount of information
even from a relatively small bandwidth. At a minimum, when applying line ratio diagnostics
it must be clear both in the model and the data whether these contaminants have been taken into account.
\item Few, if any, sources in the Universe will be in perfect equilibrium, either collisional or photo-ionized.
The present spectrum of the Perseus cluster is a good example of such complexities.
While dominated by a 4-keV temperature component, the possibility of multi-temperature cannot be eliminated
based on the data (section~\ref{sect:multit}), and is certainly expected theoretically.
Depending upon their excitation mechanism, each emission line will be affected differently by these effects,
rendering the use of just one or two diagnostic ratios precise but quite inaccurate.
Using many lines, including upper limits to non-detections, will avoid this problem,
but at some point the distinction between a many-line vs a global fit will become blurred.
\end{enumerate}

Despite the above issues, line ratios may be preferred over global fits when the source spectrum is
either too complex to be fully understood, or when calibration uncertainties dominate the broad-band spectra.
Of course, the accuracy of the physical parameters derived from global fits also relies upon
complete and accurate atomic databases.
In the case of completeness, global models contain potentially millions of atomic transitions, most of which
have not been experimentally verified. While ``spot" checks do exist, in most cases the accuracy of the data is not well known,
i.e., estimates of uncertainties are determined by comparing results from different theoretical calculations,
or by using uncertainties from portions where experimental results do exist.

In our case, we have shown that the calibration and completeness of the spectral models is not perfect
but good enough to yield a very good description of the Perseus spectrum.  
Ultimately, local and global fits must be used in a complementary way.
The broad bandwidth coupled with the high spectral resolution of the SXS makes it possible to take advantage of
the strengths of both methods, improving the reliability of the derived physical parameters of the source.

\section{An improved model\label{sect:improve}}

\begin{figure}[!htbp]
\includegraphics[width=8cm]{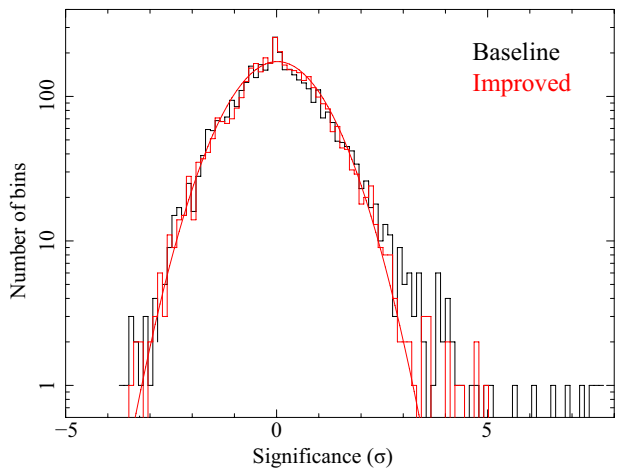}
\caption{Diagrams showing the significance of an additional \textsl{line} model, on the top of the best-fit baseline model (black)
and improved model (red), with the central energy moving across the entire band. The red curve shows a $\sigma=$1 Gaussian function
that fits the diagram of the improved model. }
\label{fig:weakdiagram}
\end{figure}

To fit the Perseus spectrum, we introduced a baseline model (section~\ref{sect:baseline})
which mainly consists of a $\sim$4-keV CIE plasma and an AGN component. Obviously the baseline model
is merely a simple approximation (section~\ref{sect:multit}), even though it already achieves
a satisfactory fit based on the current Hitomi SXS data. Throughout the paper,
we have tested a variety of plasma codes, atomic data calculations, plasma and astrophysical modelings,
additional spectral components, and instrumental effects, and compared them with the original baseline
fit. By properly incorporating some of the atomic and astrophysical effects into the baseline model,
we are able to achieve a more advanced physical model of the Perseus spectrum.

We construct an improved model as follows. Following the baseline model, SPEX version 3.03 is used, the abundance standard
are the \citet{lodders2009} proto-solar values, and the ionization balance is set to
\citet{urdampilleta2017}. The thermal emission is modeled now by
the sum of three CIE components, with temperatures of about 2~keV, 3.5~keV and 5~keV (section~\ref{sect:multit}).
The three-temperature model is chosen since it gives the best fit of all multi-temperature modelings
(table~\ref{tab:multi-t}). The three components have the same Si, S, Ar, Ca, Cr, Mn, and Ni abundances,
while the Fe abundance and turbulent velocity are left free for the 3.5-keV and 5-keV components. For the 2-keV component,
the Fe abundance and turbulent velocity are tied to those of the 3.5-keV component (section~\ref{sect:multit}).
The AGN contribution, resonance scattering, and the Galactic absorption components are added in the same way as for the baseline model.
The possible CX component (section~\ref{sect:cx}) is included in the improved model. Following section~\ref{sect:voigt},
the Voigt function is used to describe the line profiles.
We re-fit the effective area correction factor in the same way as described in appendix~\ref{app:fudge},
and show the best-fit model in figure~\ref{fig:newareacor} (a).

The improved model achieves a so-far best C-statistics, 4779 for an expected value of 4876$\pm$99,
which is significantly better than the baseline fit (C-statistics $=$ 4926). The best-fit model is plotted
in appendix~\ref{app:speccomp} (figures~\ref{fig:spec1}--\ref{fig:spec3}), and the stacked residual diagram is shown
in figure~\ref{fig:weakdiagram}. The residual diagram is calculated by adding a \textsl{line} component,
with the central energy moving from 1.9~keV to 9.5~keV with a step of 3~eV,
on the best-fit baseline and improved models. Compared to the baseline run, the residuals at
$>$2$\sigma$ are greatly suppressed by the improved fit, and the diagram follows well the expected
Gaussian distribution. As shown in table~\ref{tab:parameters}, the new model essentially reproduces
the best-fit results of the three-temperature model (section~\ref{sect:multit}).
The best-fit temperatures are 1.92$\pm$0.21~keV, 3.61$\pm$0.33~keV, and 5.43$\pm$0.38~keV
for the three components. Note that the values are sensitive to the details of the spectral modeling 
as well as the calibration of the instrumental response (see T~paper for further details).
All the Si, S, Ar, Cr, Mn, and Ni abundances become roughly 0.8~solar,
which are much more uniform than the baseline results. The Ca abundance remains to be about 0.9~solar.
The best-fit Fe abundances are 0.91$\pm$0.05~solar and 0.64$\pm$0.05~solar
for the 3.5-keV and 5-keV components, respectively. The turbulent velocities become
$\sigma_v=$ 117$\pm$11~km~s$^{-1}$ for the 3.5-keV component and $\sigma_v=$ 223$\pm$27~km~s$^{-1}$ for the 5-keV component.
This may suggest that the cooler ICM tends to have a lower level of turbulence than the hotter one, but the results depend
on the assumed temperature structure and are sensitive to the continuum modeling including
the effective area calibration. The further details are discussed in V~paper.
Moreover, the improved model gives a self-absorption column density of $($1.05$\pm$0.15$)\times$10$^{25}$~m$^{-2}$.
The column density of Fe\emissiontype{XXV} is thus $($2.18$\pm$0.23$)\times$10$^{20}$~m$^{-2}$,
in good agreement with the value that we derive from the simulated spectra in
section~\ref{sect:gdem} (4.02$\times$10$^{20}$~m$^{-2}$ for the semi-column of a line through the core;
2.64$\times$10$^{20}$~m$^{-2}$ for the semi-column averaged over the Hitomi SXS FOV).
Details on the derived resonance scattering are discussed in RS~paper.

\section{Discussion}

\subsection{Important factors}

We have shown in this paper the dependencies of several astrophysically interesting parameters, 
mainly focusing on the plasma modeling, i.e., plasma codes and atomic databases. 
We have also investigated the dependencies on astrophysical modeling as well as spectral fitting techniques. 
For a proper astrophysical modeling of the present Perseus cluster spectra,
as presented here but in greater detail discussed in a set of other papers (\citealt{hitomi2016,hitomi-3.5}, Z, T, RS, V, and AGN papers),
it is crucial to understand the possible systematic biases on the derived parameters.
Table~\ref{tab:parameters} provides a comprehensive list of the estimated biases,  
which enables us to inter-compare various aspects of systematic uncertainty.
The effects of some of the plasma modeling factors are comparable to or even larger than 
the statistical or instrumental uncertainty (appendix~\ref{app:instrumental}).
Because this is the only high-quality high-resolution X-ray spectrum of
a spatially extended thermal X-ray source up to now, it is also important for the preparation of future X-ray missions,
in the sense that priorities in calibration, astrophysical modeling, or data analysis can be set.

\subsubsection{Emission measure}

The emission measure $Y$ of the cluster ICM is a good representative of the absolute flux of the hot cluster gas.
First, it is clear that we need to use the latest \texttt{aharfgen} to obtain an accurate emission measure. Although the emission measures are uniformly underestimated by 20\% for the other cases with the older software, hereafter we ignore the difference and compare the relative values to find out which parameter affects $Y$.

The main contributor to the systematic uncertainty on $Y$ is the adopted flux of the central AGN.
Ignoring the AGN completely would give a 20\% higher emission measure as well as 12\% higher temperature for the hot gas.
In fact, the AGN contribution affects almost all parameters of the hot cluster gas.
Because we are not completely ignorant about the AGN flux, the true uncertainties are smaller than described above. 
One example of more realistic estimation is the difference due to updating the AGN model with the broader-band spectroscopy, which gives a 6\% higher emission measure.

Other important factors for $Y$ are the effective area correction (up to 4\%), the ionization balance (3\%),
and the assumption of isothermality ($\sigma_T$ free, 3\%). The differences between the plasma codes,
for which we consider here SPEX version 3.03 and AtomDB version 3.0.8 (SPEX and AtomDB briefly hereafter) to be the most sophisticated, are not very important for the emission measure: differences are less than 1\%.

\subsubsection{Temperature}

Ignoring from now on the AGN contribution uncertainty, the most important factor affecting
the temperature of the dominant 4-keV component is the fitting techniques.
Using unbinned spectra and $\chi^2$ statistics biases the temperature by 5\%.
Forcing isothermality (i.e., putting $\sigma_T$ to be zero) gives 3.5\% bias.
Using the official effective area correction based on the on-ground calibration instead of the additional effective area correction using a thinner Be-filter and \texttt{knak} gives 3\% bias.
Both plasma codes (SPEX and AtomDB) agree relatively well in their derived temperature (better than 2\%).

\subsubsection{Turbulent velocity}

While the temperature agreement between the plasma codes is good, they result in a 10\% difference in the derived amount of turbulence in the plasma.
This uncertainty is almost as large as the uncertainty introduced by ignoring completely the resonance scattering or ignoring the position-dependent bulk velocity field (both 9\%).

\subsubsection{Cluster velocity}

This bulk velocity field obviously also affects the derived velocity centroid of the cluster (23~km\,s$^{-1}$).
Obviously the gain correction is also important (14~km\,s$^{-1}$). Finally, the use of SPEX or AtomDB also results in a difference of 6~km\,s$^{-1}$.

\subsubsection{Resonance scattering}

The plasma codes result in an even bigger difference of 40\% in the derived column density
of the resonantly scattering plasma, six times larger than the statistical uncertainty on this quantity.
This relatively large difference is likely associated to the systematic uncertainties
in the line emissivities, because in the comparison we use the same resonance scattering model
(the SPEX \textsl{hot} model). However, also precise modeling of the temperature structure
(see our improved model) is important: this can also produce a difference of 35\%.

\subsubsection{Abundances}

Finally, we discuss here the uncertainties on the abundances.
Most striking is the difference in the Fe abundance associated to the plasma code:
AtomDB gives a 16\% lower abundance than SPEX. This is 17 times higher than
the small statistical uncertainty on the Fe abundance.
The differences can be attributed mostly to differences in the adopted
collisional excitation and DR rates of the strongest spectral lines (sections~\ref{sect:lineer} and \ref{sect:satellite}).
Other factors affecting the Fe abundance are the inclusion of resonance scattering (11\%) and CX (5\%).
On the other hand, the Ni abundance is almost bias-free between the latest SPEX and APEC/AtomDB
(at least within its 7\% statistical uncertainty; however the bias between SPEX versions 2 and 3
is still significant. See also Z~paper).
This is not the case for other elements. The Si and S abundance can be biased by 6--14\%
depending on each of the following four factors: the plasma code, the isothermality assumption,
the gain correction and the fitting method ($\chi^2$ fitting on unbinned data).
For Ar and Ca the main systematic uncertainties are associated to the plasma model (6--8\%).
Finally, for Cr and Mn both the isothermality assumption and the fitting method are
the main sources of systematic uncertainty.

\subsubsection{Implications to other observations}

So far we have reviewed the state-of-the-art knowledge, mostly on the K-shell transitions, 
for modeling the hot (several keV) tenuous plasma in collisional ionization equilibrium. 
We caution that the atomic uncertainties derived from the Perseus data cannot simply be copied 
to observations of other sources, as the accuracy of atomic data depends strongly on the types of
transitions (tables~\ref{tab:lya} and \ref{tab:trip}), as well as on the plasma conditions, 
such as electron temperature (figures \ref{fig:lya} and \ref{fig:fe25col}) 
and ion charge states (figures \ref{fig:tr1} and \ref{fig:tr2}). 
For instance, X-ray emission from a stellar corona (or an elliptical galaxy) is dominated by 
transitions in the Fe-L complex, which are known to be computationally more intricate than 
those in the Fe-K (e.g., \citealt{bernitt2012}), and hence less accurate (e.g., \citealt{deplaa2012}). 

A more important issue is to discuss the atomic uncertainties by the science cases. 
The doppler measurement of line-of-sight velocities would be subject to 
the reference-wavelength accuracy of the dominant transitions except for the cases 
with large bulk velocities for example in young supernova remnants (SNRs). 
The precise characterization of turbulence velocity structures, i.e., search for non-Gaussianity, 
would primarily be limited by the accuracy of atomic and astrophysical modeling of the RS effect. 
This could be avoided by making use of local fittings of optically thin emission lines, 
and in this case relative line energies and emissivities of satellite lines as well as the calibration of 
the line spread function are the dominant source of uncertainty. 
The detection of a small departure from CIE (e.g., for merger clusters; \citealt{inoue2016}) 
would mainly be limited by the uncertainties in the charge-state distribution calculation 
and thus ionization and recombination rates adopted therein (section~\ref{sect:nei}). 
Revealing detailed time evolution of NEI plasma by measuring the charge-state distribution 
(e.g., for recombining plasma in SNRs; \citealt{sawada2012}) would require 
an even higher level of accuracy for these transition rates including multiple ionization 
due to inner-shell processes followed by Auger ejections.
The elemental abundance measurement is affected mostly by the errors of the line excitations 
and branching ratios for individual transitions including those for the satellite lines, 
although only the error of the total emissivity of a line complex (e.g., He$\alpha$) 
would matter for a system with a large intrinsic line broadening ($\sim$100~eV) like young SNRs 
where the ion temperature is considerably high ($\sim$MeV).

The atomic uncertainty for each science case can be evaluated by the Monte-Carlo approach introduced in 
section~\ref{sect:lineerhe}. Ultimately, the atomic error calculation should be implemented as 
a standard analysis procedure in the spectral modeling packages. 
This would require substantial work in the code development by assessing the accuracy of detailed atomic data.

\subsection{Atomic data needs}

As shown throughout this paper, the reliability of a spectral modeling
package lies not only in the accuracy and completeness of its atomic data,
but also in its ability to properly synthesize the atomic data as
a function of physical parameters, i.e., the plasma conditions,
such as temperature and density. Synthesizing the data is tedious and
computationally taxing owing to the fact that the databases employed are large,
containing millions of data points, including transition energies,
excitation and ionization cross sections, resonant (multi-electronic) and 
non-resonant (radiative) recombination cross sections,
and non-thermal processes, such as CX recombination.
Different models use atomic databases of varying levels of completeness
and accuracy as well as different synthesis methods in their calculations.
Estimates of a model's accuracy is often given by comparison to other models.
However, a true measure of a model's accuracy  can only be determined by comparing to
laboratory benchmark measurements.

Benchmark measurements, generally, come in two forms:
as isolated experiments where a single ionic species or atomic process is studied,
or as integrated experiments, where emission or absorption is measured from
several simultaneous ions and atomic processes as a function of temperature or density.
Isolated experiments include those conducted at electron beam ion traps,
advanced light sources, or storage rings. Integrated experiments include experiments using,
for example, tokamaks or laser-produced plasmas.
Isolated experiments generally test portions of atomic databases,
and integrated experiments test synthesis models.
Examples of isolated experiments include measurements of
absolute electron-impact excitation cross sections as a function of electron energy,
transition energies, natural line widths, and oscillator strengths \citep{beiersdorfer1992a,brown:gv2006a,rudolph2013a}.
Examples of integrated experiments include the spectral signature of the He$\alpha$ complex
as a function of electron temperature \citep{bitter2008a,gu2012a,rosen2014,rice2015},
or full Fe-K and Fe-L shell spectral signatures as a function of temperature and density.

Providing laboratory benchmarks for the atomic database in all physical regimes
for all astrophysically relevant ions is not tractable.
Hence, models are tested by comparing to measurements where available.
Typically, models agree with measurements at the 10--20\% level in the cases of excitation and ionization processes.
Transition energies, however, are of much higher accuracy.  In the case of H- and He-like ions,
measurement of the transition energies have tested theory at the level of
a few to a few tens of parts per million \citep{johnson:wr1985a,beiersdorfer2009,beiersdorfer2015}.
In the case of ions with more bound electrons, i.e., L-shell ions,
the accuracy of the models is not as well known, as experimental benchmarks are 
more sparse and agreement with theory varies.

The inability of the standard X-ray astrophysics models to accurately fit a significant fraction
of the lines in the SXS Perseus spectrum \citep{hitomi2016} not only uncovered some of the limits of SPEX and APEC,
it also showed the limits of the high-accuracy laboratory measurements.
For example, laboratory measurements of relative line intensities in the Fe He$\alpha$ complex,
in particular the strength of the forbidden line (z), still introduce a limit to our ability to take full
advantage of the line complex's diagnostic power. The high-quality SXS Perseus spectrum provides the impetus
for more complete and higher accuracy calculations and systematic measurements of all the processes involved
in exciting, not only the forbidden line (z), but all of the lines found in the He$\alpha$ complex,
and not only for He-like Fe \emissiontype{XXV}, but also for other astrophysically relevant He-like ions.
Measurements such as these will be paramount to
interpreting high-resolution spectra to be returned by the future high-resolution X-ray spectroscopy missions
(see section~\ref{sect:prospects}). Largely driven by their large bandwidths, high energy resolution,
and large collecting areas, high accuracy measurements of a plethora of atomic parameters will be required.

While providing a complete list of required measurements is beyond the scope of this paper, 
a few necessary measurements, in addition to the studies of He$\alpha$,
should be mentioned. For example, a more complete study of the excitation cross sections and spectral signatures of
CX recombination should be completed.  Many CX studies have been completed, however,
at present, theory has not matured to a point of consistently predicting experimental results, and hence,
the diagnostic capability of CX emission is limited. Absolute cross-section measurements
for electron-impact excitation followed by cascades, especially in the case of high-$n$ transitions,
with accuracies on the order of 5--10\%, should also be a high priority as they determine the line strengths,
and in turn, relative ion abundances (ionic fractions) and elemental abundances from a variety of celestial sources.
High-accuracy measurements of DR-resonance strengths and of ionization cross sections should also be pursued.
Similar laboratory measurements of photo-excitation and ionization processes should also be conducted,
as these are the basis for determining column densities and scattering effects (RS~paper).

One of the most sought after and challenging integrated  laboratory experiments is an accurate measurement
of the ion charge balance as a function of electron temperature and density. 
This is a universal goal throughout plasma physics,
spanning nearly all temperature and density regimes. Integrated experiments such as these are challenging
because it is hard to know to high accuracy what the systematics are of the source plasma,
i.e., it is often hard to quantify or experimentally discount gradient and non-uniformity effects.
Regardless of these challenges,  integrated experiments where the plasma parameters have been independently
well diagnosed have been successfully conducted \citep{rosen2014}.

\subsection{Prospects for XARM, Athena and other missions}\label{sect:prospects}

The Hitomi SXS observation of Perseus, with its high-resolution spectrum in the 1.9--9.5~keV band,
showed both the strengths and weaknesses of existing plasma codes.
Pre-launch versions of both the SPEX and AtomDB codes provided generally plausible fits to the observation,
matching the continuum and many of the strong lines well.
While neither fit was formally statistically acceptable (see table~\ref{tab:parameters}),
the two codes agreed (to within $\pm$0.3~keV) on the best-fit temperature, 
and to within $\pm$0.2 on elemental abundances.
At CCD resolution, these discrepancies could easily be understood as calibration issues
or inadequacies of the collisional isothermal model; only at the resolution of the SXS were 
the clear problems with both codes apparent.
As described above, many of these disagreements could be addressed by updating 
wavelengths and cross sections for a few weaker lines and by fixing minor code bugs.
As a result, both SPEX and AtomDB are in close agreement about the emission from 
a 4-keV collisional plasma in 1.9--9.5~keV.

That an SXS observation was required to discover and address these problems may seem odd,
as gratings on both Chandra and XMM-Newton have provided high-resolution X-ray spectra of 
point sources since 1999.
Unfortunately, most X-ray point sources have intrinsically complex and variable spectra;
stellar coronae include plasmas with a broad range of temperatures,
while any model of the absorbed photo-ionized spectra of X-ray binaries and AGN must include 
a range of different geometries and source spectra.
The only truly simple point-source spectra are isolated neutron stars or white dwarfs,
which have no features in the X-ray band and are therefore used as calibration sources.
As a result, few grating observations could be used to test details of the plasma models beyond the strong lines,
since any differences in weaker features could reasonably be due to issues in the source model and not the code.

Substantial work remains, therefore, to ensure that current plasma codes will be ready
to face the challenges of data from the X-ray Astronomy Recovery Mission (XARM), ESA's Athena mission,
and proposed missions such as the Arcus grating spectrometer or the Lynx observatory.
These missions will have resolutions similar to or better than the Hitomi SXS,
and will observe a large range of sources, including collisional plasmas with temperatures between
10$^4$--10$^9$~K and photo-ionized plasmas with a similarly broad range of source flux,
either in ionization equilibrium or non-equilibrium.
These missions will cover a bandpass of $\approx$0.1--10~keV, a range that includes strong lines from
Fe L-shell ions (Fe\emissiontype{XVII}--Fe\emissiontype{XXIV}) as well as M-shell lines from many abundant elements.

The Hitomi SXS data have shown that accurate atomic models are just as important as calibration.
Preparing for these missions will require a multifaceted approach of plasma-code testing,
theoretical calculations, and laboratory measurements.
The process will begin with systematic testing of existing atomic models against
(1) each other to determine where discrepancies exist, (2)  laboratory measurements
from electron beam ion traps and synchrotrons,
and (3) deep targeted observations with existing observatories.
When areas of unresolvable disagreement are identified,
new theoretical calculations may be required or targeted laboratory measurements made.
The plasma-code community has already begun this work, starting with a set of agreed-upon standard tests developed
at a meeting at the Lorentz Center\footnote{See $\langle$https://lorentzcenter.nl/lc/web/2016/830/report.pdf$\rangle$.}.
However, a consistent and continuous effort will be required to ensure that the community is ready for this next generation of high-resolution X-ray spectra.

\begin{contribution}

M.~Sawada, L.~Gu, and J.~Kaastra led this study and wrote the final manuscript along with
R.~K.~Smith, A.~R.~Foster, G.~V.~Brown, H.~Odaka, H.~Akamatsu, and T.~Hayashi.
L.~Gu, J.~Kaastra, and M.~Sawada performed the final SXS data reduction and spectral analysis.
A.~R.~Foster and R.~K.~Smith, A.~J.~J.~Raassen, and G.~V.~Brown provided
inputs on APEC/AtomDB, FAC calculations, and discussion on the atomic data needs, respectively.
L.~Gu performed the Monte-Carlo simulations for the evaluations of the systematics of the atomic data.
C.~A.~Kilbourne contributed to the estimation of the SXS energy-scale uncertainty.
R.~K.~Smith contributed to the discussion on the future prospects.
N.~Ota and M.~A.~Leutenegger respectively contributed to the estimations of
the atomic data and plasma modeling systematics.
H.~Akamatsu and S.~Nakashima contributed to the systematics on astrophysical models.
K.~Sato, H.~Noda, and Y.~Fukazawa contributed to the estimations of non-ICM components.
S.~Nakashima and M.~Tsujimoto provided the data for the evaluations of the effective-area systematics.
T.~Hayashi confirmed the in-orbit calibration and performance of the X-ray mirrors of the SXS.
N.~Hell, M.~Tsujimoto, and R.~F.~Mushotzky contributed valuable comments on the manuscript.

The science goals of Hitomi were discussed and developed over more than
10~years by the ASTRO-H Science Working Group,
all members of which are authors of this manuscript.
All the instruments were prepared by joint efforts of the team.
The manuscript was subject to an internal collaboration-wide review process.
All authors reviewed and approved the final version of the manuscript.

\end{contribution}

\begin{ack}

We acknowledge reviewer Paul Bryans for the constructive comments and advices.
We thank the support from the JSPS Core-to-Core Program.
We acknowledge all the JAXA members who have contributed to the ASTRO-H (Hitomi)
project.
All U.S. members gratefully acknowledge support through the NASA Science Mission
Directorate. Stanford and SLAC members acknowledge support via DoE contract to SLAC
National Accelerator Laboratory DE-AC3-76SF00515. Part of this work was performed under
the auspices of the U.S. DoE by LLNL under Contract DE-AC52-07NA27344.
Support from the European Space Agency is gratefully acknowledged.
French members acknowledge support from CNES, the Centre National d'\'{E}tudes Spatiales.
SRON is supported by NWO, the Netherlands Organization for Scientific Research.  Swiss
team acknowledges support of the Swiss Secretariat for Education, Research and
Innovation (SERI).
The Canadian Space Agency is acknowledged for the support of Canadian members.
We acknowledge support from JSPS/MEXT KAKENHI grant numbers 
JP15H00773,
JP15H00785,
JP15H02070,
JP15H02090,
JP15H03639,
JP15H03641,
JP15H03642,
JP15H05438,
JP15H06896,
JP15J02737,
JP15K05107,
JP15K17610,
JP15K17657,
JP16H00949,
JP16H03983,
JP16H06342,
JP16J00548,
JP16J02333,
JP16K05295,
JP16K05296,
JP16K05296,
JP16K05296,
JP16K05300,
JP16K05309,
JP16K13787,
JP16K17667,
JP16K17672,
JP16K17673,
JP21659292,
JP23340055,
JP23340071,
JP23540280,
JP24105007,
JP24244014,
JP24540232,
JP25105516,
JP25109004,
JP25247028,
JP25287042,
JP25400236,
JP25800119,
JP26109506,
JP26220703,
JP26400228,
JP26610047,
and JP26800102.
The following NASA grants are acknowledged: NNX15AC76G, NNX15AE16G, NNX15AK71G,
NNX15AU54G, NNX15AW94G, and NNG15PP48P to Eureka Scientific.
H. Akamatsu acknowledges support of NWO via Veni grant.
C. Done acknowledges STFC funding under grant ST/L00075X/1.
A. Fabian and C. Pinto acknowledge ERC Advanced Grant 340442.
P. Gandhi acknowledges JAXA International Top Young Fellowship and UK Science and
Technology Funding Council (STFC) grant ST/J003697/2.
Y. Ichinohe, K. Nobukawa, and H. Seta are supported by the Research Fellow of JSPS for Young
Scientists.
N. Kawai is supported by the Grant-in-Aid for Scientific Research on Innovative Areas
``New Developments in Astrophysics Through Multi-Messenger Observations of Gravitational
Wave Sources''.
S. Kitamoto is partially supported by the MEXT Supported Program for the Strategic
Research Foundation at Private Universities, 2014-2018.
B. McNamara and S. Safi-Harb acknowledge support from NSERC.
T. Dotani, T. Takahashi, T. Tamagawa, M. Tsujimoto and Y. Uchiyama acknowledge support
from the Grant-in-Aid for Scientific Research on Innovative Areas ``Nuclear Matter in
Neutron Stars Investigated by Experiments and Astronomical Observations''.
N. Werner is supported by the Lend\"ulet LP2016-11 grant from the Hungarian Academy of
Sciences.
D. Wilkins is supported by NASA through Einstein Fellowship grant number PF6-170160,
awarded by the Chandra X-ray Center, operated by the Smithsonian Astrophysical
Observatory for NASA under contract NAS8-03060.

We thank contributions by many companies, including in particular, NEC, Mitsubishi Heavy
Industries, Sumitomo Heavy Industries, and Japan Aviation Electronics Industry. Finally,
we acknowledge strong support from the following engineers.  JAXA/ISAS: Chris Baluta,
Nobutaka Bando, Atsushi Harayama, Kazuyuki Hirose, Kosei Ishimura, Naoko Iwata, Taro
Kawano, Shigeo Kawasaki, Kenji Minesugi, Chikara Natsukari, Hiroyuki Ogawa, Mina Ogawa,
Masayuki Ohta, Tsuyoshi Okazaki, Shin-ichiro Sakai, Yasuko Shibano, Maki Shida, Takanobu
Shimada, Atsushi Wada, Takahiro Yamada; JAXA/TKSC: Atsushi Okamoto, Yoichi Sato, Keisuke
Shinozaki, Hiroyuki Sugita; Chubu U: Yoshiharu Namba; Ehime U: Keiji Ogi; Kochi U of
Technology: Tatsuro Kosaka; Miyazaki U: Yusuke Nishioka; Nagoya U: Housei Nagano;
NASA/GSFC: Thomas Bialas, Kevin Boyce, Edgar Canavan, Michael DiPirro, Mark Kimball,
Candace Masters, Daniel Mcguinness, Joseph Miko, Theodore Muench, James Pontius, Peter
Shirron, Cynthia Simmons, Gary Sneiderman, Tomomi Watanabe; ADNET Systems: Michael
Witthoeft, Kristin Rutkowski, Robert S. Hill, Joseph Eggen; Wyle Information Systems:
Andrew Sargent, Michael Dutka; Noqsi Aerospace Ltd: John Doty; Stanford U/KIPAC: Makoto
Asai, Kirk Gilmore; ESA (Netherlands): Chris Jewell; SRON: Daniel Haas, Martin Frericks,
Philippe Laubert, Paul Lowes; U of Geneva: Philipp Azzarello; CSA: Alex Koujelev, Franco
Moroso.

\end{ack}

\begin{appendix}

\section{Empirical corrections in energy scale and effective area and data binning\label{app:datacor}}

\subsection{Energy-scale correction\label{app:escalecorr}}

\begin{figure}[!htbp]
\includegraphics[width=8cm]{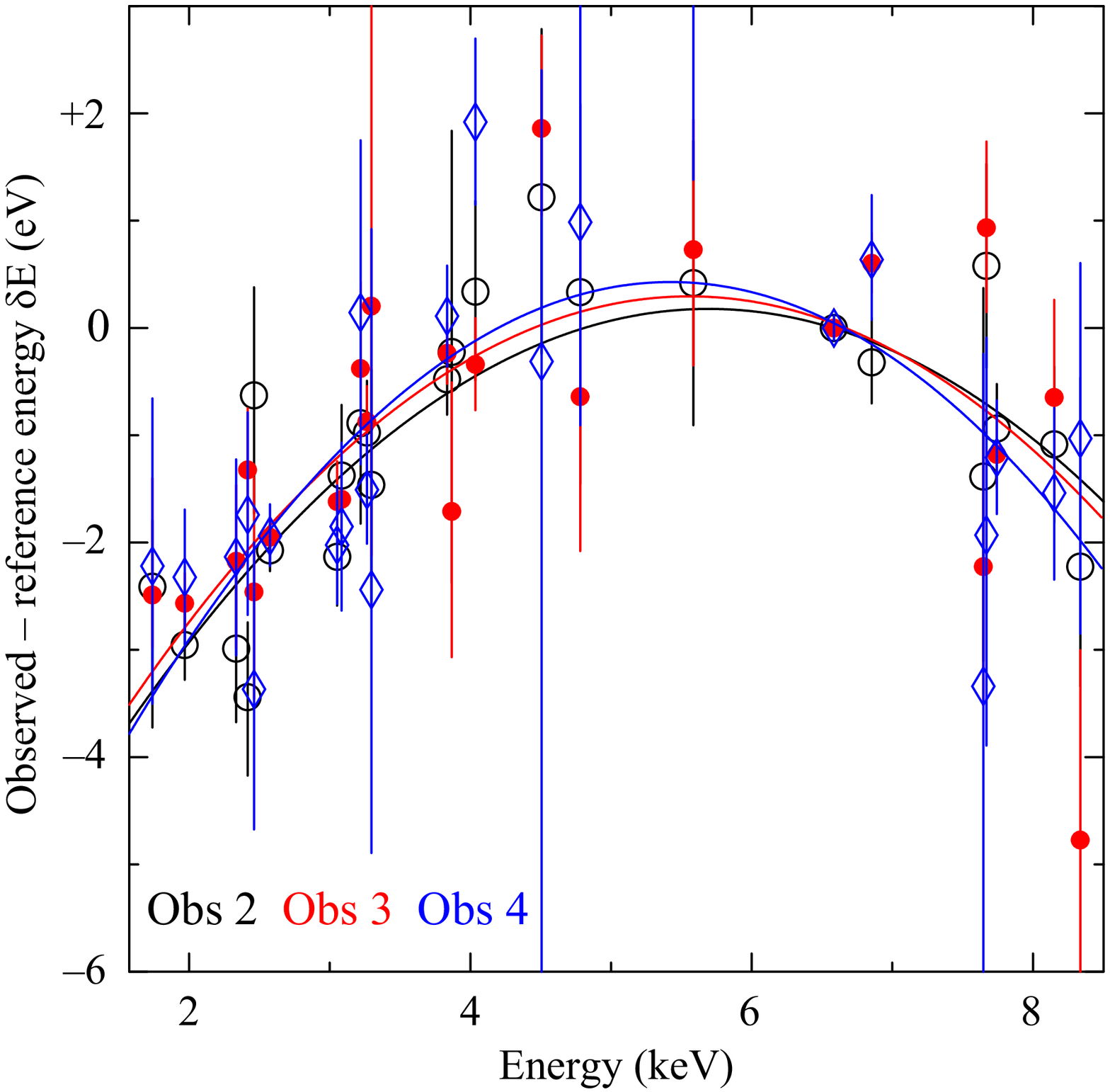}
\caption{The differences between measured and reference energies of the brightest lines (see table~\ref{tab:es} for details). The curves show the best-fit parabolic functions of the three observations.}
\label{fig:gain}
\end{figure}

\begin{table*}[!htbp]
\caption{Emission lines used for energy-scale correction.}
\label{tab:es}
\centerline{
\begin{tabular}{@{\extracolsep{2pt}}lllcccccc}
\hline
Ion                    & \multicolumn{2}{l}{Principal line}                             &\multicolumn{2}{c}{Reference energy (eV)} & \multicolumn{3}{c}{Observed shift $\delta E$ (eV)} & Ref.\footnotemark[$*$] \\ \cline{2-3} \cline{4-5} \cline{6-8}
                       & Name          & Transition        & Rest frame & Observer frame & Obs\,2 & Obs\,3 & Obs\,4 & \\
\hline
\multicolumn{9}{c}{\dotfill Instrumental\dotfill} \\
Si\emissiontype{I}     & K$\alpha_{1}$ & 2p$^{-1}$ ($^2$P$_{3/2}$) -- 1s$^{-1}$ ($^2$S$_{1/2}$)\footnotemark[$\dagger$] & 1739.99 &   1739.99  & $-$2.41 & $-$2.49 & $-$2.22 & 1 \\
\multicolumn{9}{c}{\dotfill Astronomical\dotfill} \\
Si\emissiontype{XIV}   & Ly$\alpha_1$  & 1s ($^2$S$_{1/2}$) -- 2p ($^2$P$_{3/2}$)  & 2006.08 &   1971.44  & $-$2.96 & $-$2.57 & $-$2.32 & 2 \\
Si\emissiontype{XIV}   & Ly$\beta_1$   & 1s ($^2$S$_{1/2}$) -- 3p ($^2$P$_{3/2}$)  & 2376.62 &   2335.58  & $-$2.99 & $-$2.17 & $-$2.13 & 2 \\
S\emissiontype{XV}     & He$\alpha$ w  & 1s$^2$ ($^1$S$_0$) -- 1s.2p ($^1$P$_{1}$) & 2460.63 &   2418.14  & $-$3.44 & $-$1.32 & $-$1.74 & 3 \\
Si\emissiontype{XIV}   & Ly$\gamma_1$  & 1s ($^2$S$_{1/2}$) -- 4p ($^2$P$_{3/2}$)  & 2506.37 &   2463.09  & $-$0.63 & $-$2.46 & $-$3.37 & 2 \\
S\emissiontype{XVI}    & Ly$\alpha_1$  & 1s ($^2$S$_{1/2}$) -- 2p ($^2$P$_{3/2}$)  & 2622.69 &   2577.40  & $-$2.07 & $-$1.95 & $-$1.94 & 2 \\
S\emissiontype{XVI}    & Ly$\beta_1$   & 1s ($^2$S$_{1/2}$) -- 3p ($^2$P$_{3/2}$)  & 3106.74 &   3053.10  & $-$2.13 & $-$1.62 & $-$2.02 & 2 \\
Ar\emissiontype{XVII}  & He$\alpha$ w  & 1s$^2$ ($^1$S$_0$) -- 1s.2p ($^1$P$_1$)   & 3139.77 &   3085.56  & $-$1.38 & $-$1.60 & $-$1.85 & 4 \\
S\emissiontype{XVI}    & Ly$\gamma_1$  & 1s ($^2$S$_{1/2}$) -- 4p ($^2$P$_{3/2}$)  & 3276.26 &   3219.69  & $-$0.89 & $-$0.38 & $+$0.14 & 2 \\
Ar\emissiontype{XVIII} & Ly$\alpha_1$  & 1s ($^2$S$_{1/2}$) -- 2p ($^2$P$_{3/2}$)  & 3322.98 &   3265.60  & $-$0.97 & $-$0.88 & $-$1.51 & 2 \\
S\emissiontype{XVI}    & Ly$\delta_1$  & 1s ($^2$S$_{1/2}$) -- 5p ($^2$P$_{3/2}$)  & 3354.73 &   3296.81  & $-$1.46 & $+$0.20 & $-$2.44 & 2 \\
Ca\emissiontype{XIX}   & He$\alpha$ w  & 1s$^2$ ($^1$S$_0$) -- 1s.2p ($^1$P$_1$)   & 3902.26 &   3834.88  & $-$0.48 & $-$0.24 & $+$0.11 & 5 \\
Ar\emissiontype{XVIII} & Ly$\beta_1$   & 1s ($^2$S$_{1/2}$) -- 3p ($^2$P$_{3/2}$)  & 3935.71 &   3867.75  & $-$0.22 & $-$1.71 &    ---\footnotemark[$\ddagger$]  & 2 \\
Ca\emissiontype{XX}    & Ly$\alpha_1$  & 1s ($^2$S$_{1/2}$) -- 2p ($^2$P$_{3/2}$)  & 4107.48 &   4036.56  & $+$0.34 & $-$0.34 & $+$1.92 & 2 \\
Ca\emissiontype{XIX}   & He$\beta_1$   & 1s$^2$ ($^1$S$_0$) -- 1s.3p ($^1$P$_1$)   & 4582.81 &   4503.68  & $+$1.22 & $+$1.86 & $-$0.31 & 5 \\
Ca\emissiontype{XX}    & Ly$\beta_1$   & 1s ($^2$S$_{1/2}$) -- 3p ($^2$P$_{3/2}$)  & 4864.08 &   4780.09  & $+$0.33 & $-$0.64 & $+$0.99 & 2 \\
Cr\emissiontype{XXIII} & He$\alpha$ w  & 1s$^2$ ($^1$S$_0$) -- 1s.2p ($^1$P$_1$)   & 5682.05 &   5583.94  & $+$0.42 & $+$0.73 & $+$3.53 & 6 \\
Fe\emissiontype{XXV}   & He$\alpha$ w  & 1s$^2$ ($^1$S$_0$) -- 1s.2p ($^1$P$_1$)   & 6700.42 &   6584.73  & \multicolumn{3}{c}{0\footnotemark[\S]}  & 6 \\
Fe\emissiontype{XXVI}  & Ly$\alpha_1$  & 1s ($^2$S$_{1/2}$) -- 2p ($^2$P$_{3/2}$)  & 6973.07 &   6852.67  & $-$0.32 & $+$0.60 & $+$0.64 & 2 \\
Fe\emissiontype{XXIV}  & j$_3$\footnotemark[$\|$] & 2p ($^2$P$_{3/2}$) -- 1s.2p($^3$P).3p ($^2$D$_{5/2}$) & 7782.52 &   7648.14  & $-$1.39 & $-$2.23 & $-$3.34 & 7 \\
Ni\emissiontype{XXVII} & He$\alpha$ w  & 1s$^2$ ($^1$S$_0$) -- 1s.2p ($^1$P$_1$)   & 7805.14 &   7670.37  & $+$0.58 & $+$0.94 & $-$1.93 & 5 \\
Fe\emissiontype{XXV}   & He$\beta_1$   & 1s$^2$ ($^1$S$_0$) -- 1s.3p ($^1$P$_1$)   & 7881.12 &   7745.04  & $-$0.94 & $-$1.18 & $-$1.21 & 5 \\
Fe\emissiontype{XXV}   & He$\gamma_1$  & 1s$^2$ ($^1$S$_0$) -- 1s.4p ($^1$P$_1$)   & 8295.39 &   8152.16  & $-$1.09 & $-$0.65 & $-$1.54 & 5 \\
Fe\emissiontype{XXV}   & He$\delta_1$  & 1s$^2$ ($^1$S$_0$) -- 1s.5p ($^1$P$_1$)   & 8487.22 &   8340.67  & $-$2.22 & $-$4.77 & $-$1.03 & 5 \\
\hline
\end{tabular}
}
\begin{tabnote}
\footnotemark[$*$] References:
(1)~\citet{bearden1967};
(2)~\citet{erickson1977};
(3)~\citet{kaufman1993};
(4)~\citet{kelly1987};
(5)~\citet{sugar1985};
(6)~\citet{shirai2000};
(7)~Calculations with the Flexible Atomic Code \citep{gu2008}, A.J.J. Raassen, private comminication (2017). \\
\footnotemark[$\dagger$] A vacancy is denoted as a negative index of the electron configurations. \\
\footnotemark[$\ddagger$] Poor fit. Ignored in derivation of the correction curve. \\
\footnotemark[\S] The energy shift at Fe\emissiontype{XXV} He$\alpha$ is assumed to be zero as it is already adjusted by the removal of the spatial velocity gradient. \\
\footnotemark[$\|$] The 1s--3p analogous to the 1s--2p dielectronic satellite line, j: 2p ($^2$P$_{3/2}$) -- 1s.2p$^2$ ($^2$D$_{5/2}$), labeled by \citet{phillips2008}. \\
\end{tabnote}
\end{table*}

To obtain the energy-dependent residual energy-scale errors, we fit the strongest emission
lines in the 1.9--9.5~keV band. 
For each line, we define an adjacent band with a width of 0.1--0.2~keV,
and perform a local fit of the Hitomi SXS spectrum.
Table~\ref{tab:es} lists the principal lines for the individual bands.
A collisional ionization equilibrium (CIE) model affected by redshift is used to fit the astronomical lines,
whereas a redshifted double-Gaussian model is used for the instrumental Si K$\alpha$ lines.
For the CIE model, the temperature is fixed to 4~keV, while the abundance, turbulent velocity, and redshift are left free.
The redshift which is obtained from the fit is then compared with the known Perseus redshift
\citep[$z=$0.01756 or $cz=$5264~km\,s$^{-1}$: ][]{ferruit1997} to obtain the best-fit energy shifts.
The rest-frame reference energies implemented in the CIE model in SPEX version 3.03 are calculated values except for Ar\emissiontype{XVII} He$\alpha$,
each retrieved from the references shown in table~\ref{tab:es}.
Some are not the most commonly used calculations or measurements for calibration,
but the differences are usually much smaller than the statistical uncertainties in the present analysis
and thus do not affect the correction results. Detailed comparisons of the reference energies are given in appendix~\ref{app:esdiff} (table~\ref{tab:esdiff}).
For the instrumental Si lines, the relative normalization of the double Gaussians is fixed at the known value \citep{scofield1974},
and the obtained redshift is directly converted to the energy shift.
As shown in figure~\ref{fig:gain} and table~\ref{tab:es}, these shifts appear to be $-$(1--3)~eV below 4~keV and above 7~keV,
and $+$(0--2)~eV in 4--7~keV. These differences cannot be justified by an astrophysical model --- the $\sim$2~eV differences between the
Si Ly$\alpha$ and Fe He$\alpha$ lines correspond to 300 km\,s$^{-1}$, while they are partially formed at similar temperatures.
More importantly, at the high-energy side, there is a several~eV difference in the Rydberg series of Fe\emissiontype{XXV}, which is even harder
to explain with a realistic astrophysical model. Furthermore, the energy-scale shifts at the instrumental Si K$\alpha$ lines
are in good agreement with the parabolic trend of the astrophysical lines, providing further support for a non-astrophysical explanation.

The behavior of these deviations is consistent with calibration issues \citep{eckart2017}. As shown in figure~\ref{fig:gain},
we perform an empirical fit using a parabolic function to the observed deviations ($\delta E$). The correction to the original energy $E$ (keV) is given as
\begin{equation}
c_1 \cdot (E - E_0) + c_2 \cdot (E - E_0)^{2} \mbox{~~eV}.
\end{equation}
$E_0$ is the original energy at the Fe He$\alpha$ line, where the shift is fixed at zero as it has already been corrected by removal of the spacial velocity gradient.
The best-fit values of the coefficients are ($c_1$, $c_2$) = (0.4062, 0.2281), (0.4882, 0.2360), and (0.6525, 0.2793) for Obs~2, 3, and 4, respectively.
Similar corrections have been applied in other Hitomi Collaboration papers (Z, T, AGN, and RS~papers).

We caution that this empirical correction is not to be used outside of
the range of the fit or trusted at the extremes of that range.  Because
there is no mechanism for an offset in the energy scale, the error must
eventually tend to 0 at the lowest energies.

\subsection{Effective-area correction factor\label{app:fudge}}

\begin{figure}[!htbp]
\includegraphics[width=8cm]{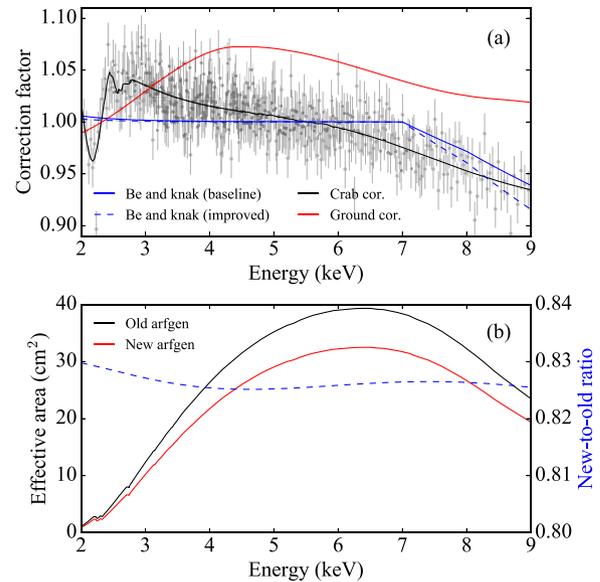}
\caption{(a) A comparison of effective area correction factors: the original correction
(blue) for each of the baseline and improved models respectively in solid and dashed curves,
the on-ground calibration correction (red), and the Crab correction (black) derived by
smoothing the SXS-to-canonical ratio (grey crosses: private communication with M.~Tsujimoto, 2017).
For the Crab analysis, the high-primary as well as mid-primary grade events are used.
(b) Effective area with the old and the new \texttt{aharfgen} in black and red, respectively,
as well as the ratio of the two calculations
with the dashed blue curve.}
\label{fig:newareacor}
\end{figure}

To identify and remove any possible residual calibration errors on the effective area,
which affects mostly the continuum spectrum, we incorporate two correction functions
in the broad-band spectral analysis. One represents uncertainty in the thickness of
the Be window of the gate valve, another the uncertainty in the effective area of the X-ray mirrors of the SXS.
To estimate the size of these factors, the Hitomi SXS spectrum is rebinned into 100-eV bins to enhance
the continuum features. We then fit it with the baseline model described later in
section~\ref{sect:baseline} incorporating a \textsl{knak} component which determines
the correction function using piece-wise power-laws in energy-correction factor
space, together with a neutral-Be absorption model.

By making several iterations between a fit with 100-eV wide bins and a fit with the optimal
binning (see appendix~\ref{app:binning}), the best-fit corrections and
Be model are determined as shown with the solid curve in figure~\ref{fig:newareacor} (a).
The fit prefers a negative absorption column of the Be model, which indicates that the actual
thickness of the Be window might be slightly lower than the value (262~$\mu$m) used in the current calibration.
The correction, however, approaches to unity with a more realistic spectral modeling of the ICM
(an improved model; see section~\ref{sect:improve}), as shown with the dashed curve in figure~\ref{fig:newareacor} (a).
Therefore the thinner Be window preferred with the baseline model is most likely due to incomplete modeling of the ICM emission.

The best-fit effective-area correction function is consistent with unity at $\leq$7~keV, and
decreases to 0.9 at $\sim$9~keV. This means that the current calibration might
be underestimated by $\leq$10\% at the high-energy end of the standard SXS bandpass.
This correction above 7~keV is more significant with an improved model.
We discuss the effective-area corrections in more detail in section~\ref{sect:arfsys}.
Note that the sharp change at 7~keV is caused by the model grids;
changing the grids has a negligible effect on the fitted parameters.

\subsection{Binning of the data\label{app:binning}}

For the binning of our X-ray data, we have followed the approach of
\citet{kaastra2016} for optimal binning. The optimal bin size depends on the
spectral resolution, number of resolution bins and local intensity of the
spectrum and is different for each energy. It is achieved by issuing the
\textsl{obin} command in SPEX. Since we started with spectrum with 0.5-eV
resolution bin, our optimal bin size is a multiple of 0.5~eV. In practice, for most
energies below 8.2~keV we use a bin size of 1.5~eV, and for higher energies 2~eV.
The exceptions are near the S\emissiontype{XV} He$\alpha$ complex where we use 1~eV,
and near the high-$n$ transition lines of Fe\emissiontype{XXV},
including He$\gamma$, He$\delta$, and He$\varepsilon$ lines where we also use 1.5-eV data bins.

\section{Reference line energies for the energy-scale correction\label{app:esdiff}}

Table~\ref{tab:esdiff} compares the reference energies of
emission lines in SPEX used in the SXS energy-scale correction
(appendix~\ref{app:escalecorr}) to the NIST Atomic Spectra Database version 5.3
\citep{kramida2016} and other available measurements and calculations.

In the case of hydrogenic ions, we use the calculations of \citet{erickson1977}.
This is in contrast to \cite{hitomi2016},
where for H-like Fe\emissiontype{XXVI} the calculations of \cite{johnson:wr1985a} were used.
The up-to-date calculations of \cite{yerokhin2015} agree well with \cite{johnson:wr1985a}.
Although \cite{johnson:wr1985a} are the accepted standard for $n=$2 to 1 transitions
in H-like ions (Ly$\alpha$) and have been well tested \citep{beiersdorfer2009},
they do not include transitions from higher Rydberg states with $n\ge$3.
The calculations are in good agreement between \cite{erickson1977} and \cite{johnson:wr1985a}
within 0.02~eV for $Z\leq$20.
For Ly$\alpha_{1}$ of Fe\emissiontype{XXVI}, \cite{erickson1977} is 0.11~eV less.
For consistency, we use the values from \cite{erickson1977}.

For $n=$2 to 1 transitions in He-like ions (He$\alpha$) of Cr and Fe, Ca and Ni, and S,
we respectively use the calculated values of \citet{shirai2000}, \citet{sugar1985},
and \citet{kaufman1993}. For Ar He$\alpha$, we use the measurement of \citet{kelly1987}.
These values are in good agreements within 0.2~eV with the calculations of
\citet{artemyev2005} as well as \citet{cheng1994}.
An exception is found in the Ni He$\alpha$ line, whose deviation is $+$0.47~eV.
Recent update by \citet{natarajan2013} gives a better agreement for Ni He$\alpha$ with the SPEX value.
These calculations have also been compared to many measured values
\citep{beiersdorfer2015}, and good agreement is found.
We also note the work of \cite{drake1988} which has been used often as a calibration standard.

For the 1s--3p transitions of Fe\emissiontype{XXV} (He$\beta_1$) and Fe\emissiontype{XXIV} \citep[j$_3$ satellite in ][]{phillips2008},
we respectively use the calculation of \citet{sugar1985} and a FAC calculation (private communication with A.~J.~J. Raassen, 2017).
\cite{smith1993} performed both calculations and measurements of these 1s--3p lines of Fe ions.
The calculated values are in good agreements with those used in SPEX within 0.1~eV.
On the other hand, the measured values have relatively large deviations ($-$0.45 and $-$0.9~eV, respectively)
from the calculations, which may be due to the limited wavelength calibration
of the crystal spectrometer, as noted in \citet{smith1993}.

For the other high-$n$ Rydberg series (Ca\emissiontype{XIX} He$\beta$ and Fe\emissiontype{XXV} He$\beta$--$\delta$),
we use the calculated values from \citet{sugar1985}.
The measurements of the high-$n$ lines of He-like Fe\emissiontype{XXV} have been conducted by
\cite{indelicato1986} and \cite{beiersdorfer1989}, respectively.
These agree well with the SPEX values within the measurement errors
($\pm$0.15~eV for He$\gamma_1$ and $\pm$0.22~eV for He$\delta_1$).

\section{Systematic factors due to instrumental effects\label{app:instrumental}}

In this section we discuss the effects of several aspects of the
instrumental calibrations on the derived parameters.

\subsection{Velocity-gradient correction\label{sect:novelcorr}}

The line broadening due to spatial bulk velocity of the ICM is removed by applying
an energy-scale correction to each pixel (section~\ref{sect:reduction}). Without
this correction, the C-statistic obtained with the baseline model increases by $\delta C_{\rm stat}=$62,
and the LOS turbulent velocity dispersion becomes larger by 13~km~s$^{-1}$ (``No vel. cor.'' in table~\ref{tab:parameters}).
The best-fit line center shifts towards shorter wavelength by 23~km~s$^{-1}$.

\subsection{Response matrices\label{sect:sxlrmf}}

We also test how much the fit changes by using a small RMF with only the
Gaussian core component, as well as by using an extra-large RMF with the electron-loss continuum\footnote{The current version of SPEX (3.03.00) is not fully compatible with the extra-large--size RMF because of its complexity. Here we apply a local fix to the incompatibility, which will be publicly available in the next release (version 3.03.01 or later).}.
As shown in table~\ref{tab:parameters}, a small RMF improves the baseline fit by $\delta C_{\rm stat}=$4,
while an extra-large RMF (listed as ``XL RMF'') gives instead a poorer fit with $\delta C_{\rm stat}=$12.
The changes on the best-fit temperature and abundances by the RMF-type selection are 1--2\%.

\subsection{Non-X-ray background\label{sect:nonxb}}

The NXB rate depends on the orbital history of the satellite.
Although this effect is already taken into account in \texttt{sxsnxbgen},
the systematic uncertainty could be large if the orbital history is biased.
For the case of the Perseus observations (Obs\,2--4), this systematic is expected to be small
as the on-source time ($\approx$290~ks) is much longer than the satellite orbital period.
Indeed, the estimated NXB rate is 3.0$\times$10$^{-2}$~counts~s$^{-1}$~cm$^{-2}$ in 1.0--10~keV,
consistent with the orbit-averaged value \citep{kilbourne2017}.
This converts to 0.4\% of the total count rate of the source events in 1.9--9.5~keV.
Here, we consider an extreme case where we completely ignore the NXB contribution.
As shown in table~\ref{tab:parameters}, the baseline run without the NXB component gives
a larger C-statistic value ($\delta C_{\rm stat}=$9) than the original run,
and the impact on the fitted parameters is minor, $\leq$1\% on temperature and abundances.

\subsection{Effective area\label{sect:arfsys}}

\subsubsection{Point-source ARFs\label{sect:psarf}}

The spatial extent of the target has an impact on the instrumental response. As shown in table~\ref{tab:parameters}
(labeled as ``PS ARF''), the use of the point-source ARFs not only on the AGN component
but also on the ICM component of the Perseus cluster gives a larger C-statistic value ($\delta C_{\rm stat}=$30)
than the original baseline fit. The improper ARF would lead to a 1\% bias on temperature and up to 5\% biases on abundances.

\subsubsection{No effective-area correction factor\label{sect:fudgeeff}}

The correction factor (appendix~\ref{app:fudge}) is included to remove potential
calibration uncertainties on the effective area. As shown in
table~\ref{tab:parameters} (labeled as ``No cor.''), ignoring the correction factor yields a poorer fit
($\delta C_{\rm stat}=$38). The temperature shifts by 0.08~keV from the original value,
and several abundances are underestimated by 0.03 times solar. The emission measure
changes by 1.3\%, larger than the statistical error by a factor of 5. This indicates that
the correction factor, despite being small, is still needed for the current calibration.

\subsubsection{Correction factor based on on-ground calibration\label{sect:offfudge}}

The baseline effective area correction is done with the SPEX model \textsl{knak}.
Alternatively, the correction can be achieved by setting \texttt{auxtransfile=CALDB}
in the \texttt{aharfgen} run, which applies an additional empirical transmission on the
original ARFs. With this correction, the discrepancy in the mirror effective area between
on-ground calibration measurements and ray-tracing simulations are removed.
We hence re-run the baseline fit by including the new correction factor while
turning off the \textsl{knak} and Be-filter fine-tuning (appendix~\ref{app:fudge}).
The original and new correction factors are respectively shown with the blue and red curves in figure~\ref{fig:newareacor} (a).
A poorer fit ($\delta C_{\rm stat}=$191; ``Ground cor.'' in table~\ref{tab:parameters}) is obtained with
the new correction, and the best-fit CIE emission measure changes by 4\%.
The temperature decreases by 0.12~keV and the abundances changes by $\leq$0.06~solar from the original values.

\subsubsection{Correction using the Crab observation}

The third method to evaluate and correct systematic uncertainty in effective area is to use standard candles.
The Crab is one of the broadly used reference sources for effective area calibration.
Using a 9.7-ks observation, \citet{tsujimoto2017b} found that the SXS spectrum of the Crab showed a systematic deviation from
the canonical model. The deviation, defined as the SXS-to-canonical ratio, behaves differently
depending on the energy bands, but in the 3--9~keV band it shows a monotonically decreasing trend
on energy within $\pm$5\%. This is reminiscent of our effective area correction with \textsl{knak} as
shown in figure~\ref{fig:newareacor} (a). Therefore, we re-run the baseline fit with the Crab ratio as
an effective-area correction instead of the original factor in appendix~\ref{app:fudge}.

As shown table~\ref{tab:parameters} (``Crab cor.''), the Crab correction does improve the fit
from the no-correction case (``No cor.'') by $\delta C_{\rm stat}=$25. This accounts for two-third of
the improvement by using the original correction factor ($\delta C_{\rm stat}=$38).
The slightly worse fit than the baseline model might be attributable to the different observation
configurations between Perseus and Crab: different pixel contributions and event-grade selections due to
different spatial extents of the sources and incoming photon rates.

\subsubsection{ARF with the latest \textsl{aharfgen}\label{sect:newarf}}

There has been released a newer version (006) of the Hitomi software
including an updated ray-tracing ARF generator \texttt{aharfgen},
in which a bug in the coordinates calculation for an input image is corrected.
A comparison of effective areas between the old and new tools,
as well as the new-to-old area ratio are shown in figure~\ref{fig:newareacor} (b).
The ratio curve has an almost constant, smooth structure over the fitting range without any line- or edge-like features.
The notable difference is rather in the total effective area.
The $\sim$20\% lower area results in a comparable amount of increase in the ICM emission measure (``New arfgen'' in table~\ref{tab:parameters}).
Although the new ARF marginally improves the fit, any changes in the best-fit values of the other parameters are less than 0.1\%,
justifying the use of the old ARF in the baseline model for the current purpose.

\subsection{Effects of the gain correction factor \label{sect:gaineff}}

The correction on energy scale is crucial for fitting the emission lines. Once
it is removed, the fitting with the baseline model becomes worse by $\delta C_{\rm stat} =
627$, and  the Si, S, and Fe abundances are affected up to 15\%. The
temperature and line broadening are not affected by the energy-scale correction.

\section{Hitomi SXS spectral fits with different codes\label{app:speccomp}}

\begin{figure*}[!htbp]
 \begin{center}
\includegraphics[width=16cm]{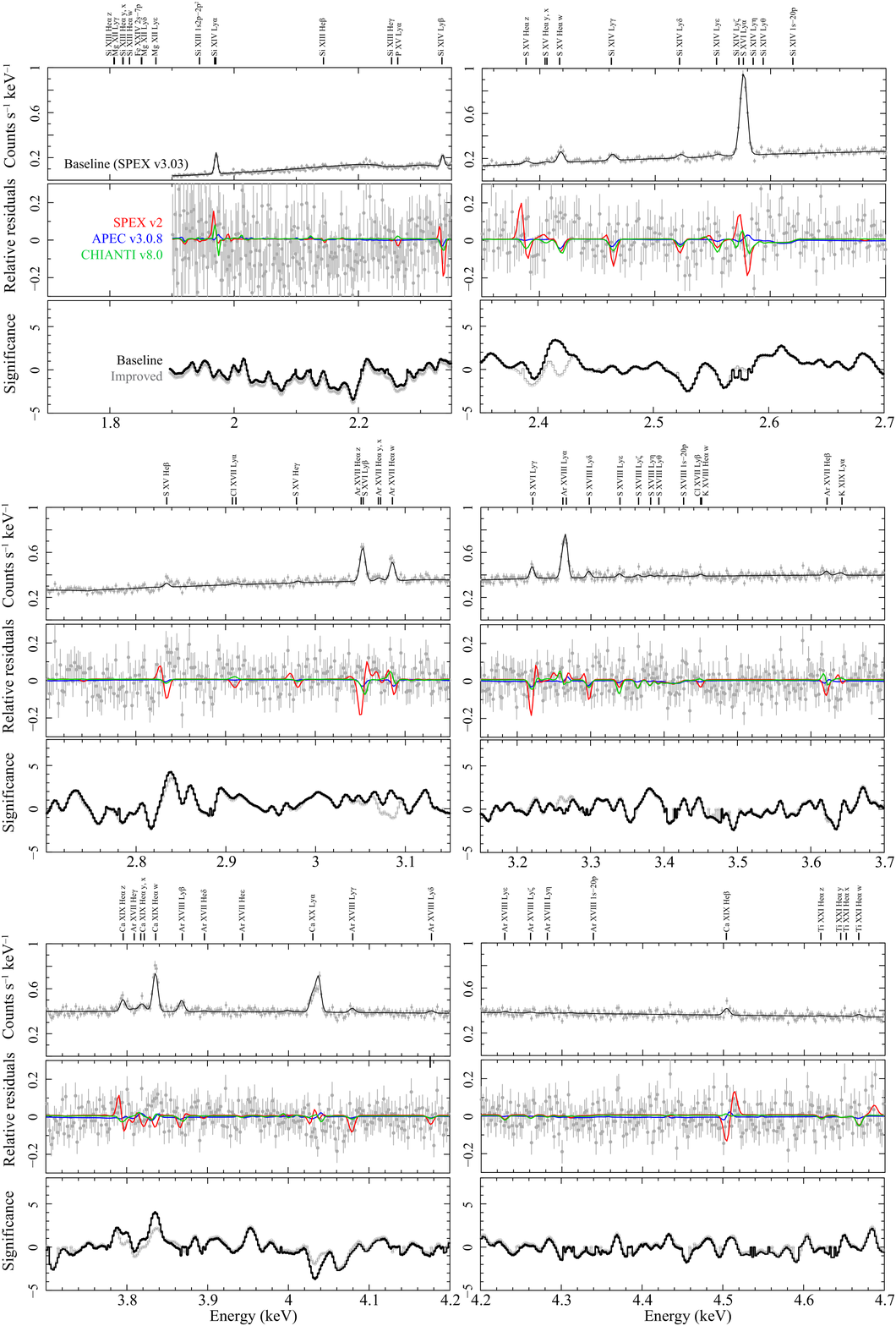}
 \end{center}
\caption{The SXS spectrum of the Perseus cluster in 1.9--4.7~keV fitted with different models.
Panels in each subfigure show (upper) fit to the data with the baseline model,
(middle) relative difference between the baseline model and the best-fit models with APEC v3.0.8 in blue,
SPEX v2 in red, and CHIANTI v8.0 in green, and (lower) significances in $\sigma$ of an additional \textsl{line} model
at each energy on top of the baseline model in black and improved model in grey.}
\label{fig:spec1}
\end{figure*}

\begin{figure*}[!htbp]
 \begin{center}
\includegraphics[width=16cm]{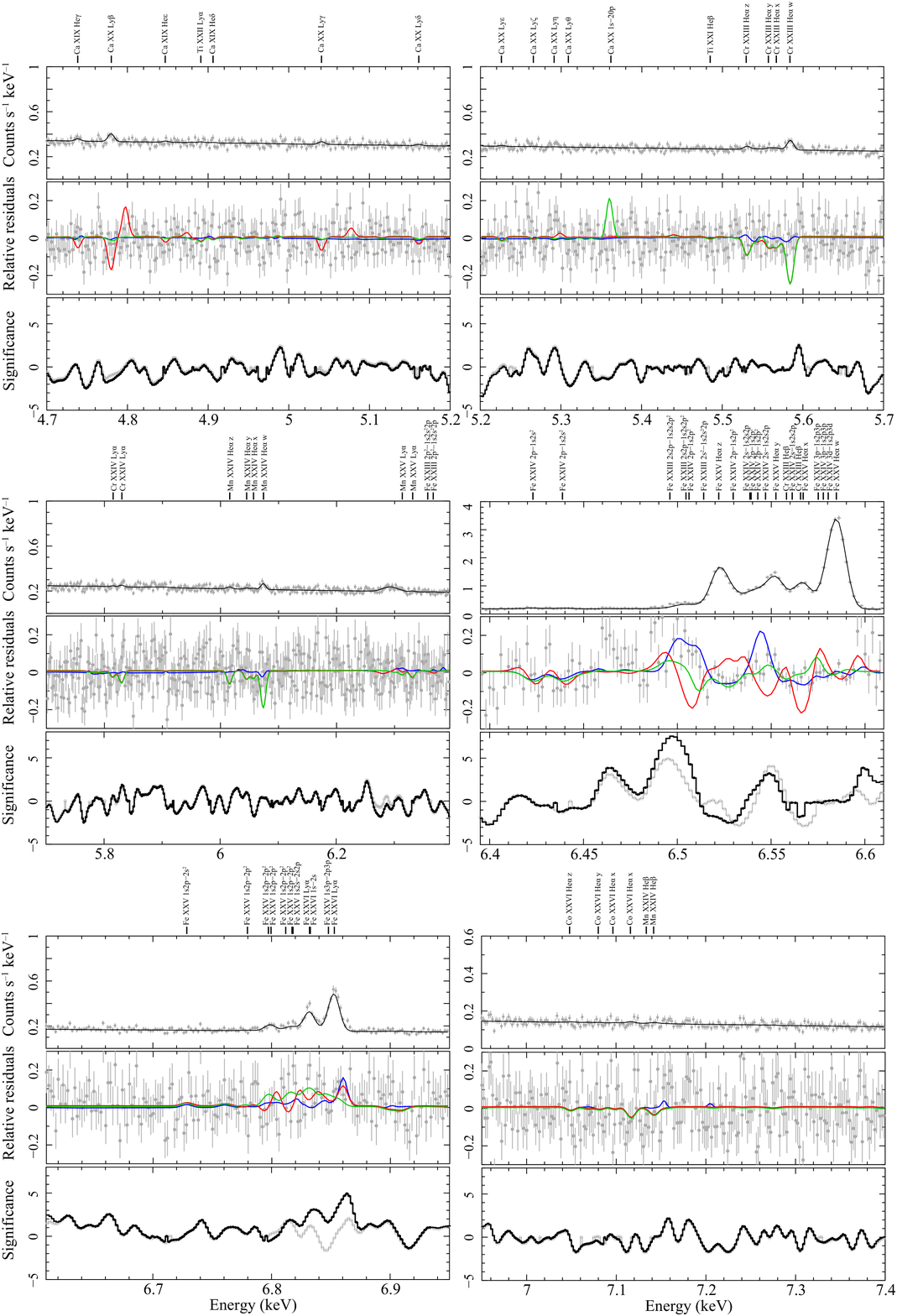}
 \end{center}
\caption{The same as figure~\ref{fig:spec1}, but in 4.7--7.4~keV.}
\label{fig:spec2}
\end{figure*}

\begin{figure*}[!htbp]
 \begin{center}
\includegraphics[width=16cm]{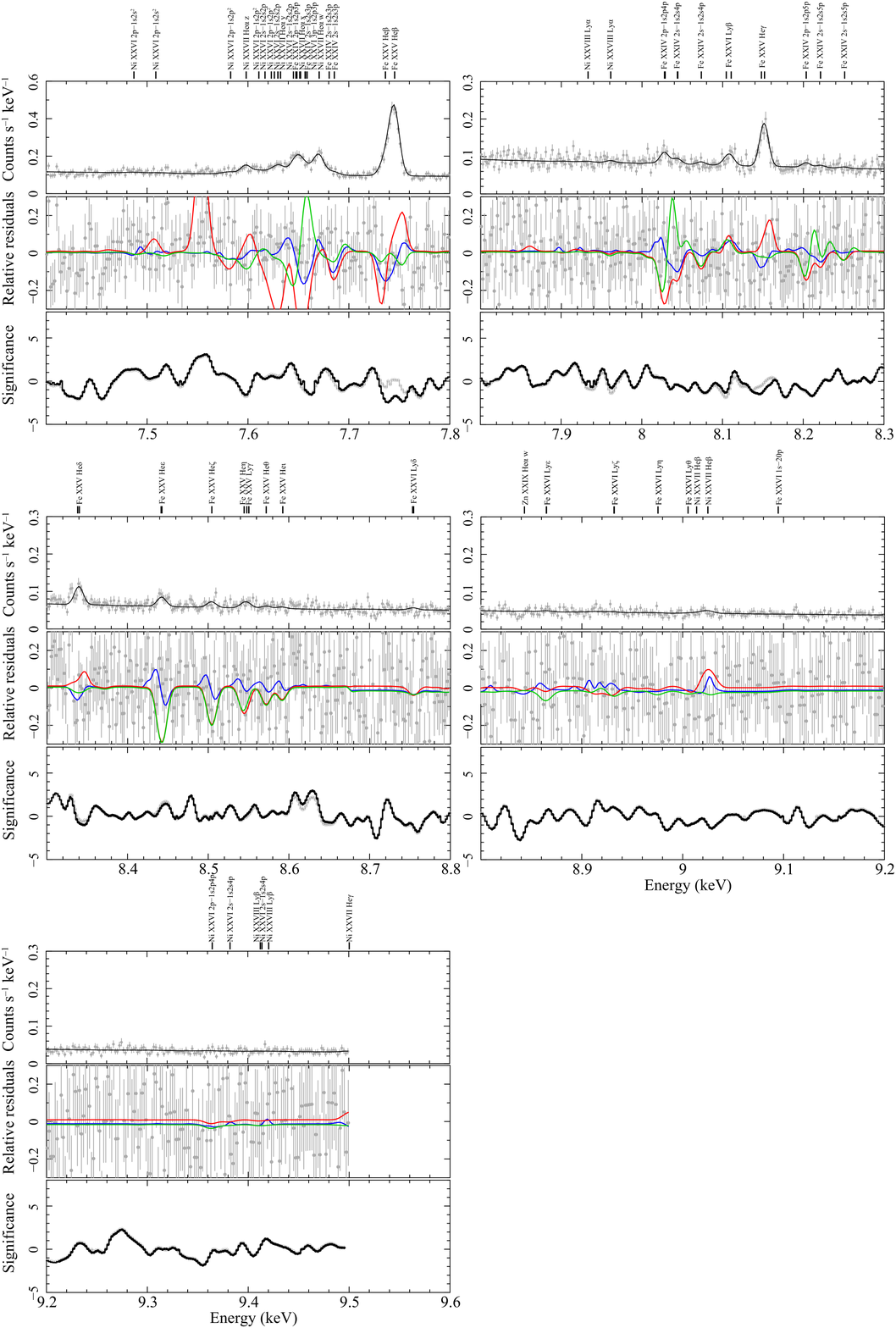}
 \end{center}
\caption{The same as figure~\ref{fig:spec1}, but in 7.4--9.5~keV.}
\label{fig:spec3}
\end{figure*}

Figures~\ref{fig:spec1}--\ref{fig:spec3} show the full-band (1.9--9.5~keV) Hitomi SXS spectrum
with the best-fit baseline model using SPEX version 3.03 and the relative differences
of the best-fit models obtained with various other plasma models.
See section~\ref{sect:plasmacode} for details.

\section{List of emission lines\label{app:linelist}}

Tables~\ref{tab:wavelength} and \ref{tab:satellite} respectively compare
the line energies $E$ of the strongest lines and satellite lines in the observed spectrum
between different plasma codes.
Einstein coefficient values $A$ and emissivities $\varepsilon$ are also compared.
The emissivity thresholds are 10$^{-26}$~photons~m$^3$~s$^{-1}$ and
10$^{-25}$~photons~m$^3$~s$^{-1}$, respectively.
See section~\ref{sect:plasmacode} for details.

\begin{landscape}
\setlength{\tabcolsep}{2pt}
\begin{table*}[!htbp]
\setlength{\textwidth}{230mm}
\caption{Comparison of energies of emission lines used in the energy-scale correction.}
\label{tab:esdiff}
\rightline{
\begin{tabular}{@{\extracolsep{2pt}}lllccccccccccccccc}
\hline
Ion                    & \multicolumn{2}{l}{Principal line}                                       & \multicolumn{5}{c}{Rest-frame energy (eV)\footnotemark[$*$]} & \multicolumn{4}{c}{Deviation from SPEX (eV)}  & $\sigma_{\rm SXS}$\footnotemark[$\dagger$] & \multicolumn{2}{c}{Ref.\footnotemark[$\ddagger$]} \\ \cline{2-3} \cline{4-8} \cline{9-12} \cline{14-15}
                       & Name          & Transition                                          &  SPEX\footnotemark[\S] & \multicolumn{2}{c}{NIST v5} &\multicolumn{2}{c}{Others} &\multicolumn{2}{c}{NIST v5} &\multicolumn{2}{c}{Others}&   (eV)   & NIST v5 & Others \\ \cline{5-6} \cline{7-8} \cline{9-10} \cline{11-12}
                       &               &                                                     &    v3.03 &      Meas. &       Calc. &      Meas. &       Calc. &        Meas. &         Calc. &     Meas. &        Calc. &          &     M, C     &    M, C  \\
\hline
Si\emissiontype{XIV}   & Ly$\alpha_1$  & 1s ($^2$S$_{1/2}$) -- 2p ($^2$P$_{3/2}$)              &  2006.08 &        --- &         --- &        --- &     2006.08 &          --- &           --- &       --- &         0.00 &     0.44 &     ---, --- & ---, (1) \\
Si\emissiontype{XIV}   & Ly$\beta_1$   & 1s ($^2$S$_{1/2}$) -- 3p ($^2$P$_{3/2}$)              &  2376.62 &        --- &         --- &        --- &         --- &          --- &           --- &       --- &          --- &     0.79 &     ---, --- & ---, --- \\
S\emissiontype{XV}     & He$\alpha$ w  & 1s$^2$ ($^1$S$_0$) -- 1s.2p ($^1$P$_{1}$)             &  2460.63 &    2460.66 &     2460.62 &    2460.63 &    2460.629 &         0.03 &       $-$0.01 &      0.00 &         0.00 &     0.76 &     (2), (3) & (4), (5) \\
Si\emissiontype{XIV}   & Ly$\gamma_1$  & 1s ($^2$S$_{1/2}$) -- 4p ($^2$P$_{3/2}$)              &  2506.37 &        --- &         --- &        --- &         --- &          --- &           --- &       --- &          --- &     1.15 &     ---, --- & ---, --- \\
S\emissiontype{XVI}    & Ly$\alpha_1$  & 1s ($^2$S$_{1/2}$) -- 2p ($^2$P$_{3/2}$)              &  2622.69 &     2622.6 &      2622.7 &   2622.704 &      2622.7 &       $-$0.1 &           0.0 &      0.01 &          0.0 &     0.22 &     (2), (6) & (4), (1) \\
S\emissiontype{XVI}    & Ly$\beta_1$   & 1s ($^2$S$_{1/2}$) -- 3p ($^2$P$_{3/2}$)              &  3106.74 &        --- &     3106.75 &        --- &         --- &          --- &          0.01 &       --- &          --- &     0.47 &     ---, (7) & ---, --- \\
Ar\emissiontype{XVII}  & He$\alpha$ w  & 1s$^2$ ($^1$S$_0$) -- 1s.2p ($^1$P$_1$)               &  3139.77 &   3139.585 &    3139.583 &    3139.58 &    3139.582 &      $-$0.19 &       $-$0.19 &   $-$0.19 &      $-$0.19 &     0.63 &     (8), (9) & (4), (5) \\
S\emissiontype{XVI}    & Ly$\gamma_1$  & 1s ($^2$S$_{1/2}$) -- 4p ($^2$P$_{3/2}$)              &  3276.26 &        --- &     3276.27 &        --- &         --- &          --- &          0.01 &       --- &          --- &     1.12 &     ---, (7) & ---, --- \\
Ar\emissiontype{XVIII} & Ly$\alpha_1$  & 1s ($^2$S$_{1/2}$) -- 2p ($^2$P$_{3/2}$)              &  3322.98 &   3322.990 &    3322.993 &   3322.993 &   3322.9932 &         0.01 &          0.01 &      0.01 &         0.01 &     0.44 &    (10), (1) & (4), (11)\\
S\emissiontype{XVI}    & Ly$\delta_1$  & 1s ($^2$S$_{1/2}$) -- 5p ($^2$P$_{3/2}$)              &  3354.73 &        --- &    3354.736 &        --- &         --- &          --- &          0.01 &       --- &          --- &     2.42 &     ---, (7) & ---, --- \\
Ca\emissiontype{XIX}   & He$\alpha$ w  & 1s$^2$ ($^1$S$_0$) -- 1s.2p ($^1$P$_1$)               &  3902.26 &        --- &         --- &     3902.2 &    3902.378 &          --- &           --- &       0.1 &         0.12 &     0.37 &     ---, --- & (12), (5) \\
Ar\emissiontype{XVIII} & Ly$\beta_1$   & 1s ($^2$S$_{1/2}$) -- 3p ($^2$P$_{3/2}$)              &  3935.71 &        --- &    3935.722 &        --- &         --- &          --- &          0.01 &       --- &          --- &     1.74 &    ---, (13) & ---, --- \\
Ca\emissiontype{XX}    & Ly$\alpha_1$  & 1s ($^2$S$_{1/2}$) -- 2p ($^2$P$_{3/2}$)              &  4107.48 &        --- &         --- &     4107.5 &      4107.5 &          --- &           --- &       0.0 &          0.0 &     0.70 &     ---, --- & (12), (1) \\
Ca\emissiontype{XIX}   & He$\beta_1$   & 1s$^2$ ($^1$S$_0$) -- 1s.3p ($^1$P$_1$)               &  4582.81 &        --- &         --- &        --- &         --- &          --- &           --- &       --- &          --- &     3.44 &     ---, --- & ---, --- \\
Ca\emissiontype{XX}    & Ly$\beta_1$   & 1s ($^2$S$_{1/2}$) -- 3p ($^2$P$_{3/2}$)              &  4864.08 &        --- &         --- &        --- &         --- &          --- &           --- &       --- &          --- &     1.79 &     ---, --- & ---, --- \\
Cr\emissiontype{XXIII} & He$\alpha$ w  & 1s$^2$ ($^1$S$_0$) -- 1s.2p ($^1$P$_1$)               &  5682.05 &        --- &      5681.9 &    5682.32 &    5682.068 &          --- &        $-$0.2 &      0.27 &         0.02 &     1.61 &    ---, (9)  & (14), (5) \\
Fe\emissiontype{XXV}   & He$\alpha$ w  & 1s$^2$ ($^1$S$_0$) -- 1s.2p ($^1$P$_1$)               &  6700.42 &        --- &      6700.0 &    6700.55 &    6700.435 &          --- &        $-$0.4 &      0.13 &         0.02 &     0.09 &    ---, (9)  & (15), (5) \\
Fe\emissiontype{XXVI}  & Ly$\alpha_1$  & 1s ($^2$S$_{1/2}$) -- 2p ($^2$P$_{3/2}$)              &  6973.07 &        --- &    6973.179 &  6972.7348 &     6973.18 &          --- &          0.11 &   $-$0.34 &         0.11 &     0.44 &    ---, (16) & (17), (1) \\
Fe\emissiontype{XXIV}  & j$_3$\footnotemark[$\|$] & 2p ($^2$P$_{3/2}$) -- 1s.2p($^3$P).3p ($^2$D$_{5/2}$) &  7782.52 &        --- &         --- &     7781.6 &      7782.6 &          --- &           --- &    $-$0.9 &          0.1 &     2.18 &    ---, ---  & (18), (18) \\
Ni\emissiontype{XXVII} & He$\alpha$ w  & 1s$^2$ ($^1$S$_0$) -- 1s.2p ($^1$P$_1$)               &  7805.14 &        --- &      7805.1 &     7804.6 &      7805.1 &          --- &           0.0 &    $-$0.5 &          0.0 &     1.31 &     ---, (9) & (19), (20) \\
Fe\emissiontype{XXV}   & He$\beta_1$   & 1s$^2$ ($^1$S$_0$) -- 1s.3p ($^1$P$_1$)               &  7881.12 &        --- &     7881.17 &    7880.67 &      7881.0 &          --- &          0.05 &   $-$0.45 &       $-$0.1 &     0.42 &   ---, (16)  & (18), (18) \\
Fe\emissiontype{XXV}   & He$\gamma_1$  & 1s$^2$ ($^1$S$_0$) -- 1s.4p ($^1$P$_1$)               &  8295.39 &        --- &     8295.48 &    8295.64 &         --- &          --- &          0.09 &      0.25 &          --- &     0.81 &   ---, (16)  & (21), --- \\
Fe\emissiontype{XXV}   & He$\delta_1$  & 1s$^2$ ($^1$S$_0$) -- 1s.5p ($^1$P$_1$)               &  8487.22 &        --- &      8487.3 &    8487.36 &         --- &          --- &           0.1 &      0.14 &          --- &     1.67 &   ---, (16)  & (14), --- \\
\hline
\end{tabular}
}
\begin{tabnote}
\footnotemark[$*$] Line energies in SPEX v3.03 and measured and calculated values in the NIST Atomic Spectra Database v5.3 \citep[NIST v5:][]{kramida2016} and those in other calibration standards. \\
\footnotemark[$\dagger$] Statistical errors in line energy shift measurements with the Hitomi SXS. Those of the three observations (Obs\,2, 3, and 4) are averaged by taking root mean squares. \\
\footnotemark[$\ddagger$] References for the measured (M) and calculated (C) values in NIST v5 and those in other results:
(1)~\cite{johnson1985};
(2)~\cite{schleinkofer1982};
(3)~\cite{aglitsky1988};
(4)~\cite{kubicek2014};
(5)~\cite{artemyev2005}; 
(6)~\cite{aglitskii1974};
(7)~\cite{kaufman1993};
(8)~\cite{bruhns2007};
(9)~\cite{kramida2016};
(10)~\cite{beyer1985};
(11)~\cite{yerokhin2015};
(12)~\cite{rice2014};
(13)~\cite{erickson1977};
(14)~\cite{beiersdorfer1989};
(15)~\cite{rudolph2013a};
(16)~\cite{shirai2000};
(17)~\cite{chantler2007};
(18)~\cite{smith1993};
(19)~\cite{bombarda1988};
(20)~\cite{natarajan2013};
(21)~\cite{indelicato1986}. \\
\footnotemark[\S] For the references for the line energies in SPEX version 3.03, see table~\ref{tab:es}. \\
\footnotemark[$\|$] The 1s--3p analogous to the 1s--2p DR satellite line, j: 2p ($^2$P$_{3/2}$) -- 1s.2p$^2$ ($^2$D$_{5/2}$), labeled by \citet{phillips2008}. \\
\end{tabnote}
\setlength{\textwidth}{170mm}
\end{table*}
\end{landscape}

\setlength{\tabcolsep}{3pt}

\scriptsize
\begin{longtable}{@{}lccccccccccccc}
\caption{Comparisons of energies, transition probabilities, and emissivities of Lyman- and Helium-series lines for flux calculation.}\label{tab:wavelength}
\label{tab:linelist}
\hline
Name & \multicolumn{4}{c}{$E$ (keV)\footnotemark[$*$]} &  $f_E$\footnotemark[$\dagger$] & \multicolumn{3}{c}{$\log_{10} A$~(s$^{-1}$)\footnotemark[$*$]} & $f_A$\footnotemark[$\dagger$] & \multicolumn{3}{c}{$\varepsilon$ (10$^{-23}$~m$^3$~s$^{-1}$)\footnotemark[$*$]} &  $f_{\varepsilon}$\footnotemark[$\dagger$] \\ \cline{2-5} \cline{7-9} \cline{11-13}
 & S3 &  A &  S2 &  C &  (\%) &            S3 &             A &             C &  (\%) &  S3 &  A &  C &  (\%) \\
\hline
\endfirsthead
\hline
Name & \multicolumn{4}{c}{$E$ (keV)\footnotemark[$*$]} &  $f_E$\footnotemark[$\dagger$] & \multicolumn{3}{c}{$\log_{10} A$~(s$^{-1}$)\footnotemark[$*$]} & $f_A$\footnotemark[$\dagger$] & \multicolumn{3}{c}{$\varepsilon$ (10$^{-23}$~m$^3$~s$^{-1}$)\footnotemark[$*$]} &  $f_{\varepsilon}$\footnotemark[$\dagger$] \\ \cline{2-5} \cline{7-9} \cline{11-13}
 & S3 &  A &  S2 &  C &  (\%) &            S3 &             A &             C &  (\%) &  S3 &  A &  C &  (\%) \\
\hline
\endhead
\hline
\multicolumn{14}{l}{\footnotemark[$*$] Energies $E$, Einstein $A$ values, and emissivities $\varepsilon$ in SPEX v3.03 (S3), AtomDB/APEC v3.0.8 (A), SPEX v2 (S2), and CHIANTI v8.0 (C).} \\
\multicolumn{14}{l}{\footnotemark[$\dagger$] Coefficient of variation (standard deviation divided by the mean) on energies $E$, $A$ values, and emissivities $\varepsilon$.} \\
\endfoot
Si\emissiontype{XIV} Ly$\alpha_2$       &       2.00432 &      2.00432 &      2.00395 &      2.00434 &  0.008 &                13.3827 &               13.3811 &               13.3831 &    0.2 &                   0.905 &                  0.856 &                  1.140 &               12.8 \\
Si\emissiontype{XIV} Ly$\alpha_1$       &       2.00608 &      2.00608 &      2.00589 &      2.00609 &  0.004 &                13.3836 &               13.3822 &               13.3845 &    0.2 &                   1.808 &                  1.706 &                  2.270 &               12.7 \\
Si\emissiontype{XIV} Ly$\beta_2$        &       2.37610 &      2.37610 &      2.37518 &      2.37609 &  0.017 &                12.8096 &               12.8061 &               12.8098 &    0.4 &                   0.147 &                  0.132 &                  0.150 &                5.5 \\
Si\emissiontype{XIV} Ly$\beta_1$        &       2.37662 &      2.37662 &    (2.37518) &      2.37663 &  0.026 &                12.8099 &               12.8094 &               12.8106 &    0.1 &                   0.294 &                  0.264 &                  0.300 &                5.5 \\
Si\emissiontype{XIV} Ly$\gamma_2$       &       2.50615 &      2.50615 &      2.50473 &      2.50615 &  0.025 &                12.4201 &               12.4156 &               12.4201 &    0.5 &                   0.054 &                  0.044 &                  0.048 &                8.4 \\
Si\emissiontype{XIV} Ly$\gamma_1$       &       2.50637 &      2.50637 &    (2.50473) &      2.50635 &  0.028 &                12.4200 &               12.4198 &               12.4209 &    0.1 &                   0.108 &                  0.089 &                  0.095 &                8.1 \\
\end{longtable}
\normalsize

\begin{table*}[!htbp]
\caption{The same as table~\ref{tab:linelist}, but for satellite lines.}
\label{tab:satellite}
\centerline{
\scriptsize
\begin{tabular}{lcccccccccccccc}
\hline
\multicolumn{2}{l}{Line}                           & \multicolumn{4}{c}{$E$ (keV)\footnotemark[$*$]} &  $f_E$\footnotemark[$\dagger$] & \multicolumn{3}{c}{$\log_{10} A$~(s$^{-1}$)\footnotemark[$*$]} & $f_A$\footnotemark[$\dagger$] & \multicolumn{3}{c}{$\varepsilon$ (10$^{-23}$~m$^3$~s$^{-1}$)\footnotemark[$*$]} &  $f_{\varepsilon}$\footnotemark[$\dagger$] \\ \cline{3-6} \cline{8-10} \cline{12-14}
Transition                                         & Key\footnotemark[$\ddagger$] & S3 &  A &  S2 &  C &  (\%) &            S3 &             A &             C &  (\%) &  S3 &  A &  C &  (\%) \\\hline
\textsl{~L-shell lines} \\
\multicolumn{1}{c}{\dotfill He-like Fe\emissiontype{XXV} \dotfill}\\
1s.2s $^3$S$_1$ -- 1s.5p $^3$P$_2$                 &            &      1.84960 &      1.84966 &           --- &      1.84971 &  0.002 &                12.2967 &               12.1875 &               12.2487 &   10.2 &                    0.014 &                   0.011 &                   0.011 &              11.8 \\
\multicolumn{1}{c}{\dotfill Li-like Fe\emissiontype{XXIV}\dotfill}\\
2p $^2$P$_{3/2}$ -- 7s $^2$S$_{1/2}$               &            &      1.81905 &      1.81893 &           --- &      1.81990 &  0.024 &                11.1216 &               11.1216 &               10.8681 &   24.4 &                    0.012 &                   0.009 &                    --- &               14.3 \\
2p $^2$P$_{3/2}$ -- 7d $^2$D$_{5/2}$               &            &      1.82117 &      1.82679 &           --- &      1.82107 &  0.147 &                12.0434 &               12.0430 &               12.0043 &    4.2 &                    0.028 &                   0.030 &                   0.006 &              51.0 \\
\hline
\end{tabular}
\normalsize
}
\begin{tabnote}
\footnotemark[$*$] Energies $E$, $A$ values, and emissivities $\varepsilon$ in SPEX v3.03 (S3), AtomDB/APEC v3.0.8 (A), SPEX v2 (S2), and CHIANTI v8.0 (C). \\
\footnotemark[$\dagger$] Coefficient of variation (standard deviation divided by the mean) on energies $E$, $A$ values, and emissivities $\varepsilon$. \\
\footnotemark[$\ddagger$] Key letters for K-shell lines. For 1s--2p and 1s--2s Li-like satellite lines
(1s$^2.nl$--1s.2p.$nl$ and 1s$^2.nl$--1s.2s.$nl$), notations are by \citet{gabriel1972}, \citet{bely-dubau1979}, and \citet{bely-dubau1982}
for $n= $2 (a--v), 3 (a1--i11), and 4 (j1--u30), respectively.
Those for Li-like satellite lines but with $\Delta n\geq $2 transitions are by \citet{phillips2008} (a--v with the upper $n$ subscripted).
Key letters for He-like (upper cases) and Be-like (Greeks) satellites respectively follow Safronova's notation and \citet{doschek1981}. \\
\end{tabnote}
\end{table*}

\end{appendix}

\clearpage

\bibliographystyle{aa}
\bibliography{atomic_pasj}

\end{document}